# Low-dimensional magnetocaloric materials for energy-efficient magnetic refrigeration: Does size matter?

Nguyen Thi My Duc [a,b], Hariharan Srikanth [a], and Manh-Huong Phan [a,c]*

[a] Department of Physics, University of South Florida, Tampa, FL, 33620, USA

[b] The University of Danang, University of Science and Education, 459 Ton Duc Thang, Danang, Viet Nam

[c] Center for Materials Innovation and Technology, VinUniversity, Gia Lam district, Hanoi 14000, Viet Nam

The magnetocaloric effect (MCE) provides a promising foundation for the development of solid-state refrigeration technologies that could replace conventional gas compression-based cooling systems. Current research efforts primarily focus on identifying cost-effective magnetic materials that exhibit large MCEs under low magnetic fields across broad temperature ranges, thereby enhancing cooling efficiency. However, practical implementation of magnetic refrigeration requires more than bulk materials; real-world devices demand efficient thermal management and compact, scalable architectures, often achieved through laminate designs or miniaturized geometries. Magnetocaloric materials with reduced dimensionality, such as ribbons, thin films, microwires, and nanostructures, offer distinct advantages, including improved heat exchange, mechanical flexibility, and integration potential. Despite these benefits, a comprehensive understanding of how size, geometry, interfacial effects, strain, and surface phenomena influence the MCE remains limited. This review aims to address these knowledge gaps and provide guidance for the rational design and engineering of magnetocaloric materials tailored for high-performance, energy-efficient magnetic refrigeration systems.

*Corresponding author: phanm@usf.edu

**Table of Contents:**

## 1. Introduction

Cooling technology is essential in modern life, supporting comfort, safety, technological performance, and environmental sustainability [1-3]. Contemporary cooling systems are increasingly focused on reducing energy consumption and minimizing the environmental impact of refrigerants to help mitigate climate change [1,2]. Vapor-compression cooling, based on gas compression and expansion, has long been the dominant method used in applications ranging from household air conditioners and supermarket refrigerators to refrigerated transport [1]. Its popularity stems from its cost-effectiveness and adaptability. However, these systems are energy-intensive, particularly in large-scale applications such as HVAC (Heating, Ventilation, and Air Conditioning) systems in commercial buildings or data centers, contributing significantly to electricity consumption, grid stress, and operational costs. Furthermore, the refrigerants used (e.g., CFCs, HCFCs) either deplete the ozone layer or possess a high global warming potential (GWP), raising serious environmental concerns. In addition, conventional cooling systems often require large, bulky components such as compressors, condensers, and evaporators, limiting their applicability in compact or mobile devices.

These limitations have driven significant research into alternative solid-state cooling technologies [1-5]. One of the most promising among them is magnetic refrigeration, which is based on the magnetocaloric effect (MCE) - a phenomenon in which a magnetic material heats up or cools down when subjected to a changing magnetic field [2,6]. When an external magnetic field is applied, the magnetic moments in the material align, reducing magnetic entropy and increasing the material's temperature. Upon removal of the field, the moments become disordered again, increasing magnetic entropy and leading to cooling. This thermodynamic principle forms the foundation of magnetic refrigeration, offering a potential path toward environmentally friendly and energy-efficient cooling. Figure 1 illustrates the working principle of a magnetic cooling cycle and its advantages over traditional gas-compression systems.

MCE-based refrigeration systems offer several benefits, including the potential to operate without harmful refrigerants and with improved energy efficiency [6]. Since the efficiency of heat exchange in such systems depends on the magnetic entropy change ($\Delta S_M$) of the refrigerant material, materials exhibiting large $\Delta S_M$ are highly desirable [6-12]. This entropy change may be induced via magnetic or magneto-structural phase transitions, provided there is a significant change in magnetization between the two phases. Current research efforts focus on identifying materials that are both cost-effective and exhibit large $\Delta S_M$ under relatively low magnetic fields across a broad temperature range, resulting in a high refrigerant capacity ($RC$) [6-10], a key metric that quantifies the amount of heat transferred between the cold and hot reservoirs in an ideal refrigeration cycle [7].

Among the materials studied, gadolinium (Gd) is widely considered a benchmark for magnetic refrigeration near room temperature. It exhibits a large MCE and a second-order magnetic phase transition around 294 K [7,11]. Gd has been used in proof-of-concept devices demonstrating that magnetic refrigeration is a viable alternative with the potential for up to 30% energy savings compared to conventional methods [1,6]. However, Gd also suffers from several drawbacks that limit its scalability and commercial viability. As a rare-earth element, Gd is expensive and subject to supply chain vulnerabilities. Its moderate thermal conductivity may hinder heat transfer efficiency in packed-bed configurations. Additionally, it is prone to oxidation when exposed to air or moisture, which degrades its long-term performance [7].

In response to these challenges, a wide range of alternative magnetocaloric materials has been developed, including $Gd_5(Ge_{1-x}Si_x)_4$ [13,14], $La(Fe_{1-x}Si_x)_{13}$ [16], $MnAs_{1-x}Sb_x$ [16],

$MnFeP_{1-x}As_x$ [17,18], $Ni_{50}Mn_{50-x}Sn_x$ [19], and $R_{1-x}T_xMnO_3$ ($R$ = La, Pr, Nd; $T$ = Ca, Sr, Ba…) [8,20-22], among others [6,9,10]. While some of these materials show enhanced entropy changes or tunable transition temperatures [6-8], comparative studies have revealed that Gd still remains one of the most suitable candidates for sub-room temperature magnetic refrigeration due to its favorable balance of MCE performance, low hysteresis, and operational simplicity [9].

To realize magnetic cooling in practical devices, however, materials cannot always be used in bulk form [10,23-26]. Real-world systems require efficient thermal management and compact architectures, often implemented through laminate structures or miniaturized geometries. Magnetocaloric materials with reduced dimensionality, including ribbons, thin films, microwires, and nanostructures, offer significant advantages over their bulk counterparts in this regard [23-26]. For example, magnetocaloric ribbons, typically fabricated via rapid solidification (e.g., melt spinning), are thin (tens of microns) and exhibit high surface-area-to-volume ratios, which facilitate rapid heat exchange and efficient coupling with heat transfer fluids [23,24]. Their geometry also enables uniform exposure to magnetic fields and reduced demagnetization effects. Ribbons can be cut, stacked, or shaped to suit various device designs, offering excellent mechanical and processing flexibility. Similarly, thin films offer unique advantages for integration into micro- and nanoscale devices [25,27]. They can be deposited directly onto substrates for on-chip cooling in microelectronics, MEMS, or lab-on-chip platforms. Due to their low thermal mass and fast thermal response, thin films are especially promising for high-frequency cooling cycles. Theoretically, reducing the dimensions of magnetic refrigerants increases the cooling power by enabling higher operational frequencies [28,29]. Meanwhile, magnetocaloric wires, such as Gd alloy-based microwires [30-32], provide mechanical robustness and better control of fluid dynamics compared to spherical or irregular particles. Arrays of aligned wires have been shown to reduce viscous losses, improve temperature span, and enhance heat transfer performance [29]. The ability to assemble wire bundles into laminate configurations makes them suitable for compact and efficient magnetic cooling in MEMS (Micro-electro-mechanical systems) and NEMS (Nano-electro-mechanical systems) applications.

Despite these promising features, a comprehensive understanding of how geometrical constraints, interfaces, strain, and surface effects influence the magnetocaloric response is still lacking. While earlier reviews have focused largely on bulk magnetocaloric materials [6-9] or isolated studies of size effects [10,24-26], this review aims to provide a critical and comparative analysis of advanced magnetocaloric materials in reduced-dimensional forms (ribbons, thin films, nanoparticles, and microwires), highlighting the interplay between structural characteristics and magnetocaloric performance. This discussion is intended to guide the rational design and engineering of advanced magnetocaloric materials for next-generation, energy-efficient magnetic refrigeration technologies.

## 2. Criteria for selecting magnetocaloric materials

### *2.1. Magnetocaloric figures of merit*

Selecting suitable magnetocaloric materials for use in magnetic refrigeration technologies involves balancing thermodynamic performance, physical properties, and practical considerations to ensure optimal efficiency, reliability, and scalability [6,7,23]. The primary requirement for a magnetic refrigerant is a large magnetocaloric effect (MCE) - quantified by the isothermal magnetic entropy change ($\Delta S_M$) and/or the adiabatic temperature change ($\Delta T_{ad}$) - under a relatively low magnetic field over a broad temperature range, resulting in a large refrigerant capacity ($RC$). These parameters determine the material's capacity to transfer heat between thermal reservoirs during a magnetic refrigeration cycle (Fig. 1).

The change in magnetic entropy ($\Delta S_M(T,\mu_0 H)$) induced by varying the external magnetic field from $H = 0$ to $H = H_0$ at a constant temperature, is commonly used to evaluate the MCE and is derived using Maxwell's relation: [7]:

$$\Delta S_M(T,H_0) = S_M(T,H_0) - S_M(T,0) = \mu_0 \int_0^{H_0} \left( \frac{\partial M(T,H)}{\partial T} \right)_H dH. \tag{1}$$

Here, $\mu_0$ is the vacuum permeability and $(\partial M/\partial T)_H$ is the temperature derivative of magnetization at constant field. A material exhibiting a steep change in magnetization near its transition temperature will have a large $(\partial M/\partial T)_H$, and consequently a large $\Delta S_M$ - a desirable feature in magnetic cooling materials.

Moreover, $\Delta S_M$ can be achieved from calorimetric measurements of the field dependence of the heat capacity and subsequent integration:

$$\Delta S_\mathrm{M}(T,H_0) = \int_0^T \frac{C(T,H) - C(T,0)}{T} dT \tag{2}$$

where $C(T, H_0)$ and $C(T,0)$ are the values of the heat capacity measured in the field $H_0$ and in zero magnetic field $H = 0$, respectively.

The adiabatic temperature change $\Delta T_{ad}$ in magnetocaloric materials can be formally described using the thermodynamic Maxwell relation as:

$$\Delta T_{ad} = -\int_0^H \frac{T}{C_P(T,H)} \left( \frac{\partial M(T,H)}{\partial T} \right)_H dH \tag{3}$$

This fundamental equation indicates that $\Delta T_{ad}$ depends on both the heat capacity $C_p(T,H)$ and the temperature derivative of the magnetization $\left(\frac{\partial M}{\partial T}\right)_H$, integrated over the range of applied magnetic field. It provides a complete thermodynamic description of the adiabatic process during magnetization or demagnetization.

The adiabatic temperature change, $\Delta T_{ad}$, at a given temperature $T_0$ can be estimated as:

$$\Delta T_{ad}(T_0, H_0) \cong -\Delta S_M(T_0, H_0)\frac{T_0}{C(T_0, H_0)} \tag{4}$$

where $C(T,H)$ is the specific heat under field. While a large $\Delta S_M$ contributes to a large $\Delta T_{ad}$, the specific heat can vary significantly between materials, meaning that high $\Delta S_M$ does not always guarantee high $\Delta T_{ad}$.

To more comprehensively assess the utility of a magnetocaloric material, its refrigerant capacity (*RC*) is typically considered as [7]:

$$RC = \int_{\mathrm{T_{hot}}}^{\mathrm{T_{cold}}} -\Delta S_{\mathrm{M}}(T)\mathrm{d}T, \tag{5}$$

Another related metric is the relative cooling power (*RCP*), defined as:

$$RCP = -\Delta S_{\mathrm{M}}^{\max}\, \delta T_{\mathrm{FWHM}}, \tag{6}$$

where $\delta T_{\mathrm{FWHM}} = T_{\mathrm{hot}} - T_{\mathrm{cold}}$ is the full width at half maximum of the $\Delta S_M(T)$ peak. Both *RC* and *RCP* provide insight into the effectiveness of a material across a practical temperature span. It is important to note that the *RC* does not correspond to the mechanical work performed during a thermodynamic cycle. Instead, it serves as a thermodynamic performance metric that quantifies the total heat transferred between the cold and hot reservoirs during a magnetization–demagnetization cycle of a magnetocaloric material. Description (Eq. 5) refers to modern active magnetic refrigeration (AMR) cycles, which often utilize stacked magnetic refrigerants arranged in parallel. In these systems, magnetocaloric materials are layered so that each operates optimally within a specific segment of the overall temperature span. Heat transfer fluid flows in parallel through the stack, facilitating efficient thermal exchange across the entire bed. This configuration enhances the regenerative heat transfer process, increasing both the cooling span and the overall efficiency of the system.

Magnetocaloric materials are broadly classified by their magnetic phase transitions. *First-order magnetic transition (FOMT)* materials exhibit an abrupt change in magnetization, often coupled with structural or volume changes. These materials typically show a large $\Delta S_M$ but within a narrow temperature range and often with magnetic or thermal hysteresis, which can reduce efficiency and reversibility [12,19]. *Second-order magnetic transition (SOMT)* materials, on the other hand, undergo a continuous magnetization change near the Curie temperature ($T_C$), without structural changes. Although $\Delta S_M$ is lower than in FOMT materials, SOMT materials such as Gd offer broader operating ranges, minimal hysteresis, and superior thermal and mechanical stability—favorable traits for cyclic refrigeration [6,7,12]. Figure 2 illustrates a general trend of the temperature dependence of magnetic entropy change for both FOMT and SOMT materials. Because hysteresis leads to energy loss and heating, materials with soft magnetic behavior and negligible hysteresis are preferable [23]. SOMT materials are particularly suitable for high-frequency and long-term refrigeration applications.

Additional practical criteria for selecting magnetocaloric materials include: (i) *high thermal conductivity*, to enable rapid and efficient heat exchange; (ii) *low electrical conductivity*, to reduce eddy current losses during dynamic magnetic field cycles, thus preserving the cooling efficiency and enabling more efficient, compact, and faster-operating refrigeration; (iii) *resistance* *to oxidation and corrosion*, to ensure durability under repeated magnetic and thermal cycling; (iv) *mechanical robustness*, essential for long-term stability and reliable device integration; (v) *environmental safety*, requiring materials to be non-toxic and free from hazardous elements; (vi) *cost-effectiveness and availability*, as abundant and low-cost materials are more suitable for large-scale commercialization; and (vii) *formability and processability*, allowing adaptation to diverse device architectures such as thin films, ribbons, and microwires.

### *2.2. Notable advantages of reduced dimensionality*

While bulk materials often suffer from limited heat exchange surface area [6,7], reduced-dimensional forms, such as nanoparticles, thin films, ribbons, or microwires, can offer enhanced heat transfer, flexibility, and integration into compact systems [10,23-30]. Nanoparticles, for example, are particularly attractive for cryogenic and localized cooling due to their scalability and adaptability, and their inherent entropy broadening can contribute to an enhanced RC. Thin films, on the other hand, exhibit strong potential for on-chip and microscale cooling applications, and can be integrated with other physical effects such as thermoelectricity. Ribbons provide high surface area, fast thermal response, and moderate mechanical flexibility, making them promising for rapid heat exchange environments. Microwires further combine a high surface-to-volume ratio with excellent mechanical flexibility, enabling effective wrapping around heat sources and fast thermal coupling with their surroundings. Kuz'min theoretically demonstrated that magnetic refrigerators have an upper operational frequency limit of approximately 200 Hz [28]. This maximum frequency is governed by the minimum delay between switching off the magnetic field and the subsequent transfer of the induced temperature change to the heat exchanger. The key limitation on operational frequency arises from a trade-off between thermal conductivity and viscous friction. Mechanical instability, often caused by flow maldistribution, can also reduce system throughput significantly. Unlike bulk materials, magnetocaloric materials in wire form, especially when arranged in wire bundles within the magnetic bed, are predicted to offer enhanced mechanical stability and lower porosity, making them more suitable for high-frequency operation [28]. D. Vuarnoz and T. Kawanami conducted an extensive analysis of pressure drop, refrigeration capacity, coefficient of performance (COP), and exergy efficiency in a reciprocating active magnetic regenerator (AMR) composed of gadolinium wires [29]. Their findings indicate that smaller wire diameters significantly improve both cooling capacity and COP. This improvement is attributed to the increased heat transfer surface area and reduced interstitial space between wires, which together enhance the convective heat transfer coefficient. For a given wire diameter, an AMR utilizing a wire stack outperforms one with a particle bed, delivering superior overall performance [29]. Nonetheless, it should be noted that the increased surface area in reduced-dimensionality materials can also lead to higher friction, which may offset the benefits of enhanced heat exchange. For practical cooling applications, the trade-off between heat transfer efficiency and pressure drop must be carefully considered, not only in terms of optimizing the size and shape of the magnetocaloric materials, but also with respect to the choice of matrix materials in which magnetocaloric components, such as magnetic nanoparticles or microwires, are embedded. These aspects will be further explored in the next section, where the role of reduced dimensionality and associated effects (e.g., strain, surface/interface phenomena) on the MCE response will be critically examined.

## 3. Magnetocaloric Materials: Reduced Dimensionality Effects

The MCE is influenced differently by reduced dimensionality across various types of magnetic ordering—ferromagnetic, antiferromagnetic, and ferrimagnetic. These effects can differ significantly when comparing bulk materials to low-dimensional forms such as nanoparticles, thin films, ribbons, and microwires. We note herein that the term "low-dimensional materials," as used in this paper, broadly refers to materials with reduced dimensionality, such as thin films, nanoparticles, ribbons, and microwires. It is not intended to be limited solely to atomically thin materials or systems exhibiting dimensional confinement at the atomic scale. Additionally, phase coexistence has been shown to markedly impact the MCE in bulk systems and may interact with reduced dimensionality effects in complex ways [31,33-35]. In this section, we examine how these factors influence the MCE response in each form of reduced-dimensional magnetic material, including nanoparticles, thin films, ribbons, and microwires.

### *3.1. Nanoparticles*

Finite size and surface effects are critical factors that significantly influence the magnetic behavior of nanoparticles compared to their bulk counterparts. Finite size effects stem from the limited number of atoms and reduced dimensions of nanoparticles, typically below 100 nm [36-38]. As particle size decreases, thermal fluctuations become more pronounced, often disrupting magnetic ordering and suppressing long-range magnetic interactions. Consequently, magnetic transition temperatures such as the Curie temperature ($T_C$) or Néel temperature ($T_N$) tend to decrease due to reduced coordination of magnetic atoms and the enhanced surface-to-volume ratio [36]. At sufficiently small sizes, nanoparticles transition to single-domain states, altering coercivity and magnetization reversal behavior, and often giving rise to superparamagnetism [39].

The high surface-to-volume ratio also means that a large fraction of atoms reside at or near the surface, where they experience altered chemical and magnetic environments. These surface atoms have fewer nearest neighbors, leading to broken magnetic exchange bonds and spin frustration or canting, which reduces overall magnetization [36,40-41]. The surface's reduced symmetry enhances magnetic anisotropy, often dominating over bulk contributions and affecting magnetization dynamics. In certain ferro/ferrimagnetic systems, the surface may become magnetically inactive (a so-called "dead layer") or develop distinct magnetic properties, forming core-shell structures—e.g., a ferro/ferrimagnetic core with a spin-glass-like shell [42,40,43]. Such surface-induced spin disorder and enhanced anisotropy can increase coercivity in single-domain nanoparticles.

As a result of these size and surface effects, the magnetocaloric response in nanoparticles often deviates markedly from that in bulk materials (see Table 1). For ferromagnets, reducing particle size generally leads to decreases in $T_C$, saturation magnetization ($M_S$), and $\Delta S_M$ [44-47]. For instance, in Gd, -$\Delta S_M^{\max}$ and *RCP* decrease from 9.45 J/kg·K and 690 J/K (bulk) to 7.73 J/kg·K and 234 J/K (100 nm), and further to 4.47 J/kg·K and 140 J/K (15 nm), with corresponding decreases in $T_C$ from 294 K to 290 K and 288 K (see Fig. 3a) [44-45]. Interestingly, ferromagnetic nanoparticles often exhibit a broader distribution of $\Delta S_M(T)$ compared to their bulk forms, which can sometimes enhance *RCP* despite a lower peak $\Delta S_M$. For example, $Gd_5Si_4$ nanoparticles produced via ball milling show a reduced peak $\Delta S_M$ and a shift to lower temperatures, but the broader $\Delta S_M$ ($T$) profile results in a 75% *RCP* increase [46].

In Co nanoparticles (~50 nm), Poddar *et al.* reported a surface spin order–disorder transition at low temperatures associated with a significant MCE, alongside a superparamagnetic transition with a smaller magnetic entropy change at higher temperatures [48]. Surface spins were further manipulated by Ag shell coatings, forming Co/Ag core-shell structures that altered the MCE response. This illustrates how surface anisotropy and exchange coupling at the core-shell interface can be engineered to tailor magnetocaloric properties for magnetic refrigeration applications. Size reduction also enhances low-temperature MCE in $Eu_8Ga_{16}Ge_{30}$ clathrate nanocrystals prepared by ball milling [49]. For 15 nm particles, -$\Delta S_M^{\max}$ reaches ~10 J/kg·K at 5

K under a 5 T field, attributed to modified interactions between $Eu^{2+}$ ions at distinct crystallographic sites.

Among magnetocaloric nanosystems, manganese oxides have been extensively studied due to their tunable magnetic and magnetocaloric properties via dopant concentration [47,50-53]. Similar to Gd, a trend of decreasing $\Delta S_M$, *RC* (*RCP*), and $T_C$ with decreasing particle size has been observed in $La_{0.6}Ca_{0.4}MnO_3$ [47] and $La_{0.7}Ca_{0.3}MnO_3$ [52]. For the latter, $-\Delta S_M^{max}$ and *RC* reduce from 7.7 J/kg·K and ~270 J/K (bulk) to 4.9 J/kg·K and ~200 J/K (35 nm), and 2.4 J/kg·K and ~150 J/K (15 nm), while $T_C$ drops from 264 K to 260 K and 241 K (see Fig. 3b). The decrease in $\Delta S_M$ correlates with reduced $M_S$, often attributed to surface spin disorder. Lampen *et al.* estimated a 1.2 nm dead magnetic layer in 15 nm $La_{0.7}Ca_{0.3}MnO_3$ nanoparticles based on geometric arguments [52]. Table 1 shows variations in $\Delta S_M$ and *RC* (*RCP*) for samples with nominally identical compositions, likely due to oxygen off-stoichiometry - an important parameter that should be accurately reported in future studies for proper comparison.

In ferrimagnets such as $Fe_3O_4$, $NiFe_2O_4$, and $CoFe_2O_4$, nanosizing typically induces superparamagnetism, with thermal energy overcoming anisotropy barriers, resulting in rapid magnetic moment fluctuations and surface spin freezing at low temperatures [54-57]. $CoFe_2O_4$ nanoparticles exhibit a small $\Delta S_M$ around the blocking temperature, while a larger entropy change occurs below the spin-freezing point [54]. However, the magnitude of $\Delta S_M$ around the blocking temperature is often insufficient for practical refrigeration. This trend is commonly observed in a wide range of ferrite nanoparticle systems.

Some ferrimagnetic nanoparticle systems benefit from surface spin freezing, which increases *M* and $\Delta S_M$ under high magnetic fields. In $Gd_3Fe_5O_{12}$ (gadolinium iron garnet), Phan *et al.* observed that $-\Delta S_M^{max}$ increased from 2.45 J/kg·K at 35 K (bulk) to 4.47 J/kg·K at 5 K for 35 nm nanoparticles under a 3 T field [58]. The enhancement is attributed to both the intrinsic magnetic frustration of the Gd sublattice and surface spin disorder. Applying sufficiently high magnetic fields effectively suppresses these disordered and frustrated spins, leading to a significant change in magnetization and, consequently, a large magnetic entropy change.

In antiferromagnetic nanoparticles, size reduction weakens AFM couplings and can induce weak ferromagnetism at the surface [59-61]. Under strong magnetic fields, AFM order may be suppressed in favor of FM alignment, increasing magnetization and magnetic entropy change. Notable examples include $Tb_2O_3$, $Dy_2O_3$, $Gd_2O_3$, and $Ho_2O_3$ nanoparticles [61]. For $Ho_2O_3$, Boutahar *et al.* reported large $\Delta S_M$ and *RCP* values of 31.9 J/kg·K and 180 J/K, respectively, near $T_N$ ~ 2 K under a 5 T field [61].

In mixed-phase systems containing coexisting FM and AFM regions, nanosizing has been reported to enhance both $\Delta S_M$ and *RC* (*RCP*) [62-63]. Unlike single-phase FM systems where $T_C$, $\Delta S_M$, and *RCP* tend to decrease with reducing particle size, Phan *et al.* observed the opposite trend in mixed-phase $La_{0.35}Pr_{0.275}Ca_{0.375}MnO_3$ nanoparticles (~50 nm) [63]. Here, nanosizing suppressed the AFM state and promoted FM ordering, enabling a large $\Delta S_M$ and *RC* (*RCP*) at relatively low magnetic fields (~2 T). For a 5 T field, *RC* increased from ~61 J/kg (bulk) to ~225 J/kg (nanoparticles), while thermal and magnetic hysteresis losses, due to the FOMT characteristics of the material, were also significantly reduced. The magnetocaloric properties in such systems can be further tuned by adjusting the FM/AFM phase volume fractions, presenting a promising strategy for developing efficient nanostructured magnetocaloric materials.

The MCE has also been investigated in ball-milled nanoparticles of austenitic alloys, such as $(Fe_{70}Ni_{30})_{99-x}Cr_{1+x}$ [64]. The addition of Cr significantly lowers the $T_C$, from 398 K at $x = 0$ to 215 K at $x = 6$. This Cr substitution slightly reduces the maximum magnetic entropy change (- $\Delta S_M^{max}$), from 1.58 J/kg·K to 1.11 J/kg·K under a magnetic field change of 5 T. Other nanoparticle systems studied for their MCE properties [65-94] are also summarized in Table 1.

### *3.2. Thin films*

Similar to magnetic nanoparticles, finite size and surface effects in magnetic thin films significantly influence their magnetic and magnetocaloric properties compared to bulk materials [25,33,95-96]. These effects arise from the reduced dimensionality (typically nanometer-scale thickness) and the high surface-to-volume ratio inherent to thin-film systems. Finite size effects become prominent when film thickness approaches characteristic magnetic length scales, such as the exchange length or domain wall width [25]. Reduced atomic coordination along the thickness direction and enhanced thermal fluctuations in two-dimensional (2D) systems generally lead to a decrease in the $T_C$ or $T_N$ temperature with decreasing film thickness [95-153].

Magnetic properties in thin films are highly sensitive to parameters such as thickness, substrate, deposition method, annealing conditions, and oxygen stoichiometry [25,97,101-106,127,137-140]. For ferromagnetic films, reductions in thickness typically lead to suppression of $T_C$, $M_S$, and $\Delta S_M$. In moderately thick films (>100 nm), these changes are often attributed to disorder from strain relaxation, which introduces defects like dislocations, vacancies, and grain boundaries [25,104,106,134]. However, in ultrathin, coherently strained films, distinguishing the effects of strain on the magnetic properties from intrinsic finite size, surface and interface phenomena remains a topic of debate.

For instance, in Gd thin films, high $\Delta S_M$ values observed in the bulk [25] are maintained in thick films [97] but diminish significantly in thinner layers [25]. Under a 1 T field, -$\Delta S_M^{max}$ drops from 2.8 J/kg·K in bulk to 2.7 J/kg·K at 17 μm thickness and to 1.7 J/kg·K at 30 nm (Fig. 4a), while $T_C$ remains nearly unchanged (~292–294 K). Interestingly, *RCP* increases from 63 J/kg (bulk) to 140 J/kg (17 μm) and 110 J/kg (30 nm), due to broadening of the $\Delta S_M(T)$ - a consequence of dimensionality, surface, and interface effects on the magnetic phase transition. Thin films inherently include two key interfaces—film/substrate and film/capping layer—where atomic coordination is broken, leading to spin canting or non-collinear spin structures that reduce net magnetization. Polarized neutron reflectometry, for example, has revealed suppressed magnetic moments at Gd/W interfaces in Gd(30 nm)/W(5 nm) multilayers, contributing to reduced $\Delta S_M$ compared to bulk Gd [98].

In alloyed thin films like $Gd_{100-x}Co_x$ (100 nm thick), varying the Gd/Co ratio significantly affects both $T_C$ and $\Delta S_M$ [99]. While increasing Co concentration generally raises $T_C$ (except at $x$ = 0), $-\Delta S_M^{max}$ and *RCP* exhibit nonlinear dependencies, peaking for $Gd_{56}Co_{44}$ (see Fig. 4b). Similarly, in $Gd_x(Fe_{10}Co_{90})_{100-x}$ films (90 nm thick), increasing Gd content shifts $T_C$ from 436 K to 558 K, with the $-\Delta S_M^{max}$ observed at $x = 50$.

In contrast to the giant $\Delta S_M$ of 18.4 J/kg·K reported for bulk $Gd_5Si_2Ge_2$ ($T_C = 276$ K) [45], its film analog, $Gd_5Si_{1.3}Ge_{2.7}$, shows a lower $-\Delta S_M^{max}$ (~8.8 J/kg·K at 194 K under 5 T) [100]. This reflects both compositional changes and size effects. Notably, thermal cycling in these films leads to degradation of magnetocaloric performance: after 1000 cycles, $-\Delta S_M^{max}$ drops from 8.1 to 1.52 J/kg·K (see Fig. 5), underscoring a key limitation of FOMT materials in magnetic refrigeration technology [101].

In Heusler alloy films (e.g., $Ni_{53.4}Mn_{33.2}Sn_{13.4}$ and $Ni_{53.2}Mn_{29.2}Co_{7.0}Sn_{10.6}$), decreasing film thickness from 1000 to 360 nm lowers $-\Delta S_M^{max}$ and $T_C$, though high $T_C$ values are retained [102]. While $-\Delta S_M^{max}$ values are modest (≤ 1.2 J/kg·K), these films are still relevant due to their tunability and potential multicaloric applications. Some Heusler alloy films of other compositions exhibit larger $-\Delta S_M^{max}$ values but relatively small *RC* values (see Table 2). It is also noteworthy from Table 2 that while some Heusler alloy films exhibit large magnetic entropy changes near their FOMT temperatures, these changes occur over narrow temperature intervals. As a result, the RC remains relatively low, particularly after accounting for magnetic and thermal hysteresis losses.

Manganese oxide thin films, like $La_{0.7}Ca_{0.3}MnO_3$ (150 nm), also exhibit reduced $T_C$ (235 K vs. 264 K bulk) and $-\Delta S_M^{max}$ (2.75 J/kg·K vs. 7.7 J/kg·K), though *RC* (*RCP*) can improve due to broadened FM-PM transitions [52]. To enhance $\Delta S_M$, Moya *et al.* exploited interfacial strain coupling with $BaTiO_3$ substrates [103]. A sharp $\Delta S_M$ peak, with $-\Delta S_M^{max}$ ~ 9 J/kg·K, was achieved at ~200 K in a 30 nm $La_{0.7}Ca_{0.3}MnO_3$ film, induced by the structural phase transition of $BaTiO_3$ from the rhombohedral (R) to orthorhombic (O) structure at $T_{R\text{-}O}$ ~ 200 K. However, due to the narrow temperature span (~2 K), *RCP* was limited (~18 J/kg). It is worth noticing here that the application of external strain to induce and control the extrinsic MCE in magnetic films underscores the multicaloric nature of the $La_{0.7}Ca_{0.3}MnO_3/BaTiO_3$ heterostructure, suggesting that the cooling efficiency of magnetic refrigerants can be significantly enhanced by simultaneously leveraging multiple external stimuli, such as magnetic fields, electric fields, and mechanical strain.

Substrate-induced epitaxial strain plays a critical role in tuning magnetic and magnetocaloric responses [104-106]. In $La_{0.8}Ca_{0.2}MnO_3$ films grown on $SrTiO_3$ substrates, tensile strain reduces $T_C$ from 210 K to 178 K with decreasing thickness, while enhancing $-\Delta S_M^{max}$ (up to 12.8 J/kg·K) and *RC* (255 J/kg) at 75 nm (Fig. 6a) [106]. Compressive strain (from $LaAlO_3$ substrates) results in reduced $-\Delta S_M^{max}$ and *RC* in $La_{0.8}Ca_{0.2}MnO_3$ films (Fig. 6b). These behaviors illustrate the complex interplay of film thickness, strain, and composition.

Oxygen non-stoichiometry is another key variable, contributing to discrepancies in reported $-\Delta S_M^{max}$ and *RC* values [107-109]. Lampen-Kelley *et al.* showed that oxygen-deficient $EuO_{1-\delta}$ films ($\delta$ = 0–0.09) exhibit altered magnetic transitions and enhanced $-\Delta S_M^{max}$ (up to 6.4

J/kg·K over 2 T) with broad refrigerant capacities (*RC* ~223 J/kg) [110]. Such tunability makes them promising for sub-liquid-nitrogen temperature applications. However, achieving precise control over the oxygen content in these and other manganese oxide thin films remains a significant challenge, particularly when tuning their magnetic and magnetocaloric properties.

In ultrathin films (few monolayers), quantum confinement effects may modify electronic states and exchange interactions. First-principles calculations by Patra *et al.* predict that 2D magnets like $GdSi_2$ and $CrX_3$ ($X$ = F, Cl, Br, and I) could exhibit substantial MCE at cryogenic temperatures, with $-\Delta S_M^{max}$ as high as 22.5 J/kg·K [111]. However, experimental studies are needed to validate this prediction. It is worth noting that in atomically thin magnetic systems, the magnetic signals are typically weak and not easily detectable using standard magnetometry techniques. Consequently, accurately evaluating the MCE performance of these 2D materials is nontrivial and requires careful measurement and analysis.

For antiferromagnetic thin films, reducing thickness and introducing strain can weaken AFM interactions, sometimes inducing ferromagnetic behavior under moderate fields [111-114]. The effect is even more pronounced in mixed-phase films where AFM and FM phases coexist and coupled with each other [112,113]. In such systems, the application of a sufficiently strong magnetic field can induce a transition from AFM to FM order, leading to a significant change in magnetization and, consequently, a large magnetic entropy change, $\Delta S_M$. Zhou *et al.* reported a large $-\Delta S_M^{max}$ of 20 J/kg·K at 320 K under a magnetic field change of 5 T for FeRh thin films, significantly outperforming their bulk counterpart (~12.6 J/kg·K) [33]. Owing to its FOMT nature, FeRh exhibits notable thermal and magnetic hysteresis losses, which can hinder its practical application. However, the incorporation of 3% and 5% Pd effectively shifts the $\Delta S_M(T)$ peaks from 319 K to 281 K and 238 K, respectively, enabling better temperature tuning. The MCE behavior of FeRh thin films can vary significantly depending on the strain induced by the underlying substrate [113,114]. Furthermore, when FeRh is grown on a $BaTiO_3$ substrate, the application of an electric field can be used to modulate both the hysteresis losses and the MCE, making this multiferroic heterostructure a promising candidate for multicaloric cooling applications [114]. Bulk $GdCoO_3$ also exhibits AFM ordering originating from the $Gd^{3+}$ magnetic moments below its $T_N$ of 3.1 K [154]. Under a magnetic field change of 7 T, it shows a large MCE with a $-\Delta S_M^{max}$ of 39.1 J/kg·K, an $\Delta T_{ad}$ of 19.1 K, and a RC of 278 J/kg. This strong MCE arises from the half-filled 4f electronic configuration of $Gd^{3+}$ ions. When a 22 nm-thick $GdCoO_3$ thin film is epitaxially grown on a $LaAlO_3$ (LAO) substrate, the $-\Delta S_M^{max}$ is further enhanced to ~59 J/kg·K, with an increased $RC$ of ~320 J/kg around an elevated $T_N$ of ~3.5 K for the same 7 T field change [115]. Similarly, 100 nm $EuTiO_3$ films demonstrate $-\Delta S_M^{max}$ ~24 J/kg·K and $RC$ ~152 J/kg at ~3 K, compared to 17 J/kg·K and 107 J/kg in bulk under $\mu_0 H$ = 2 T [112]. These enhancements are attributed to strain effects and altered magnetic interactions at the nanoscale, demonstrating how finite-size and interfacial effects can amplify the MCE in antiferromagnetic thin film systems. Both $GdCoO_3$ and $EuTiO_3$ thin films hold strong potential as active cooling materials for NEMS and MEMS operating at cryogenic temperatures, owing to their enhanced magnetocaloric response at the nanoscale.

Multilayer and heterostructure films (e.g., FM/NM or FM/AFM) also demonstrate interfacial effects like proximity-induced magnetism and exchange bias [116-117]. In $BiFeO_3$/LSMO heterostructures, increasing AFM $BiFeO_3$ layer thickness decreases LSMO's $T_C$ and $\Delta S_M$ [118] In $Ni_{80}Fe_{20}/Ni_{67}Cu_{33}/Co_{90}Fe_{10}/Mn_{80}Ir_{20}$ films, increasing spacer thickness of $Ni_{67}Cu_{33}$ decreases $T_C$ but enhances $-\Delta S_M^{max}$ and $RC$ (see Fig. 7), showing how interlayer coupling influences MCE [119].

In thin films exhibiting significant MCE anisotropy, magnetic entropy change can be triggered simply by rotating the material within a constant magnetic field, rather than switching the field on and off [120,121]. This "rotating MCE" approach reduces energy losses associated with magnetic field cycling and enables simpler, more compact device architectures [122-124]. Understanding and harnessing MCE anisotropy is essential for selecting or engineering materials, such as layered structures or textured films, where the anisotropy can be tuned to maximize $\Delta S_M$ along specific crystallographic directions. For example, in $Gd_2NiMnO_6$ thin films, although the $T_C$ remains unaffected by film orientation, $-\Delta S_M^{max}$ varies significantly from 9.84 J/kg·K (out-of-plane) to 21.82 J/kg·K (in-plane), yielding a large rotating entropy change of 11.98 J/kg·K [120]. A similar directional dependence has also been observed in epitaxial Tb films [121].

Lastly, it is worth mentioning that thermal transport in thin films differs from bulk, affecting device-level performance of magnetic refrigeration systems. While thin films offer tunable MCE through dimensionality, strain, and interface engineering, practical challenges remain, especially in maximizing $\Delta S_M$ without sacrificing thermal efficiency or cyclic durability.

### *3.3. Ribbons*

Magnetocaloric ribbons are typically fabricated using a rapid solidification technique known as melt spinning, which enables the formation of thin, amorphous and/or nanocrystalline ribbons with controlled microstructures [155-249]. In this process, molten metal is ejected onto a rotating copper wheel and solidifies almost instantaneously at cooling rates of approximately $10^6$ K/s, producing ribbons typically 20–50 µm thick and 1–5 mm wide. Depending on the targeted magnetic and structural properties, the ribbons may undergo post-annealing to induce nanocrystallization, promote the formation of desired magnetic phases, or relieve internal stresses.

Such thermal treatments are particularly critical for amorphous ribbons, which often require structural tuning to enhance their magnetocaloric performance.

The base magnetocaloric alloys, such as Gd-based, Fe-based, LaFeSi-based, Heusler, or high-entropy alloys, are initially synthesized by arc melting or induction melting of high-purity elemental constituents under an inert argon atmosphere. The key magnetocaloric properties of these alloy ribbons are summarized in Table 3.

Compared to bulk Gd ($-\Delta S_M^{max}$ ~10.2 J/kg·K and *RC* ~400 J/kg at $\mu_0 H = 5$ T) [7], its ribbon counterpart exhibits a slightly reduced magnetic entropy change ($-\Delta S_M^{max}$ ~8.7 J/kg·K) but a modestly enhanced refrigerant capacity (*RC* ~433 J/kg), while maintaining a $T_C$ near 294 K [158]. Alloying strategies have been employed to tune the magnetocaloric properties of Gd-based ribbons [158-160]. For examples, Gd–Co alloys show an increase in $T_C$, but at the expense of reduced $-\Delta S_M^{max}$, as can be seen in Fig. 8a [159]. Gd–Ni alloys retain $T_C$ close to that of pure Gd but still exhibit a reduction in $-\Delta S_M^{max}$ [160]. In $Gd_{100-x}Mn_x$ ribbons, both $T_C$ and $-\Delta S_M^{max}$ decrease with increasing Mn content [158]. Interestingly, while Gd–Mn ribbons generally display higher $\Delta S_M^{max}$ and *RC* than their Gd–Co counterparts, the latter maintain higher $T_C$ values. The incorporation of Al into Gd–Co alloys has been found to enhance the MCE, albeit with a further reduction in $T_C$ [161-162]. In Gd–Fe–Al ribbons, increasing the Fe/Al ratio leads to higher $T_C$ but a decrease in $-\Delta S_M^{max}$ [163]. Conversely, in Gd–Ni–Al systems, a higher Ni/Al ratio has been reported to simultaneously increase both $T_C$ and $-\Delta S_M^{max}$ [164]. For $(Gd_{1-x}Tb_x)_{12}Co_7$ ribbons, substituting Tb for Gd decreases $T_C$, but the highest values of $-\Delta S_M^{max}$ and *RC* are achieved at $x = 0.5$ (see Fig. 8b). Overall, alloying Gd with multiple elements tends to either raise $T_C$ while lowering $-\Delta S_M^{max}$, or vice versa. Only a limited number of Gd-based alloy ribbons maintain $T_C$ values near ambient temperature, which is a key requirement for room-temperature magnetic refrigeration.

To enable high-temperature magnetic cooling, the magnetocaloric properties of various Heusler alloy ribbon systems have been investigated (Table 3) [19,213-228]. In Heusler alloys, the magnetization and its variation associated with the martensitic transition are strongly influenced by the valence electron concentration per atom (e/a), which can be effectively modulated through chemical doping with elements such as Fe, Co, Cu, In, and Ge. As a result, both $T_C$ and $\Delta S_M$ of these alloys can be finely tuned over a broad temperature range. Most Heusler alloy ribbons exhibit SOMT ferromagnetic ordering at or above room temperature, followed by a FOMT at lower temperatures [19]. Notably, larger $\Delta S_M$ values are typically observed around the FOMT, albeit within a narrower temperature window. In contrast, $\Delta S_M$ values around the SOMT are generally smaller but extend over a wider temperature range. Consequently, some Heusler alloy systems exhibit larger *RCs* around the SOMT (see Table 3). However, significant hysteretic losses are often reported in these systems, particularly associated with the FOMT, which can substantially reduce the *RC* [19,214]. By carefully refining the chemical composition, it is possible to minimize these hysteretic losses, thereby enhancing the *RC* while retaining the high MCE values. Specialized thermal treatment is also essential for optimizing the MCE performance in these Heusler alloy ribbons [165,217,220].

Fe-based magnetocaloric ribbons have also been widely studied for their promising MCE characteristics [166-167,229-244]. For instance, $Fe_{90}Zr_{10}$ ribbons exhibit a $-\Delta S_M^{max}$ of approximately 2.7 J/kg·K and a *RC* of 497 J/kg under a 5 T magnetic field [166]. The incorporation of 1–2% boron (B) into this alloy tends to reduce both the $T_C$ and $-\Delta S_M^{max}$ [166]. However, careful adjustment of the Fe–Zr–B composition can simultaneously enhance both parameters. Notably, the addition of 1% Cu to form $Fe_{86}Zr_7B_6Cu_1$ significantly raises $T_C$ above room temperature and results in the highest observed $-\Delta S_M^{max}$ and *RC* in this alloy system [166]. In $Fe_{90-x}Ni_xZr_{10}$ ribbons ($x = 0$, 5, 10, 15), increasing the Ni content leads to a systematic rise in $T_C$ from 245 K ($x$ = 0) to 403 K ($x$ = 15), while maintaining a relatively stable -$\Delta S_M^{max}$ around 3 J/kg·K under a 4 T field change [167]. These tunable properties suggest that Fe-based ribbons, especially those incorporating Cu or Ni, could be excellent candidates for use in laminate composite structures as magnetic beds in advanced magnetic refrigeration systems.

Among intermetallic compounds reported, La(Fe,Si)$_{13}$-based alloys have garnered significant attention for MCEs and magnetic refrigeration due to the relative abundance and low cost of their constituent elements (La, Fe, and Si) compared to Gd-based alternatives [15,168]. These alloys exhibit a strong magneto-structural transition near room temperature in the La(Fe,Si)$_{13}$ (1:13) phase, which results in a large -$\Delta S_M^{max}$ (up to 30 J/kg K) and $\Delta T_{ad}$ (up to 12 K) in the temperature range of 270–300 K, making them ideal for household and commercial cooling applications. The Curie temperature of these alloys can be precisely adjusted by modifying the composition (e.g., through hydrogenation or Co substitution), enabling fine tuning of the working temperature range [169-170]. For instance, hydrogenated variants like $LaFe_{11.6}Si_{1.4}H_x$ exhibit a shift in the phase transition to higher temperatures, increasing their adaptability for various cooling applications [171]. However, hydrogenation can render these alloys brittle, leading to cracking or powdering during mechanical cycling or active AMR operation. Due to their FOMT nature, La(Fe,Si)$_{13}$ alloys typically suffer from large thermal and magnetic hysteresis, which leads to energy losses, reduces cooling efficiency, and decreases reversibility during cycling. When compared to their bulk counterparts, as-quenched ribbons of La(Fe,Si)$_{13}$-based alloys tend to exhibit reduced $T_C$ and -$\Delta S_M^{max}$ values. Therefore, specialized heat treatments are essential to optimize both the magnetic and magnetocaloric properties of these ribbons [168,172]. Huo *et al.* investigated the formation of the 1:13 phase during rapid solidification by examining the microstructures of the wheel-side and free-side surfaces of melt-spun ribbons [172]. They found that on the free-side, clusters of similarly oriented crystallites formed, with chemical segregation of La, Fe, and Si leading to nanoscale texturing of α-Fe and LaFeSi. In contrast, the wheel-side surface exhibited equiaxed 1:13 grains (~100–400 nm), with a minor α-Fe phase precipitated in the matrix. Upon annealing, the 1:13 phase grew via the dissolution of the α-Fe phase on the wheel side and a peritectoid reaction from the free side. For longer annealing times, this peritectoid reaction significantly improved the magnetic entropy change under a magnetic field change of 1.5 T, increasing the $-\Delta S_M^{max}$ from 12 J/kg·K (2 min) to 17 J/kg·K (2 h), and elevated the $T_C$ of the ribbons from 189 K to 201 K. In another case, increasing the annealing time from 10 minutes to 60 minutes for $La_{0.8}Ce_{0.2}Fe_{11.5}Si_{1.5}$ ribbons annealed at 1273 K resulted in a substantial increase in $-\Delta S_M^{max}$ from 9.7 J/kg·K to 32.8 J/kg·K, with a slight reduction in $T_C$ (193 K to 183 K) (see Fig. 9).

Additionally, $X_2Fe_{17}$ ($X$ = Nd, Y, Pr) ribbons have been shown to exhibit significant $-\Delta S_M^{max}$ values ranging from 3.7 to 4.8 J/kg·K and *RC* values between 496 and 580 J/kg around room temperature [156,173-174]. Incorporating Nd into $Pr_2Fe_{17}$ alloys to form $Pr_{2-x}Nd_xFe_{17}$ ribbons (where $x$ = 0.5 and 0.7) has resulted in enhanced $T_C$, $-\Delta S_M^{max}$, and *RC*, with the optimal values observed at $x$ = 0.7 [175].

Recently, ribbons of certain high-entropy alloys (HEAs), such as $Tm_{10}Ho_{20}Gd_{20}Ni_{20}Al_{20}$ and $Gd_{20}Dy_{20}Er_{20}Co_{20}Al_{20}$, have been explored for use in cryogenic magnetic refrigeration [155,176]. The incorporation of multiple rare-earth and transition metal elements in these alloys leads to a broadened temperature dependence of the magnetic entropy change near their magnetic ordering temperatures. This broadening effect contributes to enhanced *RC* values, while $-\Delta S_M^{max}$ values are typically reduced, as summarized in Table 3.

### *3.4. Microwires*

While Gd can be synthesized in the form of nanoparticles and thin films using chemical or sputtering techniques [7,25,44-46,65-66,97,99-101,125-127], it cannot be readily fabricated into microwires using rapid quenching methods such as melt spinning, in-rotating-water quenching, or glass-coated melt extraction. These techniques typically rely on forming amorphous or metastable phases, which require materials with high glass-forming ability. However, Gd, being a crystalline rare-earth metal, exhibits very poor glass-forming ability and crystallizes rapidly, even under extremely high cooling rates. This rapid crystallization inhibits uniform wire formation. Additionally, Gd is highly reactive, especially at elevated temperatures. During the melting or quenching process, it readily oxidizes to form $Gd_2O_3$, which degrades both its magnetic and structural properties [10]. Gd is also mechanically brittle, making it incompatible with standard wire fabrication methods [10]. These factors collectively make direct fabrication of Gd wires via rapid quenching techniques technically challenging.

To overcome these limitations, Gd has been alloyed with other elements such as Co, Fe, and Al to form compositions like Gd-Co-Al and Gd-Fe-Al [10,31,32]. These alloys possess improved glass-forming ability and can be successfully processed into high-quality microwires using melt-extraction techniques [10]. Numerous Gd-based microwires have been fabricated, and their magnetic and magnetocaloric properties have been widely investigated [7,30-32,35,161,250-254,263-271]. These microwires are produced under extremely rapid cooling rates (up to $10^6$ K/s), which results in more homogeneous amorphous structures with fewer inhomogeneities and magnetic clusters than their bulk glass counterparts. This structural uniformity leads to sharper magnetic transitions and enhanced MCEs. For example, $Gd_{55}Al_{20}Co_{25}$ amorphous microwires exhibit increased $-\Delta S_M^{max}$ and *RC*, with values of 9.69 J/kg·K and 580 J/kg, respectively, compared to 8.8 J/kg·K and 541 J/kg for their bulk glass equivalents under a 5 T field [250]. Similar improvements are observed in $Gd_{53}Al_{24}Co_{20}Zr_3$ microwires (10.3 J/kg·K and 733 J/kg) versus bulk samples (9.6 J/kg·K and 690 J/kg) [30].

Notably, most reported MCE data in the literature are obtained using magnetometry on bundles of microwires [30-32,35,161,252-282], rather than single-wire measurements [251]. Comparative studies show that multi-wire samples of $Gd_{53}Al_{24}Co_{20}Zr_3$ demonstrate superior MCE performance ($-\Delta S_M^{max}$ of 10.3 J/kg·K and *RC* of 733 J/kg) compared to a single wire (8.8 J/kg·K and 600 J/kg) (Table 4). This enhancement can be attributed to multiple factors, notably averaging effects and magnetostatic interactions. In bundled microwires, variations in diameter, composition, and internal stress among individual wires are effectively averaged out, leading to broader and more uniform magnetic transitions that improve the *RC* [10]. Additionally, the close proximity of wires facilitates dipolar (magnetostatic) interactions, which can amplify the overall magnetization change ($\Delta M$) and, consequently, the $\Delta S_M$. However, wire–wire interactions can also negatively affect performance through magnetic pinning, depending on spacing, orientation, matrix material, and applied field geometry. Therefore, detailed reporting on the number and arrangement of wires used in measurements is essential for accurate comparison.

***Annealing and structural optimization:*** Amorphous microwires often undergo thermal annealing to further improve their magnetic and magnetocaloric properties [31,251-252]. Annealing promotes structural relaxation and controlled nanocrystallization, optimizing the microstructure for magnetic ordering and energy conversion. As-quenched wires contain high levels of defects and internal stress; low-temperature, short-duration annealing relieves these stresses and facilitates atomic rearrangement while retaining the amorphous phase. For instance, $Gd_{53}Al_{24}Co_{20}Zr_3$ microwires annealed at 100 °C exhibit significant improvements, achieving a - $\Delta S_M^{max}$ of 9.5 J/kg·K and $RC$ of 689 J/kg, as shown in Fig. 10 [251]. This $RC$ is 35%–91% higher than that of bulk samples. The annealed wires show formation of nanocrystallites (5–10 nm in size) embedded in the amorphous matrix, leading to lattice distortions that alter magnetic properties and increase mechanical strength (up to 1845 MPa at 100 °C). This dual-phase (amorphous + nanocrystalline) structure is found desirable for enhancing both magnetocaloric and mechanical responses.

***Compositional engineering and melt-extraction control:*** The nanocrystalline/ amorphous structure can also be tailored during melt-extraction itself. In $Gd_{(50+5x)}Al_{(30-5x)}Co_{20}$ ($x$ = 0, 1, 2) microwires, about 20% of uniformly distributed ~10 nm nanocrystallites embedded in the amorphous matrix enhanced magnetocaloric response [31]. These microwires displayed large values of $-\Delta S_M^{max}$ (~9.7 J/kg·K), $\Delta T_{ad}$ (~5.2 K), and $RC$ (~654 J/kg) under a 5 T field. Gd enrichment significantly adjusts the $T_C$ while preserving high $\Delta S_M$ and $RC$ values. This structural configuration also broadens the operating temperature span of magnetic beds, which is critical for energy-efficient magnetic refrigeration. Additionally, novel composite microwires with embedded antiferromagnetic nanocrystals, such as $GdB_6$ in an amorphous ferromagnetic $Gd_{73.5}Si_{13}B_{13.5}$ matrix, showed promising MCE behavior ($-\Delta S_M^{max} \approx$ 6.4 J/kg·K, $RC \approx$ 890 J/kg) over wide temperature intervals (~130 K) [35]. Similar effects were reported in $Gd_3Ni/Gd_{65}Ni_{35}$ composite microwires [253]. By tailoring magnetic interactions, including RKKY ferromagnetic (Gd–Gd) and antiferromagnetic (Gd–Co, Gd–Ni) couplings, researchers have demonstrated the potential to fine-tune $T_C$ while maintaining high $RC$ in $Gd_{55}Co_{20+x}Ni_{10}Al_{15-x}$ ($x$ = 0, 5, 10) microwires, as can be seen in Fig. 11a [254].

***Toward tunable magnetic beds and room-temperature MCE:*** An important advantage of Gd-based alloy microwires is their tunable $T_C$ through compositional design, enabling the selection of wires with staggered $T_C$ and high $\Delta S_M$. This allows for the construction of engineered magnetic beds with laminate structures, achieving a table-like MCE response - ideal for Ericsson-cycle magnetic refrigeration systems [255]. However, Gd-based microwires are mostly limited to cryogenic and sub-room-temperature ranges (90–150 K). To enable ambient temperature applications, alternative systems are under investigation. Luo *et al.* reported a tunable giant MCE around room temperature in $Mn_xFe_{2-x}P_{0.5}Si_{0.5}$ ($0.7 \leq x \leq 1.2$) microwires produced via melt-extraction and thermal treatment [256]. By adjusting Mn/Fe ratios, $T_C$ was varied from 190 to 351 K, and a large $-\Delta S_M^{max}$ of 18.3 J/kg·K at 300 K was achieved for $x$ = 0.9 (see Fig. 11b). After accounting for magnetic hysteresis loss due to the FOMT nature, the *RC* was ~285 J/kg. Ongoing work focuses on reducing magnetic losses while maintaining high $\Delta S_M$. For instance, controlling the metal-to-nonmetal ratio (M/NM = $x$:1) in $(MnFe)_x(P_{0.5}Si_{0.5})$ ($x$ = 1.85-2.0) microwires can reduce thermal and magnetic hysteresis by up to 40%, with $-\Delta S_M^{max}$ and *RC* reaching optimal values at $x$ = 1.90 ($-\Delta S_M^{max}$ ~ 26.0 J/kg·K; *RC* ~ 367.4 J/kg; $T_C$ ~ 370 K) [257]. The effect of Fe content on the microstructure, magnetic, and magnetocaloric properties of $MnFe_xP_{0.5}Si_{0.5}$ ($0.9 \leq x \leq 1.05$) microwires has also been investigated [258]. As the Fe content increases, the system undergoes a transition from a FOMT for $x$ = 1.00 and 1.05 to a SOMT for $x$ = 0.90 and 0.95, leading to reduced magnetic losses but also a decrease in both the $-\Delta S_M^{max}$ and *RC*.

Heusler alloy microwires (e.g., $Ni_{50.5}Mn_{29.5}Ga_{20}$ and $Ni_{45.6}Fe_{3.6}Mn_{38.4}Sn_{12.4}$) have also shown significant $-\Delta S_M^{max}$ (up to 18.5 J/kg·K) in the sub-room and room temperature regions, though their *RC* values (60–230 J/kg) remain much lower than those of GdCo- or MnFe-based microwires [274-282]. Similar to their ribbon and thin film counterparts, Heusler alloy microwires exhibit SOMT ferromagnetic ordering at or above room temperature, followed by a FOMT at lower temperatures. Typically, larger $\Delta S_M$ values are observed near the FOMT, though within a relatively narrow temperature span. In contrast, $\Delta S_M$ values associated with the SOMT are smaller but distributed over a broader temperature range. As a result, certain Heusler microwire systems demonstrate enhanced *RC* around the SOMT. Through careful compositional tuning, hysteretic magnetic losses, particularly those associated with the FOMT, can be minimized, enabling improvements in *RC* while maintaining strong MCE performance. Additionally, targeted thermal treatments are critical for optimizing the microstructure and enhancing the overall MCE properties of Heusler alloy microwires.

Recently, the MCE in high-entropy magnetic materials has garnered increasing attention for magnetic refrigeration applications, primarily due to their excellent mechanical and magnetic properties [155,176,259]. High-entropy alloy microwires typically exhibit reduced $\Delta S_M(T)$ peaks but over significantly broader temperature ranges compared to conventional magnetocaloric materials. Notably, Yin *et al.* demonstrated that the magnetocaloric properties of high-entropy alloy microwires with the composition $(Gd_{36}Tb_{20}Co_{20}Al_{24})_{100-x}Fe_x$ can be significantly improved through current annealing of their as-cast amorphous counterparts [259]. This treatment induces the controlled precipitation of nanocrystals within the amorphous matrix, creating phase compositional heterogeneity along the microwires. The resulting microstructure broadens the temperature range of the $\Delta S_M$ and thereby enhances the *RC* in the annealed samples. While current annealing can enhance both MCE and *RC*, it is equally important to maintain the exceptional mechanical integrity characteristic of these high-entropy systems.

For cryogenic applications, microwires of rare-earth-based compositions such as HoErCo, HoErFe, DyHoCo, and $Dy_{36}Tb_{20}Co_{20}Al_{24}$ show large $-\Delta S_M^{max}$ values (~10 J/kg·K), making them attractive candidates for cryogenic magnetic cooling applications [260-262]. However, the mechanical properties of these systems remain largely unexplored.

## 4. Material Candidates for Energy-efficient Magnetic Refrigeration

Based on a comprehensive analysis of the magnetocaloric properties across various material forms, including nanoparticles, thin films, ribbons, and microwires, we propose several promising candidates for active magnetic cooling applications, categorized by temperature range: cryogenic ($T$ < 80 K), intermediate (80 K < $T$ < 300 K), and high temperature ($T$ > 300 K). These candidates are highlighted in Figures 12-15, as well as summarized in Table 5.

Since $\Delta S_M$ values are often reported under varying experimental conditions, such as different magnetic field strengths and measurement protocols, it is not straightforward to directly compare the performance of magnetocaloric materials across different studies. To address this, we define a performance coefficient as the ratio of the maximum magnetic entropy change ($-\Delta S_M^{max}$) to the corresponding maximum applied magnetic field change ($\mu_0\Delta H_{max}$). This normalized metric provides a more consistent basis for evaluating the effectiveness of magnetocaloric materials. A performance coefficient greater than one is considered indicative of a promising candidate for magnetic refrigeration. Using this criterion, we highlight a selection of high-potential magnetocaloric materials in various reduced-dimensional forms, including nanoparticles (Fig. 12), thin films (Fig. 13), ribbons (Fig. 14a-c), and microwires (Fig. 15).

As shown in Fig. 12, the majority of magnetocaloric nanoparticle candidates are oxides. Among them, $GdVO_4$ nanoparticles exhibit the highest performance coefficient in the low-temperature range ($T$ < 80 K), while $DyCrTiO_3$ nanoparticles lead in the intermediate temperature range (80 K < $T$ < 300 K). In the high-temperature range ($T$ > 300 K), $La_{0.7}Ca_{0.2}Sr_{0.1}MnO_3$ nanoparticles demonstrate the greatest performance coefficient. Although certain manganite oxide nanoparticles exhibit notable magnetic entropy changes, their inherently high heat capacities often lead to low or moderate adiabatic temperature changes, which can limit their overall cooling efficiency.

In the case of magnetocaloric thin films, various candidate materials are distributed across the three major cooling temperature regimes, as illustrated in Fig. 13. In the low-temperature range ($T$ < 80 K), $EuTiO_3$ exhibits the highest performance coefficient. Within the intermediate temperature range (80 K < $T$ < 300 K), $GdCoO_3$ shows the strongest performance. At high temperatures ($T$ > 300 K), $Ni_{51}Mn_{29}Gd_{20}$ demonstrates the highest performance coefficient among the thin film candidates. However, the performance coefficient of the $Ni_{51}Mn_{29}Gd_{20}$ thin film is relatively low compared to other magnetocaloric candidates and requires enhancement to enable its use in AMR. Additionally, the adiabatic temperature change - an even more critical parameter for evaluating magnetocaloric materials - remains largely unexplored in these thin-film systems.

As illustrated in Fig. 14, a wide range of ribbon-based magnetocaloric materials are available across the three primary cooling temperature regimes. In the low-temperature range ($T$ < 80 K), as shown in Fig. 14a, rare-earth-based ribbons are the leading candidates. In the intermediate temperature range (80 K < $T$ < 300 K), Gd and Gd-based alloy ribbons (GdCo, Gd-Co-$X$, Gd-Fe-$X$) dominate (Fig. 14b). At high temperatures ($T$ > 300 K), Heusler alloy ribbons (e.g., Ni-Mn-Ga, Ni-Co-Mn-Sn, Ni-Co-Mn-In) emerge as the principal candidates (Fig. 14c).

Similar to magnetocaloric ribbons, rare-earth-based microwires (e.g., DyHoCo) are the leading candidates in the low-temperature range ($T$ < 80 K). In the intermediate temperature range (80 K < $T$ < 300 K), Gd alloy-based microwires (e.g., $Gd_{60}Al_{20}Co_{20}$, $Gd_{36}Tb_{20}Co_{20}Al_{24}$) are the principal candidates. At high temperatures ($T$ > 300 K), Mn–Fe–P–Si and Heusler alloy-based microwires (e.g., $Ni_{45.6}Fe_{3.6}Mn_{38.4}Sn_{12.4}$) dominate. Although several Heusler alloy microwires exhibit a large magnetic entropy change and a high-performance coefficient, the $\Delta S_M(T)$ is confined to a narrow temperature range, leading to a moderate *RC*, which may limit their suitability for practical cooling applications.

## 5. Challenges and Opportunities

Despite their scientific promise, low-dimensional magnetocaloric materials face several key challenges that limit their implementation in active cooling systems. Below, we outline the primary hurdles associated with nanoparticles, thin films, ribbons, and microwires.

Magnetocaloric nanoparticles, while promising for cryogenic and localized cooling, face significant application barriers related to magnetic field requirements, thermal integration, stability, scalability, and device engineering. Ferromagnetic nanoparticles often exhibit suppressed $T_C$ and reduced $\Delta S_M$, although refrigerant capacity (*RC* or *RCP*) may improve due to entropy broadening. The diminished $\Delta S_M$ reduces the cooling power per cycle, particularly under moderate magnetic fields. Antiferromagnetic nanoparticles may show large $\Delta S_M$ but typically require high magnetic fields (3–7 T), necessitating superconducting magnets or bulky setups—hindering the development of compact and energy-efficient devices. Thermally, nanoparticles have low intrinsic conductivity and high interfacial resistance when embedded in fluids or solids, making efficient heat transfer to/from the load difficult. They are also prone to agglomeration, oxidation, and degradation under thermal cycling. Maintaining long-term operational stability under cyclic magnetic fields remains a critical challenge. Embedding nanoparticles into functional matrices (e.g., elastomers, porous scaffolds, or composite heat exchangers) without compromising magnetocaloric or thermal performance is complex. Encapsulation or binder materials may insulate thermally or magnetically, reducing system efficiency. Scalable synthesis techniques such as sol-gel, co-precipitation, or hydrothermal methods often struggle to maintain particle quality, size uniformity, and crystallinity, which are essential for consistent magnetic behavior. Poor crystallinity and wide size distributions lead to variable MCE responses. Moreover, many oxide and intermetallic nanoparticles (e.g., Gd-, LaFeSi-, and Mn-based alloys) are highly sensitive to stoichiometry and surface oxidation. This can alter magnetic properties and degrade MCE performance. Protective coatings like $SiO_2$ or polymers are often necessary but may introduce thermal or magnetic barriers [23]. Although particle bed configurations offer a large interfacial area for heat exchange, they are associated with a non-negligible pressure drop [28,29]. Ultimately, reliable integration of magnetocaloric nanoparticles into practical solid-state or fluidic refrigeration systems remains a significant technological bottleneck.

The application of magnetocaloric thin films in magnetic refrigeration, particularly for on-chip cooling, micro-refrigerators, and cryogenic devices, offers exciting opportunities but also presents considerable challenges stemming from dimensional constraints, interfacial effects, and material integration issues. Similar to ferromagnetic nanoparticles, ferromagnetic thin films typically exhibit reduced $T_C$ and $\Delta S_M$ compared to their bulk counterparts. These reductions arise from finite-size effects, epitaxial strain, and surface/interface interactions. While *RC* can be enhanced due to broadening of the transition, the strong suppression of $\Delta S_M$ limits their effectiveness for active cooling. Epitaxial strain from lattice-mismatched substrates can distort crystal symmetry, suppress magnetic ordering, or shift transition temperatures. In some systems, such strain can enhance $\Delta S_M$ through coupling with substrate structural transitions, but the effect is typically confined to a narrow temperature window, thereby reducing *RC*. Interestingly, in weakly antiferromagnetic or mixed-phase (FM + AFM) thin films, the combined influence of strain and reduced dimensionality can enhance the MCE, enabling large $\Delta S_M$ values under lower critical magnetic fields. However, thin films inherently possess low thermal mass and limited thermal conductivity, especially in multilayered or oxide-based systems, posing serious challenges for efficient heat extraction and limiting practical cooling capacity. Compared to bulk materials, thin films typically exhibit reduced electrical conductivity due to increased electron scattering at surfaces and interfaces, which leads to lower carrier mobility. Moreover, thin films undergoing FOMTs are prone to performance degradation over repeated thermal or magnetic cycling. For instance, $Gd_5Si_2Ge_2$ thin films have demonstrated significant $\Delta S_M$ loss after ~1000 cycles. From a fabrication standpoint, achieving high film quality is challenging due to variations introduced by deposition techniques (e.g., pulsed laser deposition, sputtering). Issues such as grain boundaries, off-stoichiometry, crystalline defects, and oxygen vacancies, especially in complex oxides, can cause inconsistent MCE performance across samples and devices. Only a limited set of magnetocaloric materials (e.g., Gd, Heusler alloys, and certain manganites) have been successfully deposited as high-quality thin films. Maintaining stoichiometry, crystallinity, and magnetic order during deposition remains particularly difficult for multicomponent or intermetallic systems. Furthermore, the thin geometry (typically 10–500 nm) inherently limits volumetric entropy change and cooling power. Scaling up to practical applications necessitates multilayer stacking or large-area film integration, which introduces additional thermal and magnetic engineering complexities. Integrating the MCE with other phenomena, such as the thermoelectric effect, in thin-film systems may offer a promising route to enhance overall cooling efficiency [283], though further investigation is required to validate this approach

Magnetocaloric ribbons, typically fabricated via rapid solidification techniques such as melt spinning, offer advantages like flexibility, high surface area, and fast thermal response, making them promising for use in magnetic refrigeration systems. However, their practical application faces several notable challenges. Compared to their bulk counterparts, ribbons often exhibit lower $\Delta S_M$, primarily due to microstructural disorder from rapid solidification, grain texture effects, and diminished long-range magnetic ordering. Many magnetocaloric ribbons, particularly those based on intermetallic compounds such as $La(Fe,Si)_{13}$ and Mn-based Heusler alloys, are mechanically brittle, a result of their crystalline or partially amorphous nature. This brittleness limits their durability under mechanical stress, thermal cycling, or during device integration. Additionally, ribbons generally possess low thermal conductivity, particularly in amorphous or disordered phases, which restricts efficient heat exchange and slows cooling response. Compared to their bulk counterparts, electrical conductivity tends to decrease in magnetic ribbons. The reduction mainly arises from structural disorder (amorphous or nanocrystalline phases), increased grain boundary scattering, and possible surface oxidation. However, the exact difference depends on composition, thickness, and post-processing treatments (e.g., annealing to induce crystallization). For thinner ribbons (typically $< 50$ μm), the small volume further limits the overall cooling capacity, making it difficult to scale up for higher-power applications. Some ribbons undergo FOMT, often accompanied by thermal and magnetic hysteresis, which reduces energy efficiency during cyclic operation and may impact device lifespan under repeated use. Their high surface area also makes them vulnerable to oxidation, especially in ambient or humid environments, degrading their magnetocaloric performance over time. From a fabrication standpoint, achieving compositional and structural uniformity during rapid solidification is challenging, with minor processing variations leading to significant changes in performance. Finally, despite their mechanical flexibility, practical integration of ribbons into magnetic cooling modules (e.g., regenerators or heat exchangers) requires precise alignment and mechanical support while ensuring effective thermal and magnetic coupling - an engineering challenge that remains unresolved.

Magnetocaloric microwires, with cylindrical and flexible geometries ranging from a few micrometers to sub-micron diameters, offer several advantages for magnetic refrigeration applications, including high surface-to-volume ratio, mechanical flexibility, and rapid heat exchange. Despite these benefits, significant challenges hinder their practical implementation. One major issue is the controlled fabrication of high-quality microwires with consistent diameter, uniform composition, and crystallinity. Only a narrow range of magnetocaloric materials can be processed into microwires using methods such as melt extraction, in-rotating-water quenching, or glass-coated melt spinning. Many promising magnetocaloric alloys like $La(Fe,Si)_{13}$ are brittle or chemically unstable during wire processing, limiting material options. Microwires, particularly those composed of intermetallic compounds, tend to be mechanically fragile, especially under repeated thermal or magnetic cycling. Over time, this can lead to microcracks or delamination from protective coatings or composite matrices, compromising structural integrity and performance. In magnetic microwires, the electron mean free path can approach the wire diameter, resulting in enhanced surface scattering. This increased scattering reduces carrier mobility and, consequently, lowers electrical conductivity compared to bulk materials. In practical cooling devices, bundling or aligning large numbers of microwires is required to achieve significant cooling power. However, ensuring uniform magnetic field exposure and efficient thermal contact with the working fluid (liquid or gas) across such arrays is technically demanding. The design of mechanical supports that maintain wire alignment without introducing thermal resistance remains an open engineering challenge. Thermal management is further complicated by poor thermal coupling between wires in dense bundles and between wires and their embedding matrices. Oxidation-preventing coatings may inadvertently act as thermal insulators, impeding heat exchange. Achieving uniform temperature distribution during heating and cooling cycles in densely packed microwire arrays is also difficult, reducing system efficiency. These combined challenges, ranging from fabrication constraints to integration and thermal engineering, must be addressed before magnetocaloric microwires can be effectively utilized in compact, scalable magnetic refrigeration systems. Table 6 summarizes the main advantages and key challenges in the development of low-dimensional magnetocaloric materials.

## 6. Concluding Remarks

Magnetocaloric materials with reduced dimensionality, such as nanoparticles, thin films, ribbons, and microwires, offer promising avenues for the development of energy-efficient magnetic refrigeration technologies. Compared to their bulk counterparts, ferromagnetic nanoparticles and thin films often exhibit lower Curie temperatures and reduced magnetic entropy change, though they may show enhanced refrigerant capacity due to broadened magnetic transitions. In contrast, magnetocaloric ribbons and microwires, typically produced via rapid solidification, can exhibit both enhanced magnetocaloric performance and improved mechanical properties relative to their bulk glassy forms. Particularly notable are antiferromagnetically weakened or mixed-phase FM/AFM nanoparticles and thin films, which outperform bulk antiferromagnets due to their negligible magnetic and thermal hysteresis losses and enhanced MCE performance. These effects are tunable via substrate-induced strain, finite size effects, and surface/interface modifications. In amorphous ribbons and microwires, thermal treatments such as annealing can significantly improve magnetocaloric properties by inducing nanocrystalline phases. However, this often comes at the cost of mechanical fragility, creating a trade-off between magnetocaloric performance and structural durability. Optimizing nanocrystallization conditions may simultaneously enhance both thermal and mechanical properties. This is especially crucial in Heusler alloy-based ribbons and microwires, where precise thermal processing governs performance outcomes.

From a fabrication and scalability standpoint, methods for producing ribbons and microwires are more readily adaptable to practical cooling device architectures, whereas the synthesis of high-quality nanoparticles and thin films remains largely limited to laboratory-scale methods. This underscores the need for new scalable fabrication techniques for reduced-dimensionality materials.

While much of the current research emphasizes achieving high $\Delta S_M$ and $RC$, fewer studies have rigorously addressed the adiabatic temperature change ($\Delta T_{ad}$) - a key parameter for evaluating real-world cooling performance. $\Delta T_{ad}$ is technically challenging to measure, particularly in nanoparticle and thin film systems. In some cases, ribbons and microwires exhibit significantly reduced $\Delta T_{ad}$ compared to their bulk analogs, necessitating further comprehensive investigations into this metric for all reduced-dimensionality formats.

Theoretically, magnetocaloric materials with enhanced surface areas (e.g., nanoparticles, ribbons, microwires) are predicted to offer superior heat exchange with the surrounding environment, boosting cooling efficiency. However, these models often overlook the influence of structural assembly materials, such as binders, matrices, and coatings, on overall system performance. In practice, interactions between particles or wires, as well as between layers in laminated structures, can significantly modify the magnetic and thermal behavior. These interfacial and collective effects require thorough theoretical and experimental scrutiny when engineering composite or structured cooling devices.

In summary, while magnetocaloric materials with reduced dimensionality hold great promise, numerous technical barriers, ranging from synthesis and processing to integration and thermal management, must be systematically addressed. Only through coordinated advances in materials science, device engineering, and system-level optimization can these materials be effectively utilized in compact, scalable, and high-performance magnetic refrigeration systems.

**Acknowledgements**

Research at USF was supported by the U.S. Department of Energy, Office of Basic Energy Sciences, Division of Materials Sciences and Engineering under Award No. DE-FG02-07ER46438.

## References

[1] Brown JS, Domanski PA. Review of alternative cooling technologies. *Appl Therm Eng*. 2014;64(1–2):252–262. doi: 10.1016/j.applthermaleng.2013.12.014

[2] Klinar K, Law JY, Franco V, et al. Perspectives and energy applications of magnetocaloric, pyromagnetic, electrocaloric, and pyroelectric materials. Adv Energy Mater. 2024;14(39):2401739. doi: 10.1002/aenm.202401739

[3] Chu RC, Simons RE, Ellsworth MJ, et al. Review of cooling technologies for computer products. *IEEE Trans Device Mater Reliab*. 2004;4(4):568–585. doi: 10.1109/TDMR.2004.840855

[4] Koohestani SS, Nižetić S, Santamouris M. Comparative review and evaluation of state-of-the-art photovoltaic cooling technologies. *J Clean Prod*. 2023;406:136953. doi: 10.1016/j.jclepro.2023.136953

[5] Mahek MK, Ramadan M, Dol SS, et al. A comprehensive review of thermoelectric cooling technologies for enhanced thermal management in lithium-ion battery systems. *Heliyon*. 2024;10(24):e40649. doi: 10.1016/j.heliyon.2024.e40649

[6] Franco V, Blázquez JS, Ipus JJ, et al. Magnetocaloric effect: From materials research to refrigeration devices. *Prog Mater Sci*. 2018;93:112–232. doi: 10.1016/j.pmatsci.2017.10.005

[7] Gschneidner KA, Pecharsky VK. Magnetocaloric materials. *Annu Rev Mater Sci*. 2000;30:387–429. doi: 10.1146/annurev.matsci.30.1.387

[8] Phan MH, Yu SC. Review of the magnetocaloric effect in manganite materials. *J Magn Magn Mater*. 2007;308(2):325–340. doi: 10.1016/j.jmmm.2006.07.025

[9] Law JY, Moreno-Ramírez LM, Díaz-García Á, et al. Current perspective in magnetocaloric materials research. *J Appl Phys*. 2023;133(4):040903. doi: 10.1063/5.0130035

[10] Shen HX, Duc NTM, Belliveau H, et al. Advanced magnetocaloric microwires: What does the future hold? *Viet J Sci Technol Eng*. 2023;65940:14–24. doi: 10.31276/VJSTE.65(4).14-24

[11] Dan'kov SY, Tishin AM, Pecharsky VK, et al. Magnetic phase transitions and the magnetothermal properties of gadolinium. *Phys Rev B*. 1998;57(6):3478–3490. doi: 10.1103/PhysRevB.57.3478

[12] Bingham NS, Phan MH, Srikanth H, et al. Magnetocaloric effect and refrigerant capacity in charge-ordered manganites. *J Appl Phys*. 2009;106(2):023909. doi: 10.1063/1.3174396

[13] Pecharsky VK, Gschneidner KA Jr. Giant magnetocaloric effect in $Gd_5(Si_2Ge_2)$. *Phys Rev Lett*. 1997;78(23):4494–4497. doi: 10.1103/PhysRevLett.78.4494

[14] Pecharsky VK, Gschneidner KA Jr. Tunable magnetic regenerator alloys with a giant magnetocaloric effect for magnetic refrigeration from ~20 to ~290 K. *Appl Phys Lett*. 1997;70(24):3299–3301. doi: 10.1063/1.119206

[15] Hu FX, Shen BG, Sun JR, et al. Influence of negative lattice expansion and metamagnetic transition on magnetic entropy change in the compound $LaFe_{11.4}Si_{1.6}$. *Appl Phys Lett*. 2001;78(23):3675–3677. doi: 10.1063/1.1375836

[16] Wada H, Tanabe Y. Giant magnetocaloric effect of $MnAs_{1-x}Sb_x$. *Appl Phys Lett*. 2001;79(20):3302–3304. doi: 10.1063/1.1419048

[17] Tegus O, Brück E, Buschow KHJ, et al. Transition-metal-based magnetic refrigerants for room-temperature applications. *Nature*. 2002;415(6868):150–152. doi: 10.1038/415150a

[18] Tegus O, Brück E, Zhang L, et al. Magnetic-phase transitions and magnetocaloric effects. *Physica B*. 2002;319(1–4):174–192. doi: 10.1016/S0921-4526(02)01119-5

[19] Phan TL, Zhang P, Dan NH, et al. Coexistence of conventional and inverse magnetocaloric effects and critical behaviors in $Ni_{50}Mn_{50-x}Sn_x$ (x = 13 and 14) alloy ribbons. *Appl Phys Lett*. 2012;101(21):212403. doi: 10.1063/1.4767453

[20] Phan MH, Yu SC, Hur NH. Excellent magnetocaloric properties of $La_{0.7}Ca_{0.3-x}Sr_xMnO_3$ ($0.05 \leq x \leq 0.25$) single crystals. *Appl Phys Lett*. 2005;86(7):072504. doi: 10.1063/1.1867564

[21] Phan MH, Peng HX, Yu SC. Large magnetocaloric effect in single crystal $Pr_{0.63}Sr_{0.37}MnO_3$. *J Appl Phys*. 2005;97(10):10M306. doi: 10.1063/1.1849554

[22] Zhang YD, Lampen P, Phan TL, et al. Tunable magnetocaloric effect near room temperature in $La_{0.7-x}Pr_xSr_{0.3}MnO_3$ ($0.02 < x < 0.30$) manganites. *J Appl Phys*. 2012;111(6):063918. doi: 10.1063/1.3698346

[23] Kitanovski A. Energy applications of magnetocaloric materials. *Adv Energy Mater*. 2020;10(10):1903741. doi: 10.1002/aenm.201903741

[24] Khovaylo VV, Rodionova VV, Shevyrtalov SN, et al. Magnetocaloric effect in "reduced" dimensions: Thin films, ribbons, and microwires of Heusler alloys and related compounds. *Phys Status Solidi B*. 2014;251(10):2104–2113. doi: 10.1002/pssb.201451217

[25] Miller CW, Belyea DD, Kirby BJ. Magnetocaloric effect in nanoscale thin films and heterostructures. *J Vac Sci Technol A*. 2014;32(4):040802. doi: 10.1116/1.4882858

[26] Belo JH, Pires AL, Araújo JP, et al. Magnetocaloric materials: From micro- to nanoscale. *J Mater Res*. 2019;34(1):134–157. doi: 10.1557/jmr.2018.352

[27] Scheunert G, Heinonen O, Hardeman R, et al. A review of high magnetic moment thin films for microscale and nanotechnology applications. *Appl Phys Rev*. 2016;3(1):011301. doi: 10.1063/1.4941311

[28] Kuzmin MD. Factors limiting the operation frequency of magnetic refrigerators. *Appl Phys Lett*. 2007;90(25):251916. doi: 10.1063/1.2750540

[29] Vuarnoz D, Kawanami T. Numerical analysis of a reciprocating active magnetic regenerator made of gadolinium wires. *Appl Therm Eng*. 2012;37:388–395. doi: 10.1016/j.applthermaleng.2011.11.053

[30] Bingham NS, Wang H, Qin FX, et al. Excellent magnetocaloric properties of melt-extracted Gd-based amorphous microwires. *Appl Phys Lett*. 2012;101(10):102407. doi: 10.1063/1.4751038

[31] Shen HX, Xing DW, Sánchez Llamazares JL, et al. Enhanced refrigerant capacity in Gd–Al–Co microwires with a biphase nanocrystalline/amorphous structure. *Appl Phys Lett*. 2016;108:092403. doi: 10.1063/1.4943137

[32] Duc NTM, Shen HX, Clements E, et al. Enhanced refrigerant capacity and Curie temperature of amorphous $Gd_{60}Fe_{20}Al_{20}$ microwires. *J Alloys Compd*. 2019;807:151694. doi: 10.1016/j.jallcom.2019.151694

[33] Zhou T, Cher MK, Shen L, et al. On the origin of giant magnetocaloric effect and thermal hysteresis in multifunctional α-FeRh thin films. *Phys Lett A*. 2013;377(42):3052–3059. doi: 10.1016/j.physleta.2013.09.027

[34] Mosca DH, Vidal F, Etgens VH. Strain engineering of the magnetocaloric effect in MnAs epilayers. *Phys Rev Lett*. 2008;101(12):125503. doi: 10.1103/PhysRevLett.101.125503

[35] Duc NTM, Shen HX, Thiabgoh O, et al. Melt-extracted $Gd_{73.5}Si_{13}B_{13.5}/GdB_6$ ferromagnetic/antiferromagnetic microwires with excellent magnetocaloric properties. *J Alloys Compd*. 2020;818:153333. doi: 10.1016/j.jallcom.2019.153333

[36] Phan MH, Alonso J, Khurshid H, et al. Exchange bias effects in iron oxide based nanoparticle systems. *Nanomaterials*. 2016;6(11):221. doi: 10.3390/nano6110221

[37] Iglesias O, Labarta A. Finite-size and surface effects in maghemite nanoparticles: Monte Carlo simulations. *Phys Rev B*. 2001;63(18):184416. doi: 10.1103/PhysRevB.63.184416

[38] Mandal S, Menon KSR, Mahatha SK, et al. Finite size versus surface effects on magnetic properties of antiferromagnetic particles. *Appl Phys Lett*. 2011;99(23):232507. doi: 10.1063/1.3668091

[39] Xiao Y, Du J. Superparamagnetic nanoparticles for biomedical applications. *J Mater Chem B*. 2020;8(3):354–367. doi: 10.1039/C9TB01955C

[40] Dutta P, Pal S, Seehra MS, et al. Size dependence of magnetic parameters and surface disorder in magnetite nanoparticles. *J Appl Phys*. 2009;105(7):07B501. doi: 10.1063/1.3055272

[41] Demortière A, Panissod P, Pichon BP, et al. Size-dependent properties of magnetic iron oxide nanocrystals. *Nanoscale*. 2011;3(1):225–232. doi: 10.1039/C0NR00521E

[42] Curiale J, Granada M, Troiani HE, et al. Magnetic dead layer in ferromagnetic manganite nanoparticles. *Appl Phys Lett*. 2009;95(4):043106. doi: 10.1063/1.3187538

[43] Unni M, Uhl AM, Savliwala S, et al. Thermal decomposition synthesis of iron oxide nanoparticles with diminished magnetic dead layer by controlled addition of oxygen. *ACS Nano*. 2017;11(2):2284–2303. doi: 10.1021/acsnano.7b00609

[44] Wang GF, Li LR, Zhao ZR, et al. Structural and magnetocaloric effect of $Ln_{0.67}Sr_{0.33}MnO_3$ (Ln = La, Pr and Nd) nanoparticles. *Ceram Int.* 2014;40(10):16449–16454. doi: 10.1016/j.ceramint.2014.07.154

[45] Gschneidner KA Jr, Pecharsky VK, Tsokol AO. Recent developments in magnetocaloric materials. *Rep Prog Phys.* 2005;68(6):1479–1539. doi: 10.1088/0034-4885/68/6/R04

[46] Harstad SM, El-Gendy AA, Gupta S, et al. Magnetocaloric Effect of Micro- and Nanoparticles of $Gd_5Si_4$. *JOM.* 2019;71(9):3159–3163. doi: 10.1007/s11837-019-03626-1

[47] Andrade VM, Caraballo Vivas RJ, Pedro SS, et al. Magnetic and magnetocaloric properties of $La_{0.6}Ca_{0.4}MnO_3$ tunable by particle size and dimensionality. *Acta Mater.* 2016;102:49–55. doi: 10.1016/j.actamat.2015.08.080

[48] Poddar P, Srinath S, Gass J, et al. Magnetic transition and large magnetocaloric effect associated with surface spin disorder in Co and CoAg core–shell nanoparticles. *J Phys Chem C.* 2007;111(38):14020–14024. doi: 10.1021/jp073274i

[49] Biswas A, Chandra S, Stefanoski S, et al. Enhanced cryogenic magnetocaloric effect in $Eu_8Ga_{16}Ge_{30}$ clathrate nanocrystals. *J Appl Phys*. 2015;117(3):033903. doi: 10.1063/1.4906280

[50] Lu WJ, Luo X, Hao CY, et al. Magnetocaloric effect and Griffiths-like phase in $La_{0.67}Sr_{0.33}MnO_3$ nanoparticles. *J Appl Phys.* 2008;104(11):113908. doi: 10.1063/1.3037236

[51] Lampen P, Puri A, Phan MH, et al. Structure, magnetic, and magnetocaloric properties of amorphous and crystalline $La_{0.4}Ca_{0.6}MnO_{3+}$ nanoparticles. *J Alloys Compd.* 2012;512:94–98. doi: 10.1016/j.jallcom.2011.09.027

[52] Lampen P, Bingham NS, Phan MH, et al. Impact of reduced dimensionality on the magnetic and magnetocaloric response of $La_{0.7}Ca_{0.3}MnO_3$. *Appl Phys Lett.* 2013;102(6):062414. doi: 10.1063/1.4792239

[53] Xi S, Lu W, Sun Y. Magnetic properties and magnetocaloric effect of $La_{0.8}Ca_{0.2}MnO_3$ nanoparticles tuned by particle size. *J Appl Phys.* 2012;111(6):063922. doi: 10.1063/1.3699037

[54] Poddar P, Gass J, Rebar DJ, et al. Magnetocaloric effect in ferrite nanoparticles. *J Magn Magn Mater*. 2006;307(2):227–231. doi: 10.1016/j.jmmm.2006.04.007

[55] Gass J, Srikanth H, Kislov N, et al. Magnetization and magnetocaloric effect in ball-milled zinc ferrite powder. *J Appl Phys*. 2008;103(7):07B309. doi: 10.1063/1.2829754

[56] Lee KD, Kambale RC, Hur N. Magnetocaloric effect in Ni-Zn ferrite nanoparticles prepared by using solution combustion. *J Korean Phys Soc*. 2014;65(11):1930–1934. doi: 10.3938/jkps.65.1930

[57] Yamamoto TA, Tanaka M, Nakayama T, et al. Magnetocaloric effect of superparamagnetic nanocomposite composed of iron oxide and silver. *Jpn J Appl Phys*. 2000;39(8):4761–4763. doi: 10.1143/JJAP.39.4761

[58] Phan MH, Morales MB, Chinnasamy CN, et al. Magnetocaloric effect in bulk and nanostructured $Gd_3Fe_5O_{12}$ materials. *J Phys D: Appl Phys.* 2009;42(11):115007. doi: 10.1088/0022-3727/42/11/115007

[59] Das K, Banu N, Das I, et al. Investigation of size-dependent magnetic ordering in charge ordered antiferromagnetic nanoparticles via magnetocaloric effect. *J Magn Magn Mater.* 2023;573:170925. doi: 10.1016/j.jmmm.2019.165309

[60] Bamana B, Prinsloo A, Mohanty P, et al. Magnetic phase transitions and magnetocaloric effect in $DyCrTiO_5$ nanoparticles. *AIP Advances.* 2023;13(2):025049. doi: 10.1063/9.0000552

[61] Shinde KP, Tien VM, Huang L, et al. Magnetocaloric effect in $Tb_2O_3$ and $Dy_2O_3$ nanoparticles at cryogenic temperatures. *J Appl Phys.* 2020;127(5):054903. doi: 10.1063/1.5120350

[62] Biswas A, Chandra S, Phan MH, et al. Magnetocaloric properties of nanocrystalline $LaMnO_3$: Enhancement of refrigerant capacity and relative cooling power. *J Alloys Compd.* 2013;545:157–161. doi: 10.1016/j.jallcom.2012.08.001

[63] Phan MH, Chandra S, Bingham NS, et al. Collapse of charge ordering and enhancement of magnetocaloric effect in nanocrystalline $La_{0.35}Pr_{0.275}Ca_{0.375}MnO_3$. *Appl Phys Lett.* 2010;97(24):242506. doi: 10.1063/1.3526380

[64] Chaudhary V, Ramanujan RV. Magnetocaloric properties of Fe–Ni–Cr nanoparticles for active cooling. *Sci Rep.* 2016;6:35156. doi: 10.1038/srep35156

[65] Rajkumar DM, Manivel Raja M, Gopalan R, et al. Magnetocaloric effect in high-energy ball-milled $Gd_5Si_2Ge_2$ and $Gd_5Si_2Ge_2$/0.1 wt% Fe nanopowders. *J Magn Magn Mater.* 2008;320(8):1479–1484. doi: 10.1016/j.jmmm.2007.12.005

[66] Li J, Wang Y, Wang H, et al. Enhanced cryogenic magnetocaloric effect induced by small size $GdNi_5$ nanoparticles. *J Mater Sci Technol.* 2014;30(10):973–978. doi: 10.1016/j.jmst.2014.01.009

[67] Biswas A, Samanta T, Banerjee S, et al. Magnetocaloric properties of nanocrystalline $La_{0.125}Ca_{0.875}MnO_3$. *Appl Phys Lett.* 2009;94(23):233109. doi: 10.1063/1.3152785

[68] Pekala M, Drozd V, Fagnard JF, et al. Magnetocaloric effect in nano- and polycrystalline manganites $La_{0.5}Ca_{0.5}MnO_3$. *J Alloys Compd.* 2010;507(2):350–355. doi: 10.1016/j.jallcom.2010.07.165

[69] Nasri M, Trik M, Dhahri E, et al. Investigation of structural, magnetocaloric and electrical properties of $La_{0.6}Ca_{0.4-x}Sr_xMnO_3$ compounds. *Phys B Condens Matter.* 2013;408:104–109. doi: 10.1016/j.physb.2012.09.003

[70] Hueso LE, Sande P, Miguens DR, et al. Tuning of the magnetocaloric effect in δ nanoparticles synthesized by sol–gel techniques. *J Appl Phys.* 2002;91(10):9943–9945. doi: 10.1063/1.1476972

[71] Tang W, Lu WJ, Luo X, et al. Particle size effects on $La_{0.7}Ca_{0.3}MnO_3$: size-induced changes of magnetic phase transition order and magnetocaloric study. *J Magn Magn Mater.* 2010;322(17):2360–2363. doi: 10.1016/j.jmmm.2010.02.038

[72] Yang H, Zhu YH, Xian T, et al. Synthesis and magnetocaloric properties of $La_{0.7}Ca_{0.3}MnO_3$ nanoparticles with different sizes. *J Alloys Compd.* 2013;555:150–155. doi: 10.1016/j.jallcom.2012.11.200

[73] Xu Y, Meier M, Das P, et al. Perovskite manganites: potential materials for magnetic cooling at or near room temperature. *Cryst Eng.* 2002;5(6):383–392. doi: 10.1016/S1463-0184(02)00049-7

[74] Mira J, Rivas J, Hueso LE, et al. Drop of magnetocaloric effect related to the change from first- to second-order magnetic phase transition in $La_{2/3}(Ca_{1-x}Sr_x)_{1/3}MnO_3$. *J Appl Phys.* 2002;91(10):8903–8905. doi: 10.1063/1.1451892

[75] Morelli DT, Mance AM, Mantese JV, et al. Magnetocaloric properties of doped lanthanum manganites films. *J Appl Phys.* 1996;79(1):373–375. doi: 10.1063/1.360840

[76] Rostamnejadi A, Venkatesan M, Kameli P, et al. Magnetocaloric effect in $La_{0.67}Sr_{0.33}MnO_3$ manganite above room temperature. *J Magn Magn Mater.* 2011;323(17):2214–2217. doi: 10.1016/j.jmmm.2011.03.036

[77] Pekala M, Drozd V. Magnetocaloric effect in nano- and polycrystalline $La_{0.8}Sr_{0.2}MnO_3$ manganites. *J Non-Cryst Solids*. 208;354(47-51):5308-5314. doi: 10.1016/j.jnoncrysol.2008.06.112

[78] Mahato RN, Sethupathi K, Sankaranarayanan V, et al. Large magnetic entropy change in nanocrystalline $Pr_{0.7}Sr_{0.3}MnO_3$. *J Appl Phys.* 2010;107(9):09A943. doi: 10.1063/1.3359810

[79] Bhatt RC, Singh SK, Srivastava PC, et al. Impact of sintering temperature on room temperature magneto-resistive and magneto-caloric properties of $Pr_{2/3}Sr_{1/3}MnO_3$. *J Alloys Compd.* 2013;580:377–382. doi: 10.1016/j.jallcom.2013.06.040

[80] Chen P, Du YW, Ni G. Low-field magnetocaloric effect in $Pr_{0.5}Sr_{0.5}MnO_3$. *Europhys Lett.* 2000;52(5):589–594. doi: 10.1209/epl/i2000-00478-8

[81] Biswas A, Samanta T, Banerjee S, et al. Observation of large low field magnetoresistance and large magnetocaloric effects in polycrystalline $Pr_{0.65}(Ca_{0.7}Sr_{0.3})_{0.35}MnO_3$. *Appl Phys Lett.* 2008;92(1):012502. doi: 0.1063/1.2828980

[82] Biswas A, Samanta T, Banerjee S, et al. Magnetotransport properties of nanocrystalline $Pr_{0.65}(Ca_{1-y}Sr_y)_{0.35}MnO_3$ (y = 0.4, 0.3): Influence of phase coexistence *Appl Phys Lett.* 2007;91(1):013107. doi: 10.1063/1.2753709

[83] Manchón-Gordón AF, Ipus-Bados JJ, Kowalczyk M, et al. Milling effects on the distribution of Curie temperatures and magnetic properties of Ni-doped $La_{0.7}Ca_{0.3}MnO_3$ compounds. *J Alloys Compd.* 2020;848:156566. doi: 10.1016/j.jallcom.2020.156566

[84] Bolarín-Miró AM, Taboada-Moreno CA, Cortés-Escobedo CA, et al. Effect of high-energy ball milling on the magnetocaloric properties of $La_{0.7}Ca_{0.2}Sr_{0.1}MnO_3$. *Appl Phys A.* 2020;126(5):369. doi: 10.1007/s00339-020-03555-w

[85] Ruan MY, Yang CQ, Wang L, et al. Size-dependent magnetocaloric effect in $GdVO_4$ nanoparticles. *J Alloys Compd.* 2022;894:162351. doi: 10.1016/j.jallcom.2021.162351

[86] Chaudhary V, Ramanujan RV. Magnetic and structural properties of high relative cooling power $(Fe_{70}Ni_{30})_{92}Mn_8$ magnetocaloric nanoparticles. *J Phys D Appl Phys.* 2015;48(30):305003. doi: 10.1088/0022-3727/48/30/305003

[87] Bohra M, Alman V, Khan MA, et al.. Analytical modeling of magnetocaloric effect in dense nanoparticle systems. *Nano Select.* 2024;5(6):2300196. doi: 10.1002/nano.202300196

[88] Álvarez-Alonso P, Sánchez Llamazares JL, Sánchez-Valdés CF, et al. On the broadening of the magnetic entropy change due to Curie temperature distribution. *J Appl Phys.* 2014;115(17):17A929. doi: 10.1063/1.4867346

[89] Ahmad A, Mitra S, Srivastava SK, et al. Giant magnetocaloric effect in $Co_2FeAl$ Heusler alloy nanoparticles. *J Phys D Appl Phys.* 2021;54(38):385001. doi: 10.1088/1361-6463/ac0aba

[90] Sánchez-Alarcos V, Recarte V, Pérez-Landazábal JI, et al. Mechanically induced disorder and crystallization process in Ni–Mn–In ball-milled alloys. *J Alloys Compd.* 2016;689:983–991. doi: 10.1016/j.jallcom.2016.08.068

[91] Zeng R, Wang SQ, Du GD, et al. Abnormal magnetic behaviors and large magnetocaloric effect in $MnPS_3$ nanoparticles. *J Appl Phys.* 2012;111(7):07E144. doi: 10.1063/1.3679409

[92] Zhang F, Taake C, Huang B, et al. Magnetocaloric effect in the $(Mn,Fe)_2(P,Si)$ system: From bulk to nano. *Acta Mater.* 2022;224:117532. doi: 10.1016/j.actamat.2021.117532

[93] Lee CG, Nallathambi V, Kang T, et al. Magnetocaloric effect of $Fe_{47.5}Ni_{37.5}Mn_{15}$ bulk and nanoparticles: A cost-efficient alloy for room temperature magnetic refrigeration. *arXiv preprint*. 2024;arXiv:2410.15776. doi: 10.48550/arXiv.2410.15776

[94] Ipus JJ, Ucar H, McHenry ME. Near room temperature magnetocaloric response of an (FeNi)ZrB alloy. *IEEE Trans Magn.* 2011;47(10):2494–2497. doi: 10.1109/TMAG.2011.2159781

[95] Kurth F, Weisheit M, Leistner K, et al. Finite-size effects in highly ordered ultrathin FePt films. *Phys Rev B*. 2010;82:184404. doi: 10.1103/PhysRevB.82.184404

[96] Trassinelli M, Marangolo M, Eddrief M, et al. Suppression of the thermal hysteresis in magnetocaloric MnAs thin film by highly charged ion bombardment. *Appl Phys Lett*. 2014;104:081906. doi: 10.1063/1.4866663

[97] Nguyen BD, Zheng Y, Becerra L, et al. Magnetocaloric effect in flexible, free-standing gadolinium thick films for energy conversion applications. *Phys Rev Appl.* 2021;15(6):064045. doi: 10.1103/PhysRevApplied.15.064045

[98] Kirby BJ, Lau JW, Williams DV, et al. Impact of interfacial magnetism on magnetocaloric properties of thin film heterostructures. *J Appl Phys*. 2011;109(6):063905. doi: 10.1063/1.3555101

[99] Tadout M, Lambert CH, El Hadri MS, et al. Magnetic properties and magnetocaloric effect in $Gd_{100-x}Co_x$ thin films. *Crystals.* 2019;9(6):278. doi: 10.3390/cryst9060278

[100] Hadimani RL, Silva JHB, Pereira AM, et al. $Gd_5(Si,Ge)_4$ thin film displaying large magnetocaloric and strain effects due to magnetostructural transition. *Appl Phys Lett.* 2015;106(3):032402. doi: 10.1063/1.4906056

[101] Pires AL, Belo JH, Gomes IT, et al. Influence of thermal cycling on giant magnetocaloric effect of $Gd_5Si_{1.3}Ge_{2.7}$ thin film. *arXiv preprint* arXiv:1505.02575. doi: 10.48550/arXiv.1505.02575

[102] Modak R, Manivel RM, Srinivasan A. Enhanced magneto-caloric effect upon Co substitution in Ni-Mn-Sn thin films. *J Magn Magn Mater.* 2017;448:146–152. doi: 10.1016/j.jmmm.2017.06.063

[103] Moya X, Hueso LE, Maccherozzi F, et al. Giant and reversible extrinsic magnetocaloric effects in $La_{0.7}Ca_{0.3}MnO_3$ films due to strain. *Nat Mater.* 2013;12(1):52–58. doi: 10.1038/nmat3463

[104] Passanante SE, Goijman DY, Linares MMM, et al. Magnetocaloric effect in $La_{0.88}Sr_{0.12}MnO_3$ films. *Mater Today Proc.* 2019;14:104–108. doi: 10.1016/j.matpr.2019.05.063

[105] Kumar VS, Chukka R, Chen Z, et al. Strain dependent magnetocaloric effect in $La_{0.67}Sr_{0.33}MnO_3$ thin-films. *AIP Adv.* 2013;3(5):052127. doi: 10.1063/1.4807739

[106] Giri SK, Akram W, Bansal M, et al. Tuning the magnetocaloric effect by optimizing thickness-induced three-dimensional strain states. *Phys Rev B.* 2021;104(22):224432. doi: 10.1103/PhysRevB.104.224432

[107] Mitrofanov VYa, Estemirova SKh, Kozhina GA. Effect of oxygen content on structural, magnetic and magnetocaloric properties of $(La_{0.7}Pr_{0.3})_{0.8}Sr_{0.2}Mn_{0.9}Co_{0.1}O_{3\pm\delta}$. *J Magn Magn Mater*. 2019;476:199–206. doi: 10.1016/j.jmmm.2018.12.097

[108] Nadig PR, Murari MS, Daivajna MD. Influence of heat sintering on the physical properties of bulk $La_{0.67}Ca_{0.33}MnO_3$ perovskite manganite: role of oxygen in tuning the magnetocaloric response. *Phys Chem Chem Phys*. 2024;26:5237–5252. doi: 10.1039/D3CP04185A

[109] M'nassri R, Cheikhrouhou A. Evolution of magnetocaloric behavior in oxygen deficient $La_{2/3}Ba_{1/3}MnO_{3-\delta}$ manganites. *J Supercond Nov Magn*. 2014;27:1463–1468. doi: 10.1007/s10948-013-2459-y

[110] Lampen-Kelley P, Madhogaria R, Bingham NS, et al. Tablelike magnetocaloric effect and enhanced refrigerant capacity in $EuO_{1-\delta}$ thin films. *Phys Rev Mater.* 2021;5(9):094404. doi: 10.1103/PhysRevMaterials.5.094404

[111] Patra L, Quan Y, Liao B. Impact of dimensionality on the magnetocaloric effect in two-dimensional magnets. *J Appl Phys.* 2024;136(2):024301. doi: 10.1063/5.0218007

[112] Das R, Prabhu R, Venkataramani N, et al. Giant low-field magnetocaloric effect and refrigerant capacity in reduced dimensionality $EuTiO_3$ multiferroics. *J Alloys Compd.* 2021;850:156819. doi: 10.1016/j.jallcom.2020.156819

[113] Qiao K, Liang Y, Zhang H, et al. Manipulation of magnetocaloric effect in FeRh films by epitaxial growth. J Alloys Compd. 2022;907:164574. doi: 10.1016/j.jallcom.2022.164574

[114] Liu Y, Phillips LC, Mattana R, et al. Large reversible caloric effect in FeRh thin films via a dual-stimulus multicaloric cycle. *Nat Commun*. 2016;7:11614. doi: 10.1038/ncomms11614

[115] Jia JH, Ke YJ, Li X, et al. A large magnetocaloric effect of $GdCoO_{3-\delta}$ epitaxial thin films prepared by a polymer assisted spin-coating method. *J Mater Chem C.* 2019;7:14970–14976. doi: 10.1039/C9TC04464G

[116] Phan MH, Kalappattil V, Ortiz JV, et al. Exchange bias and interface-related effects in two-dimensional van der Waals magnetic heterostructures: open questions and perspectives. *J Alloys Compd*. 2023;937:168375. doi: 10.1016/j.jallcom.2022.168375

[117] Hellman F, Hoffmann A, Tserkovnyak Y, et al. Interface-induced phenomena in magnetism. *Rev Mod Phys*. 2017;89:025006. doi: 10.1103/RevModPhys.89.025006

[118] Hamad MA, Hemeda OM, Alamri HR, et al. $BiFeO_3$ layer thicknesses effect on magnetocaloric effect in $BiFeO_3|La_{0.7}Sr_{0.3}MnO_3$ thin films. *Phys Solid State.* 2021;63:709–713. doi: 10.1134/S1063783421050085

[119] Vdovichev SN, Polushkin NI, Rodionov ID, et al. High magnetocaloric efficiency of a NiFe/NiCu/CoFe/MnIr multilayer in a small magnetic field. *Phys Rev B.* 2018;98(1):014428. doi: 10.1103/PhysRevB.98.014428

[120] Ghosh A, Roy R, Sahoo RC, et al. Magnetic anisotropy and magnetocaloric effect in $Gd_2NiMnO_6$ thin films. *Phys Rev B.* 2023;108(21):214423. doi: 10.1103/PhysRevB.108.214423

[121] El Hadri MS, Polewczyk V, Xiao Y, et al. Large anisotropic magnetocaloric effect in all-sputtered epitaxial terbium thin films. *Phys Rev Mater.* 2020;4(12):124404. doi: 10.1103/PhysRevMaterials.4.124404

[122] Oliveira NA, Ranke PJ. Theoretical aspects of the magnetocaloric effect. *Phys Rep.* 2010;489(4-5):89–159. doi: 10.1016/j.physrep.2009.12.006

[123] Tishin AM, Spichkin YI, Zverev VI, et al. A review and new perspectives for the magnetocaloric effect: New materials and local heating and cooling inside the human body. Int J Refrig. 2016;68:177–186. doi: 10.1016/j.ijrefrig.2016.04.020

[124] Balli M, Jandl S, Fournier P, et al. Anisotropy-enhanced giant reversible rotating magnetocaloric effect in $HoMn_2O_5$ single crystals. *Appl Phys Lett.* 2014;104:232402. doi: 10.1063/1.4880818

[125] Chung KB, Choi YK, Jang MH, et al. Formation of $GdSi_2$ film on Si(111) via phase transformation assisted by interfacial $SiO_2$ layer. J Vac Sci Technol B. 2005;23(1):153–156. doi: 10.1116/1.1849222

[126] Jagadish KG, Guo Z, Gu L, et al. Broad table-like magnetocaloric effect in GdFeCo thin-films for room temperature Ericsson-cycle magnetic refrigeration. *J Appl Phys.* 2024;135(12):123904. doi: 10.1063/5.0191497

[127] Gu L, Galivarapu JK, Wang Z, et al. Interface-induced conventional and inverse magnetocaloric properties of GdFeCo thin films. ACS Appl Electron Mater. 2025;7(5):1812–1819. doi: 10.1021/acsaelm.4c02066

[128] Recarte V, Pérez-Landazábal JI, Sánchez-Alárcos V, et al. Magnetocaloric effect linked to the martensitic transformation in sputter-deposited Ni–Mn–Ga thin films. *Appl Phys Lett.* 2009;95(14):141908. doi: 10.1063/1.3246149

[129] Zhang Y, Hughes RA, Britten JF, et al. Magnetocaloric effect in Ni-Mn-Ga thin films under concurrent magnetostructural and Curie transitions. *J Appl Phys.* 2011;110(1):013910. doi: 10.1063/1.3602088

[130] Ni Y, Wang H, Karaman I, et al. Metamagnetic transitions and magnetocaloric effect in epitaxial Ni–Co–Mn–In films. *Appl Phys Lett.* 2010;97(22):222507. doi: 10.1063/1.3517443

[131] Yüzüak E, Dincer I, Elerman Y, et al. Inverse magnetocaloric effect of epitaxial Ni–Mn–Sn thin films. *Appl Phys Lett.* 2013;103(22):222403. doi: 10.1063/1.4834357

[132] Schleicher B, Niemann R, Schwabe S, et al. Reversible tuning of magnetocaloric Ni-Mn-Ga-Co films on ferroelectric PMN-PT substrates. *Sci Rep.* 2017;7:14462. doi: 10.1038/s41598-017-14525-3

[133] Campillo G, Martínez B, Labarta A. Effect of 20% Cr-doping on structural and electrical properties of $La_{0.7}Ca_{0.3}MnO_3$ thin films. *J Phys Conf Ser.* 2019;1247(1):012013. doi: 10.1088/1742-6596/1247/1/012013

[134] Lampen P, Bingham NS, Phan MH, et al. Impact of reduced dimensionality on the magnetic and magnetocaloric response of $La_{0.7}Ca_{0.3}MnO_3$. *Appl Phys Lett.* 2013;102(6):062414. doi: 10.1063/1.4792239

[135] Oumezzine M, Chirila CF, Pasuk I, et al. Magnetocaloric and giant magnetoresistance effects in La-Ba-Mn-Ti-O epitaxial thin films: Influence of phase transition and magnetic anisotropy. *Materials.* 2022;15(22):8003. doi: 10.3390/ma15228003

[136] Zhao B, Hu X, Dong F, et al. The magnetic properties and magnetocaloric effect of $Pr_{0.7}Sr_{0.3}MnO_3$ thin film grown on $SrTiO_3$ substrate. *Materials.* 2023;16(1):75. doi: 10.3390/ma16010075

[137] Matte D, de Lafontaine M, Ouellet A, et al. Tailoring the magnetocaloric effect in $La_2NiMnO_6$ thin films. *Phys Rev Appl.* 2018;9(5):054042. doi: 10.1103/PhysRevApplied.9.054042

[138] Bouhani H, Endichi A, Kumar D, et al. Engineering the magnetocaloric properties of $PrVO_3$ epitaxial oxide thin films by strain effects. *Appl Phys Lett.* 2020;117(7):072402. doi: 10.1063/5.0021031

[139] Induja S, Janani V, Jaison D, et al. Effect of annealing on magnetic and magnetocaloric properties of $SmCo_3B_2$ thin films. *Solid State Commun.* 2023;373–374:115321. doi: 10.1016/j.ssc.2023.115321

[140] Kim M, Kim JW, Lim SH. Magnetic and magnetocaloric properties of Er-Co-Al thin-film alloys. *Met Mater Int.* 2015;21(6):1101–1107. doi: 10.1007/s12540-015-5063-9

[141] Guo Z, Kumar GJ, Wang Z, et al. Magnetocaloric properties of ferrimagnetic TbFeCo thin films near compensation temperature with perpendicular anisotropy: effect of sputtering power. *Appl Phys A.* 2025;131:96. doi: 10.1007/s00339-025-08240-4

[142] Saqat RSH, Forbes A, Philip J. Magnetic properties and magnetocaloric effect of $(Fe_{70}Ni_{30})_{96}Mo_4$ thin films grown by molecular beam epitaxy. *J Vac Sci Technol A.* 2023;41(1):013404. doi: 10.1116/6.0002213

[143] Jiang T, Xie L, Yao Y, et al. Large magnetocaloric effect in $CrO_2/TiO_2$ epitaxial films above room temperature. *Mater Lett.* 2012;76:25–27. doi: 10.1016/j.matlet.2012.02.057

[144] Shaji S, Mucha NR, Giri P, et al. Magnetic and magnetocaloric properties of $Fe_2Ta$ thin films. *AIP Adv.* 2020;10(2):025222. doi: 10.1063/1.5134796

[145] Liu Q, Liu Y, Jiang X, et al. Magnetocaloric effect near room temperature in a freestanding two-dimensional non–van der Waals crystal of MnCoAs. *Phys Rev B.* 2023;108(5):054427. doi: 10.1103/PhysRevB.108.054427

[146] Sas W, Fitta M. The study of magnetocaloric effect in the electrodeposited thin films of $M_3[Cr(CN)_6]_2 \cdot zH_2O$ (M = Fe and Cr). *J Magn Magn Mater.* 2023;575:170719. doi: 10.1016/j.jmmm.2023.170719

[147] Polushkin NI, Kravtsov EA, Vdovichev SN, et al. Magnetic and magnetocaloric properties of Py/Gd/CoFe/IrMn stacks. *J Magn Magn Mater.* 2019;491:165601. doi: 10.1016/j.jmmm.2019.165601

[148] Passanante S, Granja LP, Albornoz C, et al. Magnetocaloric effect in nanocrystalline manganite bilayer thin films. *J Magn Magn Mater.* 2022;559:169545. doi: 10.1016/j.jmmm.2022.169545

[149] Kuznetsov MA, Pashenkin IY, Polushkin NI, et al. Magnetocaloric effect in exchange-coupled strong/weak/strong ferromagnetic trilayers. *J Appl Phys.* 2020;127(18):183904. doi: 10.1063/5.0003223

[150] Kulyk M, Persson M, Polishchuk D, et al. Magnetocaloric effect in multilayers studied by membrane-based calorimetry. *J Phys D Appl Phys.* 2023;56(2):025002. doi: 10.1088/1361-6463/aca67f

[151] Persson M, Kulyk MM, Kravets AF, et al. Proximity-enhanced magnetocaloric effect in ferromagnetic trilayers. *J Phys: Condens Matter.* 2023;35(7):075801. doi: 10.1088/1361-648X/ac9f95

[152] Kuznetsov MA, Karashtin EA. Exchange enhancement of magnetocaloric effect in a ferromagnet/antiferromagnet/ferromagnet layered structure. *Phys Rev B.* 2024;109(22):224432. doi: 10.1103/PhysRevB.109.224432

[153] Kuznetsov MA, Drovosekov AB, Fraerman AA. Magnetocaloric effect in nanosystems based on ferromagnets with different Curie temperatures. *J Exp Theor Phys.* 2021;132(1):79–93. doi: 10.1134/S1063776121010131

[154] Dong QY, Hou KY, Zhang XQ, et al. Giant reversible magnetocaloric effect in antiferromagnetic rare-earth cobaltide $GdCoO_3$. *J Appl Phys*. 2020;127:033904. doi: 10.1063/1.5132864

[155] Xue L, Luo Q, Shao L, et al. Magnetocaloric difference between ribbon and bulk shape of Gd-based metallic glasses. *J Magn Magn Mater.* 2020;497:166015. doi: 10.1016/j.jmmm.2019.166015

[156] Garrido Álvarez JL, Álvarez-Alonso P, Sánchez-Valdés CF, et al. Exploring the magnetic and magnetocaloric behavior of nanocrystalline melt-spun $R_2Fe_{17}$ (R = Pr, Nd) ribbons. *J Alloys Compd*. 2024;979:173575. doi: 10.1016/j.jallcom.2024.173575

[157] Deltell A, Mohamed AEA, Álvarez-Alonso P, et al. Martensitic transformation, magnetic and magnetocaloric properties of Ni–Mn–Fe–Sn Heusler ribbons. *J Mater Res Technol*. 2021;12:1091–1103. doi: 10.1016/j.jmrt.2021.03.049

[158] Jayaraman TV, Boone L, Shield JE. Magnetocaloric effect and refrigerant capacity in melt-spun Gd–Mn alloys. *J Magn Magn Mater.* 2013;345:153–158. doi: 10.1016/j.jmmm.2013.06.016

[159] Zhang CL, Wang DH, Han ZD, et al. Large magnetic entropy changes in Gd–Co amorphous ribbons. *J Appl Phys.* 2009;105(1):013912–013914. doi: 10.1063/1.3040009

[160] Zhong XC, Tang PF, Liu ZW, et al. Magnetic properties and large magnetocaloric effect in Gd–Ni amorphous ribbons for magnetic refrigeration applications in intermediate temperature range. *J Alloys Compd.* 2011;509(24):6889–6892. doi: 10.1016/j.jallcom.2011.03.173

[161] Xia L, Zhang Z, Liu G, et al. Magnetocaloric response of the $Gd_{60}Co_{25}Al_{15}$ metallic glasses. *Appl Phys Lett.* 2014;105(19):192402. doi: 10.1063/1.4901263

[162] Yu P, Zhang NZ, Cui YT, et al. Achieving an enhanced magnetocaloric effect by melt spinning a $Gd_{55}Co_{25}Al_{20}$ bulk metallic glass into amorphous ribbons. *J Alloys Compd.* 2016;655:353–356. doi: 10.1016/j.jallcom.2015.09.205

[163] Yuan F, Li Q, Shen B. The effect of Fe/Al ratio on the thermal stability and magnetocaloric effect of $Gd_{55}Fe_xAl_{45-x}$ (x = 15–35) glassy ribbons. *J Appl Phys.* 2012;111(7):07A937. doi: 10.1063/1.3677780

[164] Yuan F, Du J, Shen B. Controllable spin-glass behavior and large magnetocaloric effect in Gd–Ni–Al bulk metallic glasses. *Appl Phys Lett.* 2012;101(3):032405. doi: 10.1063/1.4738778

[165] Aliev AM, Batdalov AB, Kamilov IK, Koledov VV, et al. Magnetocaloric effect in ribbon samples of Heusler alloys Ni–Mn–M (M = In, Sn). *Appl Phys Lett.* 2010;97(21):212505. doi: 10.1063/1.3521261

[166] Ibarra-Gaytán PJ, Sánchez-Valdés CF, Sánchez Llamazares JL, et al. High-magnetic field characterization of magnetocaloric effect in FeZrB(Cu) amorphous ribbons. *J Appl Phys.* 2015;117(17):17A710. doi: 10.1063/1.4907188

[167] Thanh TD, Yu Y, Thanh PT, et al. Magnetic properties and magnetocaloric effect in $Fe_{90-x}Ni_xZr_{10}$ alloy ribbons. *J Appl Phys.* 2013;113(21):213908. doi: 10.1063/1.4809754

[168] Shen BG, Hu FX, Dong QY, et al. Magnetic properties and magnetocaloric effects in $NaZn_{13}$-type $La(Fe,Al)_{13}$-based compounds. Chin Phys B. 2013;22(1):017502. doi: 10.1088/1674-1056/22/1/017502

[169] Hu FX, Shen BG, Sun JR, et al. Very large magnetic entropy change near room temperature in $LaFe_{11.2}Co_{0.7}Si_{1.1}$. Appl Phys Lett. 2002;80(5):826–828. doi: 10.1063/1.1447592

[170] Hu FX, Shen BG, Sun JR, et al. Large magnetic entropy change in $La(Fe_{1-x}Co_x)_{11.83}Al_{1.17}$ (x = 0.06, 0.08). *Phys Rev B*. 2001;64(1):012409. doi: 10.1103/PhysRevB.64.012409

[171] Wang GF, Tan X, Yang BY, et al. Hydrogenation and magnetocaloric effect in La-excessive $LaxFe_{11.5}Si_{1.5}$ Hδ alloys. *J Alloys Compd*. 2020;816:152614. doi: 10.1016/j.jallcom.2019.152614

[172] Hou XL, Lampen-Kelley P, Xue Y, et al. Formation mechanisms of $NaZn_{13}$-type phase in giant magnetocaloric La–Fe–Si compounds during rapid solidification and annealing. *J Alloys Compd*. 2015;646:503–511. doi: 10.1016/j.jallcom.2015.05.173

[173] Sánchez-Llamazares JL, Álvarez-Alonso P, Sánchez-Valdés CF, et al. Investigating the magnetic entropy change in single-phase $Y_2Fe_{17}$ melt-spun ribbons. *Curr Appl Phys.* 2016;16(9):963–968. doi: 10.1016/j.cap.2016.05.013

[174] Fang YK, Chang CW, Yeh CC, et al. Microstructure and magnetocaloric effect of melt-spun $Y_2Fe_{17}$ ribbons. *J Appl Phys.* 2008;103(7):07B302. doi: 10.1063/1.2829031

[175] Dahal B, Kharel P, Ott T, et al. Magnetic and magnetocaloric properties of $Pr_{2-x}Nd_xFe_{17}$ ribbons. *AIP Adv.* 2019;9(3):035211. doi: 10.1063/1.5080105

[176] Fu Q, Liu ZH, Hao ZH. Magnetic properties and magnetocaloric effect (MCE) in the melt-spun $Tm_{20}Ho_{20}Gd_{20}Ni_{20}Al_{20}$ amorphous ribbon. *Solid State Commun.* 2023;364:115137. doi: 10.1016/j.ssc.2023.115137

[177] Svalov A, Andreev S, Arkhipov A, et al. Magnetic and magnetocaloric properties of Gd melt-spun ribbons. *J Phys Conf Ser.* 2019;1389(1):012100. doi: 10.1088/1742-6596/1389/1/012100

[178] Wang ZW, Yu P, Cui YT, et al. Near room temperature magneto-caloric effect of a $Gd_{48}Co_{52}$ amorphous alloy. *J Alloys Compd.* 2016;658:598–602. doi: 10.1016/j.jallcom.2015.10.293

[179] Zhang JL, Zheng ZG, Cao WH, et al. Magnetic behavior of $Gd_4Co_3$ metallic glass. *J Magn Magn Mater.* 2013;326:157–161. doi: 10.1016/j.jmmm.2012.09.002

[180] Schwarz B, Gutfleisch O. Influence of sample geometry on determination of magnetocaloric effect for $Gd_{60}Co_{30}Al_{10}$ glassy ribbons using direct and indirect methods. *J Magn Magn Mater.* 2011;323(13):1782–1786. doi: 10.1016/j.jmmm.2011.02.004

[181] Tang BZ, Guo DQ, Ding D, et al. Large adiabatic temperature rise above the water ice point of a minor Fe substituted $Gd_{50}Co_{50}$ amorphous alloy. *J Non-Cryst Solids.* 2017;464:30–33. doi: 10.1016/j.jnoncrysol.2017.03.016

[182] Tang BZ, Yu P, Ding D, et al. Improved magneto-caloric effect of the $Gd_{50}Co_{50}$ metallic glass by minor Si addition. *J Magn Magn Mater.* 2017;424:275–278. doi: 10.1016/j.jmmm.2016.10.081

[183] Yu P, Zhang NZ, Cui YT, et al. Achieving better magneto-caloric effect near room temperature in amorphous $Gd_{50}Co_{50}$ alloy by minor Zn addition. *J Non-Cryst Solids.* 2016;434:36–40. doi: 10.1016/j.jnoncrysol.2015.12.007

[184] Zhong XC, Huang XW, Shen XY, et al. Thermal stability, magnetic properties and large refrigerant capacity of ternary $Gd_{55}Co_{35}M_{10}$ (M = Mn, Fe, and Ni) amorphous alloys. *J Alloys Compd.* 2016;682:476–480. doi: 10.1016/j.jallcom.2016.04.307

[185] Gencer H, Izgi T, Kolat VS, et al. The crystallisation kinetics, magnetic and magnetocaloric properties of $Gd_{55}Co_{20}Fe_5Al_{20-x}Si_x$ (x = 0, 5, 10, 15) alloys. *J Non-Cryst Solids.* 2013;379:185–191. doi: 10.1016/j.jnoncrysol.2013.08.009

[186] Atalay S, Gencer H, Kaya AO, et al Influence of Si substitution on the structural, magnetic and magnetocaloric properties of $Gd_{55}Co_{20}Fe_5Al_{20-x}Si_x$ alloys. *J Non-Cryst Solids.* 2013;365:99–104. doi: 10.1016/j.jnoncrysol.2013.01.042

[187] Zheng Q, Zhang L, Du J. Magnetic entropy change in $Gd_{95}Fe_{2.8}Al_{2.2}$ amorphous/ nanocrystalline ribbons. *Scr Mater.* 2017;130:170–173. doi: 10.1016/j.scriptamat.2016.11.041

[188] Dong QY, Shen BG, Chen J, et al. Large magnetic refrigerant capacity in $Gd_{71}Fe_3Al_{26}$ and $Gd_{65}Fe_{20}Al_{15}$ amorphous alloys. *J Appl Phys.* 2009;105(5):053908. doi: 10.1063/1.3072631

[189] Rajivgandhi R, Arout Chelvane J, Quezado S, et al. Effect of rapid quenching on the magnetism and magnetocaloric effect of equiatomic rare earth intermetallic compounds RNi (R = Gd, Tb and Ho). *J Magn Magn Mater.* 2017;433:169–177. doi: 10.1016/j.jmmm.2017.03.011

[190] Li Z, Ding D, Xia L. Excellent magneto-caloric effect of a binary $Gd_{63}Ni_{37}$ amorphous alloy. *Intermetallics.* 2017;86:11–14. doi: 10.1016/j.intermet.2017.03.007

[191] Chang J, Hui X, Xu ZY, et al. Ni–Gd–Al metallic glasses with large magnetocaloric effect. *Intermetallics.* 2010;18(6):1132–1136. doi: 10.1016/j.intermet.2010.02.015

[192] Yu P, Wu C, Cui YT, et al. Excellent magneto-caloric effect of a low cost $Gd_{34}Ni_{22}Co_{11}Al_{33}$ metallic glass. *Mater Lett.* 2016;173:239–241. doi: 10.1016/j.matlet.2016.03.053

[193] Min JX, Zhong XC, Liu ZW, et al. Magnetic properties and magnetocaloric effects of Gd–Mn–Si ribbons in amorphous and crystalline states. *J Alloys Compd.* 2014;606:50–54. doi: 10.1016/j.jallcom.2014.04.014

[194] Zheng ZG, Zhong XC, Liu ZW, et al. Magnetocaloric effect and critical behavior of amorphous $(Gd_4Co_3)_{1-x}Si_x$ alloys. *J Magn Magn Mater.* 2013;343:184–188. doi: /10.1016/j.jmmm.2013.04.087

[195] Shishkin DA, Baranov NV, Volegov AS, et al. Substitution and liquid quenching effects on magnetic and magnetocaloric properties of $(Gd_{1-x}Tb_x)_{12}Co_7$. *Solid State Sci.* 2016;52:92–96. doi: 10.1016/j.solidstatesciences.2015.12.012

[196] Dong QY, Shen BG, Chen J, et al. Magnetic properties and magnetocaloric effects in amorphous and crystalline GdCuAl ribbons. *Solid State Commun.* 2009;149(21-22):417–420. doi: 10.1016/j.ssc.2008.12.006

[197] Zhong XC, Tang PF, Liu ZW, et al. Large magnetocaloric effect and refrigerant capacity in Gd–Co–Ni metallic glasses. *J Appl Phys.* 2012;111(7):07A919. doi: 10.1063/1.3673422

[198] Mayer C, Chevalier B, Gorsse S. Magnetic and magnetocaloric properties of the ternary Gd-based metallic glasses $Gd_{60}Mn_{30}X_{10}$, with X = Al, Ga, In. *J Alloys Compd.* 2010;507(1):370–375. doi: 10.1016/j.jallcom.2010.07.210

[199] Mayer C, Gorsse S, Ballon G, et al. Tunable magnetocaloric effect in Gd-based glassy ribbons. *J Appl Phys.* 2011;110(5):053920. doi: 10.1063/1.3632983

[200] Schwarz B, Podmilsak B, Mattern N, et al. Magnetocaloric effect in Gd-based $Gd_{60}Fe_xCo_{30-x}Al_1$ metallic glasses. *J Magn Magn Mater.* 2010;322(16):2298–2303. doi: 10.1016/j.jmmm.2010.02.029

[201] Fang YK, Chen HC, Hsieh CC, et al. Structures and magnetocaloric effects of $Gd_{65-x}RE_xFe_{20}Al_{15}$ (x = 0–20; RE = Tb, Dy, Ho, and Er) ribbons. *J Appl Phys.* 2011;109(7):07A933. doi: 0.1063/1.3561447

[202] Mo HY, Zhong XC, Jiao DL, et al. Table-like magnetocaloric effect and enhanced refrigerant capacity in crystalline $Gd_{55}Co_{35}Mn_{10}$ alloy melt-spun ribbons. *Phys Lett A.* 2018;382(22):1679–1684. doi: 10.1016/j.physleta.2018.03.053

[203] Pierunek N, Śniadecki Z, Marcin J, et al. Magnetocaloric effect of amorphous $Gd_{65}Fe_{10}Co_{10}Al_{10}X_5$ (X = Al, Si, B) alloys. *IEEE Trans Magn.* 2014;50(11):2506603. doi: 10.1109/TMAG.2014.2318595

[204] Zhong XC, Tang PF, Gao BB, et al. Magnetic properties and magnetocaloric effects in amorphous and crystalline $Gd_{55}Co_{35}Ni_{10}$ ribbons. *Sci China Phys Mech Astron.* 2013;56(6):1096–1099. doi: 10.1007/s11433-013-5082-9

[205] Liu GL, Zhang L, Du J. Table-like magnetocaloric effect in Gd–Ni–Al amorphous/nanocrystalline composites. *J Phys D Appl Phys.* 2016;49(5):055004. doi: 10.1088/1361-6463/aa7a8f

[206] Zhang YK, Guo D, Wu BB, et al. Magnetic properties, magnetocaloric effect and refrigeration performance in $RE_{60}Al_{20}Ni_{20}$ (RE = Tm, Er and Ho) amorphous ribbons. *J Appl Phys.* 2020;127(3):033905. doi: 10.1063/1.5140765

[207] Sánchez-Llamazares JL, Ibarra-Gaytán PJ, Sánchez-Valdés CF, et al. Magnetocaloric effect in $ErNi_2$ melt-spun ribbons. *J Rare Earths.* 2020;38(6):612–616. doi: 10.1016/j.jre.2019.07.011

[208] Sánchez-Llamazares JL, Sánchez-Valdés CF, Ibarra-Gaytán PJ, et al. Magnetic entropy change and refrigerant capacity of rapidly solidified $TbNi_2$ alloy ribbons. *J Appl Phys.* 2013;113(17):17A912. doi: 10.1063/1.4794988

[209] Ibarra-Gaytán PJ, Sánchez-Valdés CF, Sánchez-Llamazares JL, et al. Texture-induced enhancement of the magnetocaloric response in melt-spun $DyNi_2$ ribbons. *Appl Phys Lett.* 2013;103(15):152401. doi: 10.1063/1.4824073

[210] Sánchez-Llamazares JL, Ibarra-Gaytán PJ, Sánchez-Valdés CF, et al. Enhanced magnetocaloric effect in rapidly solidified $HoNi_2$ melt-spun ribbons. *J Alloys Compd.* 2019;774:700–705. doi: 10.1016/j.jallcom.2018.09.305

[211] Sánchez-Llamazares JL, Ibarra-Gaytán PJ, Sánchez-Valdés CF, et al. Magnetocaloric properties of rapidly solidified $Dy_3Co$ alloy ribbons. *J Appl Phys.* 2015;117(17):17A706. doi: 10.1063/1.4906764

[212] Wang X, Chan KC, Zhao L, et al. Microstructure and its effect on the magnetic, magnetocaloric and magnetostrictive properties of $Tb_{55}Co_{30}Fe_{15}$ glassy ribbons. *Materials (Basel).* 2021;14(11):3068. doi: 10.3390/ma14113068

[213] Li H, Feng S, Ren J, et al. Magnetostructural transitions in Mn-rich Heusler Mn–Ni–In melt-spun ribbons with enhanced magnetocaloric effect. *J Magn Magn Mater.* 2015;391:17–20. doi: 10.1016/j.jmmm.2015.04.098

[214] Hernando B, Sánchez-Llamazares JL, Prida VM, et al. Magnetocaloric effect in preferentially textured $Mn_{50}Ni_{40}In_{10}$ melt-spun ribbons. *Appl Phys Lett.* 2009;94(22):222502. doi: 10.1063/1.3147875

[215] Li Z, Zhang Y, Sánchez-Valdés CF, et al. Giant magnetocaloric effect in melt-spun Ni-Mn-Ga ribbons with magneto-multistructural transformation. *Appl Phys Lett.* 2014;104(4):044101. doi: 10.1063/1.4863273

[216] Zhao XG, Lee EH, Hsieh C, et al. Phase evolution and magnetocaloric effect of melt-spun $Mn_3Sn_{2-x}M_x$ (M = B, C; x = 0–0.5) ribbons. *J Appl Phys.* 2012;111(7):07A912. doi: 10.1063/1.3671789

[217] Sánchez-Llamazares JL, Flores-Zúñiga H, Sánchez-Valdés CF, et al. Refrigerant capacity of austenite in as-quenched and annealed $Ni_{51.1}Mn_{31.2}In_{17.7}$ melt-spun ribbons. *J Appl Phys.* 2012;111(7):07A932. doi: 10.1063/1.3676606

[218] Kaya M, Elerman Y, Dincer I. Effect of heat treatment procedure on magnetic and magnetocaloric properties of $Ni_{43}Mn_{46}In_{11}$ melt-spun ribbons. *Philosophical Magazine.* 2018;98(21):1929–1942. doi: 0.1080/14786435.2018.1465240

[219] Li ZB, Sánchez-Llamazares JL, Sánchez-Valdés CF, et al. Microstructure and magnetocaloric effect of melt-spun $Ni_{52}Mn_{26}Ga_{22}$ ribbon. *Appl Phys Lett.* 2012;100(17):174102. doi: 10.1063/1.4704780

[220] Yang Y, Li Z, Li Z, et al. Microstructural feature and magnetocaloric effect of $Mn_{50}Ni_{40.5}In_{9.5}$ melt-spun ribbons. *Crystals.* 2017;7(10):289. doi: 10.3390/cryst7100289

[221] Nguyen Y, Nguyen MT, Vu Q, et al. Investigation of magnetic phase transition and magnetocaloric effect of (Ni,Co)-Mn-Al melt-spun ribbons. *EPJ Web Conf.* 2018;185:05001. doi: 10.1051/epjconf/201818505001

[222] Sahoo R, Raj KDM, Arvindha BD, et al. In-plane and out-of-plane magnetic properties in $Ni_{46}Co_4Mn_{38}Sb_{12}$ Heusler alloy ribbons. *J Appl Phys.* 2013;113(17):17A940. doi: 10.1063/1.4800505

[223] Ma SC, Ge Q, Yang S, et al. Microstructure, magnetic and magnetocaloric properties in $Ni_{42.9}Co_{6.9}Mn_{38.3}Sn_{11.9}$ alloy ribbons. *AIP Adv.* 2018;8(5):056410. doi: 10.1063/1.5006245

[224] Pandey S, Quetz A, Ibarra-Gaytán PJ, et al. Magnetostructural transitions and magnetocaloric effects in $Ni_{50}Mn_{35}In_{14.25}B_{0.75}$ ribbons. *AIP Adv.* 2018;8(5):056434. doi: 10.1063/1.5006467

[225] Pandey S, Quetz A, Ibarra-Gaytán PJ, et al. Magnetic and martensitic transformations in $Ni_{48}Co_2Mn_{35}In_{15}$ melt-spun ribbons. *AIP Adv.* 2018;8(10):101410. doi: 10.1063/1.5041954

[226] Guan W, Liu QR, Gao B, et al. Large magnetocaloric effect at low magnetic field in $Ni_{50-x}Co_xMn_{35}In_{15}$ ribbons. *J Appl Phys.* 2011;109(7):07A903. doi: 10.1063/1.3540649

[227] Bruno NM, Huang YJ, Dennis CL, et al. Effect of grain constraint on the field requirements for magnetocaloric effect in $Ni_{45}Co_5Mn_{40}Sn_{10}$ melt-spun ribbons. *J Appl Phys.* 2016;120(7):075101. doi: 10.1063/1.4960353

[228] Ma SC, Cao QQ, Xuan HC, et al. Magnetic and magnetocaloric properties in melt-spun and annealed $Ni_{42.7}Mn_{40.8}Co_{5.2}Sn_{11.3}$ ribbons. *J Alloys Compd.* 2011;509(4):1111–1114. doi: 10.1016/j.jallcom.2010.09.205

[229] Peng J, Tang B, Wang Q, et al. Effect of heavy rare-earth (Dy, Tb, Gd) addition on the glass-forming ability and magneto-caloric properties of $Fe_{89}Zr_7B_4$ amorphous alloy. *J Alloys Compd.* 2022;925:166707. doi: 10.1016/j.jallcom.2022.166707

[230] Wang X, Wang Q, Tang BZ, et al. Magnetic and magneto-caloric properties of the amorphous $Fe_{92-x}Zr_8B_x$ ribbons. *Materials (Basel).* 2020;13(23):5334. doi: 10.3390/ma13235334

[231] Yu P, Zhang JZ, Xia L. Effect of boron on the magneto-caloric effect in $Fe_{91-x}Zr_9B_x$ (x = 3, 4, 5) amorphous alloys. *J Mater Sci.* 2017;52(22):13948–13955. doi: 10.1007/s10853-017-1476-9

[232] Thanh TD, Yen NH, Duc NH, et al. Large magnetocaloric effect around room temperature in amorphous Fe-Gd-Zr alloy ribbon with short-range interactions. *J Electron Mater.* 2016;45(5):2608–2614. doi: 10.1007/s11664-016-4431-7

[233] Caballero-Flores R, Franco V, Conde A, et al. Influence of Mn on the magnetocaloric effect of nanoperm-type alloys. *J Appl Phys.* 2010;108(7):073921. doi: 10.1063/1.3489990

[234] Franco V, Conde A, Kiss LF. Magnetocaloric response of FeCrB amorphous alloys: Predicting the magnetic entropy change from the Arrott–Noakes equation of state. *J Appl Phys.* 2008;104(3):033903. doi: 10.1063/1.2961310

[235] Li X, Pan Y. Magnetocaloric effect in Fe-Zr-B-M (M = Ni, Co, Al, and Ti) amorphous alloys. *J Appl Phys.* 2014;116(9):093910. doi: 10.1063/1.4895048

[236] Li X, Pan Y, Lu T. Magnetocaloric effect in Fe-based amorphous alloys and their composites with low boron content. *J Non-Cryst Solids.* 2018;487:7–11. doi: 10.1016/j.jnoncrysol.2018.02.022

[237] Wang Q, Tang B, Ding D, et al. Formation and magnetocaloric properties of the amorphous $Fe_{88}La_{7-x}Ce_xB_5$ (x = 0–7) ribbons. *J Phys Chem Solids.* 2022;169:110854. doi: 10.1016/j.jpcs.2022.110854

[238] Phan TL, Dan NH, Thanh TD, et al. Magnetic properties and magnetocaloric effect in $Fe_{90-x}Sn_xZr_{10}$ alloy ribbons. *J Korean Phys Soc.* 2015;66(8):1247–1252. doi: 10.3938/jkps.66.1247

[239] Wang GF, Li HL, Zhang XF, et al. Large magnetocaloric effect in Fe-B-Mn-Zr-Nb amorphous alloys near room temperature. *J Supercond Nov Magn.* 2016;29(7):1837–1842. doi: 10.1007/s10948-016-3464-8

[240] Zhang Z, Luo Q, Shao L, et al. Near room-temperature magnetocaloric effect in FeMnPBC metallic glasses. *J Magn Magn Mater.* 2013;347:131–135. doi: 10.1016/j.jmmm.2013.07.020

[241] Caballero-Flores R, Franco V, Conde A, et al. Influence of Co and Ni addition on the magnetocaloric effect in $Fe_{88-2x}Co_xNi_xZr_7B_4Cu_1$ soft magnetic amorphous alloys. *Appl Phys Lett.* 2010;96(18):182506. doi: 10.1063/1.3427439

[242] Liu X, Yuan J, Wang Q, et al. Excellent magnetocaloric properties near 285 K of amorphous $Fe_{88}Pr_6Ce_4B_2$ ribbon. *Metals (Basel).* 2024;14(11):1214. doi: 10.3390/met14111214

[243] Yu P, Zhang JZ, Xia L. $Fe_{87}Zr_7B_4Co_2$ amorphous alloy with excellent magneto-caloric effect near room temperature. *Intermetallics.* 2018;95:85–88. doi: 10.1016/j.intermet.2018.01.019

[244] Łukiewska A, Gębara P. Structure, magnetocaloric effect and critical behavior of the $Fe_{60}Co_{12}Gd_4Mo_3B_{21}$ amorphous ribbons. *Materials (Basel).* 2022;15(1):34. doi: 10.3390/ma15010034

[245] Sánchez-Valdés CF, Ibarra-Gaytán PJ, Llamazares JLS, et al. Enhanced refrigerant capacity in two-phase nanocrystalline/amorphous $NdPrFe_{17}$ melt-spun ribbons. *Appl Phys Lett.* 2014;104(21):212401. doi: /10.1063/1.4879544

[246] Gutfleisch O, Yan A, Müller KH. Large magnetocaloric effect in melt-spun $LaFe_{13-x}Si_x$. *J Appl Phys.* 2005;97(10):10M305. doi: 10.1063/1.1847871

[247] Yan A, Müller KH, Gutfleisch O. Structure and magnetic entropy change of melt-spun $LaFe_{11.57}Si_{1.43}$ ribbons. *J Appl Phys.* 2005;97(3):036102. doi: 10.1063/1.1844605

[248] Hou X, Han N, Xue Y, et al. Magnetocaloric properties response in high-speed melt-spun La-Ce-Fe-Si ribbons. *J Electron Mater.* 2016;45(10):4730–4735. doi: 10.1007/s11664-016-4626-y

[249] Zhong XC, Tang PF, Liu ZW, et al. Large magnetocaloric effect and refrigerant capacity in Gd–Co–Ni metallic glasses. *J Appl Phys.* 2012;111:07A919. doi: 10.1063/1.3673422

[250] Xing DW, Shen HX, Liu JS, et al. Magnetocaloric effect in uncoated $Gd_{55}Al_{20}Co_{25}$ amorphous wires. *Mater Res.* 2015;18:49–54. doi: 10.1590/1516-1439.325414

[251] Belliveau HF, Yu YY, Luo Y, et al. Improving mechanical and magnetocaloric responses of amorphous melt-extracted Gd-based microwires via nanocrystallization. *J Alloys Compd.* 2017;692:658–664. doi: 10.1016/j.jallcom.2016.08.254

[252] Yao Y, Li Z, Liu J, et al. Effect of Ni alloying on the microstructure and magnetocaloric properties of Gd-based metallic microfibers. *J Alloys Compd.* 2023;961:170979. doi: 10.1016/j.jallcom.2023.170979

[253] Wang YF, Yu YY, Belliveau H, et al. Enhanced magnetocaloric performance in nanocrystalline/amorphous $Gd_3Ni/Gd_{65}Ni_{35}$ composite microwires. *J Sci Adv Mater Devices.* 2021;6:587–593. doi: 10.1016/j.jsamd.2021.07.010

[254] Wang Y, Feng T, Li Y, et al. Designing the magnetocaloric tunability by driving ferromagnetic and antiferromagnetic interactions in Gd-based microwires. *J Magn Magn Mater*. 2025;624:173031. doi: 10.1016/j.jmmm.2025.173031

[255] Shen HX, Belliveau H, Xing DW, et al. The magnetocaloric behaviors of Gd-based microwire arrays with different Curie temperatures. *Metals*. 2022;12(9):1417. doi: 10.3390/met12091417

[256] Luo L, Shen H, Zhang L, et al. Giant magnetocaloric effect and hysteresis loss in $Mn_xFe_{2-x}P_{0.5}Si_{0.5}$ ($0.7 \leq x \leq 1.2$) microwires at ambient temperatures. *J Sci Adv Mater Devices*. 2024;9(3):100756. doi: 10.1016/j.jsamd.2024.100756

[257] Luo L, Shen HX, Zhang LY, et al. The effect of metal/non-metal ratio on the microstructure and magnetic properties of $(MnFe)_x(P_{0.5}Si_{0.5})$ microwires. *Prog Nat Sci Mater Int*. 2024;34(5):1085–1092. doi: 10.1016/j.pnsc.2024.08.004

[258] Luo L, Shen H, Zhang L, et al. Effect of Fe on the microstructure and magnetic transition of Mn-Fe-P-Si microwires. *J Alloys Compd*. 2024;1003:175579. doi: 10.1016/j.jallcom.2024.175579

[259] Yin HB, Law JY, Huang YJ, et al. Enhancing the magnetocaloric response of high-entropy metallic-glass by microstructural control. *Sci China Mater*. 2022;65:1134–1142. doi: 10.1007/s40843-021-1825-1

[260] Bao Y, Shen HX, Liang J, et al. Manufacture and characterization of HoErCo medium-entropy alloy microwires with excellent magnetic entropy change. *J Non-Cryst Solids.* 2021;556:120570. doi: 10.1016/j.jnoncrysol.2020.120570

[261] Bao Y, Shen H, Liu J, et al. Magnetocaloric effect and microstructure of amorphous/nanocrystalline HoErFe melt-extracted microwires. *Intermetallics.* 2020;127:106974. doi: 10.1016/j.intermet.2020.106974

[262] Shen HX, Luo L, Bao Y, et al. New DyHoCo medium entropy amorphous microwires of large magnetic entropy change. *J Alloys Compd.* 2020;837:155431. doi: 10.1016/j.jallcom.2020.155431

[263] Du J, Zheng Q, Li YB, et al. Large magnetocaloric effect and enhanced magnetic refrigeration in ternary Gd-based bulk metallic glasses. *J Appl Phys.* 2008;103:023918. doi: 10.1063/1.2836956

[264] Wang Y, Duc NTM, Feng T, et al. Competing ferromagnetic and antiferromagnetic interactions drive the magnetocaloric tunability in $Gd_{55}Co_{30}Ni_xAl_{15-x}$ microwires. *J Alloys Compd.* 2022;907:164328. doi: 10.1016/j.jallcom.2022.164328

[265] Xing DW, Shen HX, Jiang S, et al. Magnetocaloric effect and critical behavior in melt-extracted $Gd_{60}Co_{15}Al_{25}$ microwires. *Phys Status Solidi B.* 2015;212:1905–1910. doi: 10.1002/pssa.201532192

[266] Qin FX, Bingham NS, Wang H, et al. Mechanical and magnetocaloric properties of Gd-based amorphous microwires fabricated by melt-extraction. *Acta Mater.* 2013;61(4):1354–1362. doi: 10.1016/j.actamat.2012.11.006

[267] Duc NTM, Shen HX, Clements E, et al. Critical magnetic and magnetocaloric behavior of amorphous melt-extracted $Gd_{50}(Co_{69.25}Fe_{4.25}Si_{13}B_{13.5})_{50}$ microwires. *Intermetallics.* 2019;110:106479. doi: 10.1016/j.intermet.2019.106479

[268] Liu J, Wang Q, Wu M, et al. Improving the refrigeration capacity of Gd-rich wires through Fe-doping. *J Alloys Compd.* 2017;711:71–75. doi: 10.1016/j.jallcom.2017.03.363

[269] Wang H, Qin F, Li Z, et al. Comparable magnetocaloric properties of melt-extracted $Gd_{36}Tb_{20}Co_{20}Al_{24}$ metallic glass microwires. *J Alloys Compd.* 2020;815:150983. doi: 10.1016/j.jallcom.2019.06.085

[270] Yin H, Law J, Huang Y, et al. Design of Fe-containing GdTbCoAl high-entropy-metallic-glass composite microwires with tunable Curie temperatures and enhanced cooling efficiency. *Mater Des.* 2021;206:109824. doi: 10.1016/j.matdes.2021.109824

[271] Yin H, Wang JQ, Huang Y, et al. Relating microstructure to magnetocaloric properties in $RE_{36}Tb_{20}Co_{20}Al_{24}$ (RE = Gd, Dy or Ho) high-entropy metallic-glass microwires designed by binary eutectic clusters method. *J Mater.* 2023;149:167. doi: 10.1016/j.jmst.2022.12.008

[272] Luo L, Law JY, Shen H, et al. Enhanced magnetocaloric properties of annealed melt-extracted $Mn_{1.3}Fe_{0.6}P_{0.5}Si_{0.5}$ microwires. *Metals.* 2022;12(9):1536. doi: 10.3390/met12091536

[273] Dong JD, Yan AR, Liu J. Microstructure and magnetocaloric properties of melt-extracted La–Fe–Si microwires. *J Magn Magn Mater.* 2014;357:73–76. doi: 10.1016/j.jmmm.2014.01.031

[274] Zhukov A, Garcia C, Ilyn M, et al. Magnetic and transport properties of granular and Heusler-type glass-coated microwires. *J Magn Magn Mater.* 2012;324:3558–3562. doi: 10.1016/j.jmmm.2012.02.089

[275] Qian MF, Zhang XX, Wei L, et al. Tunable magnetocaloric effect in Ni–Mn–Ga microwires. *Sci Rep.* 2018;8:16574. doi: 10.1038/s41598-018-35028-9

[276] Liu Y, Zhang X, Xing D, et al. Magnetocaloric effect (MCE) in melt-extracted Ni–Mn–Ga–Fe Heusler microwires. *J Alloys Compd.* 2014;616:184–188. doi: 10.1016/j.jallcom.2014.07.077

[277] Zhang XX, Qian MF, Zhang Z, et al. Magnetostructural coupling and magnetocaloric effect in Ni–Mn–Ga–Cu microwires. *Appl Phys Lett.* 2016;108:052401. doi: 10.1063/1.4941232

[278] Zhang H, Zhang X, Qian M, et al. Magnetocaloric effect in Ni–Fe–Mn–Sn microwires with nano-sized $\gamma$ precipitates. *Appl Phys Lett.* 2020;116:063904. doi: 10.1063/1.5132767

[279] Zhukov A, Rodionova V, Ilyn M, et al. Magnetic properties and magnetocaloric effect in Heusler-type glass-coated NiMnGa microwires. *J Alloys Compd.* 2013;575:73–77. doi: 10.1016/j.jallcom.2013.04.083

[280] Zhang H, Qian M, Zhang X, et al. Magnetocaloric effect of Ni–Fe–Mn–Sn microwires prepared by melt-extraction technique. *Mater Des.* 2017;114:1–9. doi: 10.1016/j.matdes.2016.10.077

[281] Zhang XX, Miao SP, Sun JF. Magnetocaloric effect in Ni–Mn–In–Co microwires prepared by Taylor–Ulitovsky method. *Trans Nonferrous Met Soc China.* 2014;24:3152–3157. doi: 10.1016/S1003-6326(14)63454-3

[282] Varga R, Ryba T, Vargova Z, et al. Magnetic and structural properties of Ni–Mn–Ga Heusler-type microwires. *Scr Mater.* 2011;65:703–706. doi: 10.1016/j.scriptamat.2011.07.018

[283] Hung CM, Madhogaria RP, Muchharla B, Duong AT, Das R, Huy PT, Cho SL, Witanachchi S, Srikanth H, and Phan MH. MnP films with desired magnetic, magnetocaloric and thermoelectric properties for a perspective magneto-thermo-electric cooling device. *Physics Status Solidi* A 2022;219:2100367. doi: 10.1002/pssa.202100367

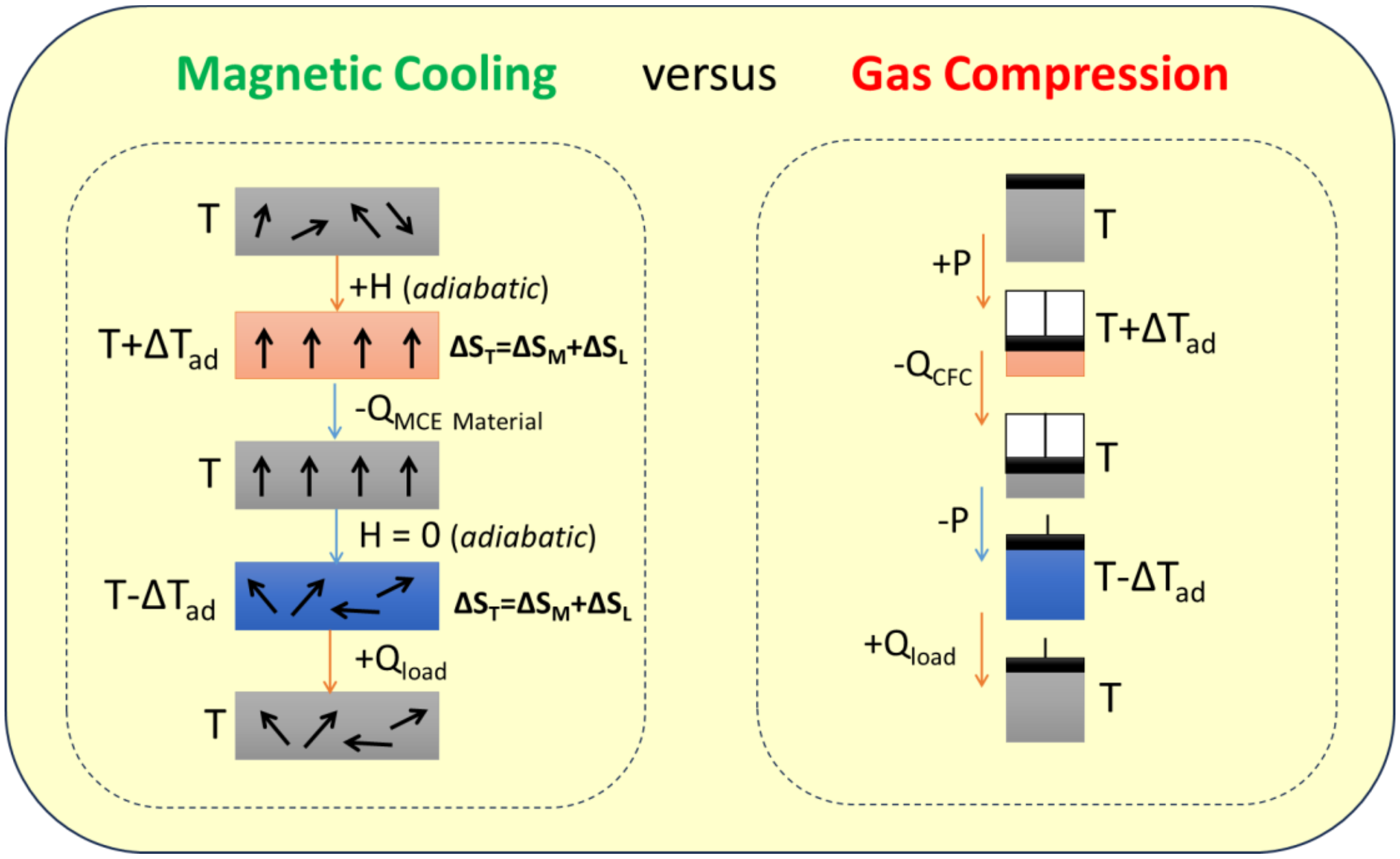

**Figure 1.** A full cooling cycle in magnetic refrigeration (left) differs fundamentally from that in conventional gas compression (right) techniques. In magnetic refrigeration, temperature changes are induced by cyclically magnetizing and demagnetizing a magnetocaloric material, which serves as the refrigerant. In contrast, gas compression refrigeration relies on compressing and expanding a gaseous refrigerant to produce pressure-induced temperature changes.

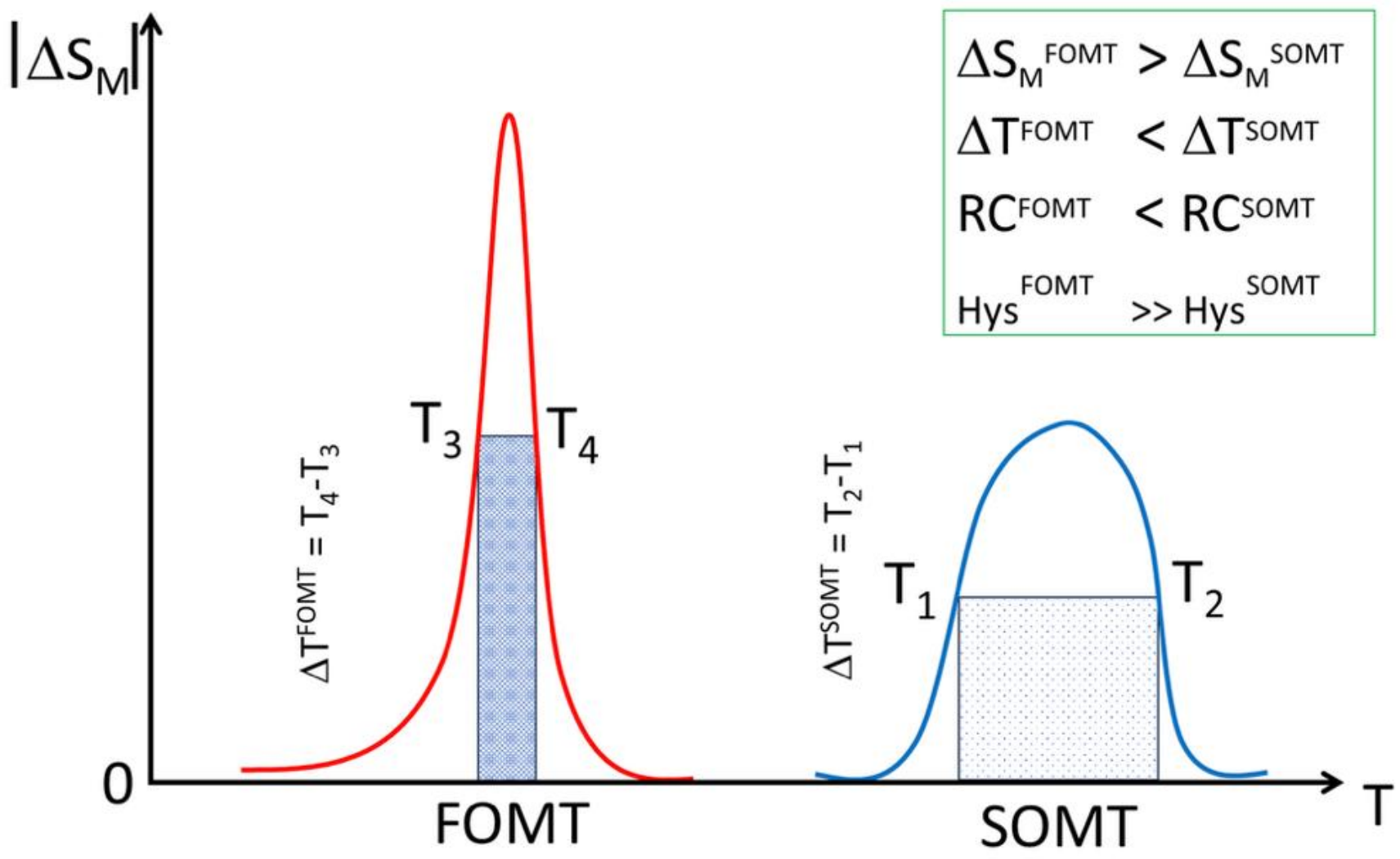

**Figure 2.** Schematic illustration of the magnetic entropy change ($|\Delta S_M|$) as a function of temperature for materials exhibiting first-order magnetic transitions (FOMT) and second-order magnetic transitions (SOMT). FOMT materials typically display a larger $\Delta S_M$ confined to a narrow temperature range ($\Delta T^{FOMT} = T_4\text{-}T_3$), often accompanied by significant magnetic and thermal hysteresis ($Hys^{FOMT}$). In contrast, SOMT materials exhibit a smaller $\Delta S_M$ over a broader temperature span ($\Delta T^{SOMT} = T_2\text{-}T_1$) with minimal hysteresis ($Hys^{SOMT}$), which can even result in a higher *RC*.

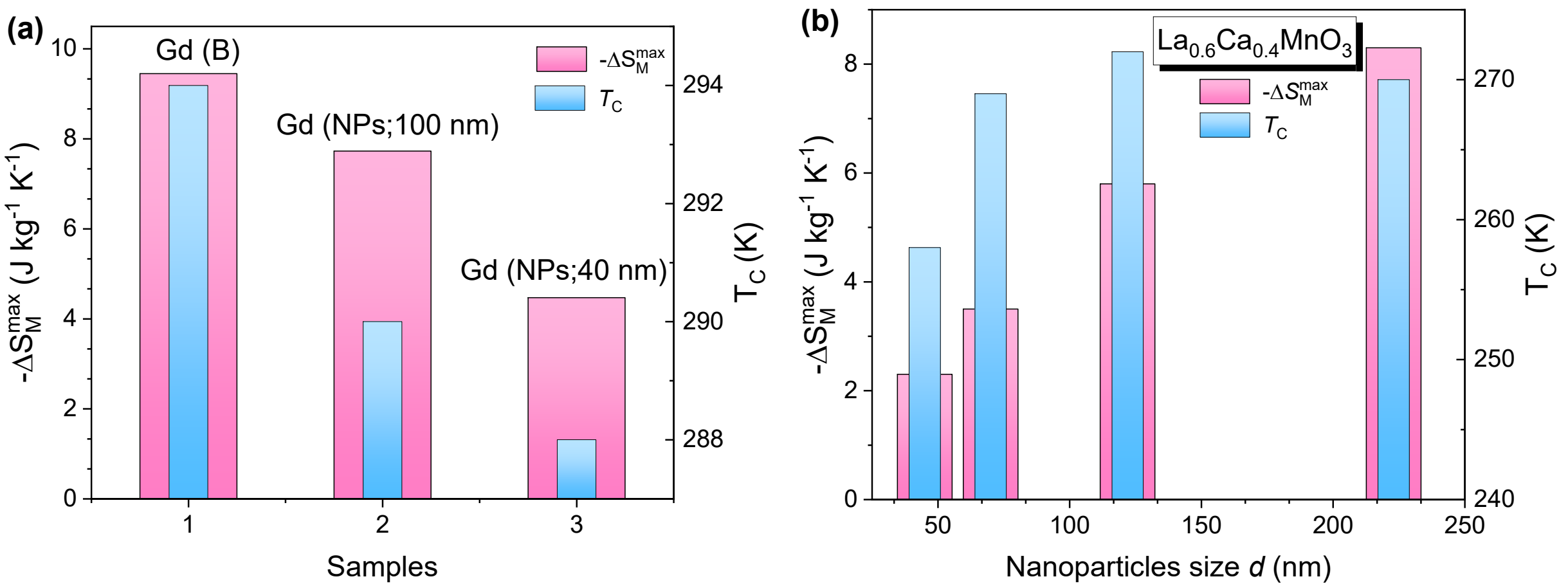

**Figure 3.** Maximum magnetic entropy change ($-\Delta S_M^{max}$) and Curie temperature ($T_C$) for bulk Gd and Gd nanoparticles with grain sizes of 100 nm and 15 nm under a magnetic field change of 5 T; (b) $-\Delta S_M^{max}$ and $T_C$ as functions of nanoparticle grain size under a magnetic field change of 5 T.

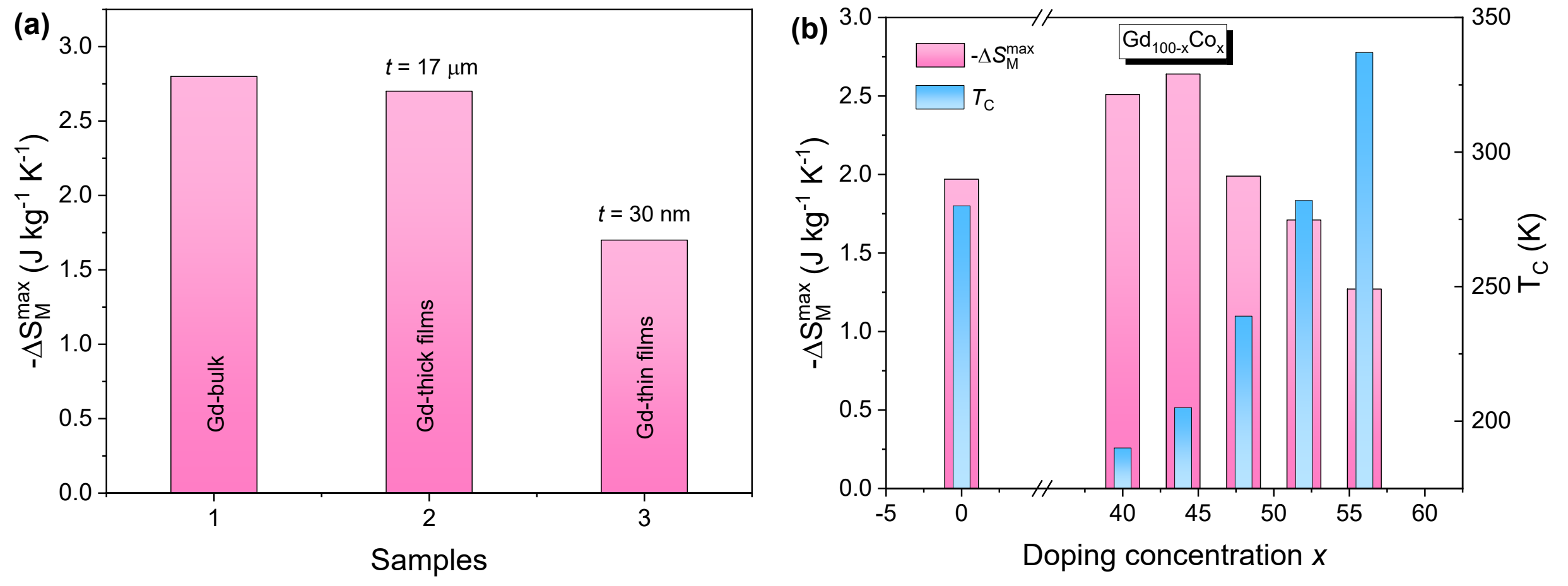

**Figure 4.** (a) Temperature dependence of the magnetic entropy change ($-\Delta S_M$) for bulk, thick films, and thin films Gd, showing a clear enhancement of $-\Delta S_M^{max}$ with increasing film thickness under a field change of 1 T; (b) Maximum magnetic entropy change ($-\Delta S_M^{max}$) and Curie temperature ($T_C$) as functions of Co-doping concentration for $Gd_{100-x}Co_x$ ($x$ = 0 – 56) thin films under a field change of 2 T.

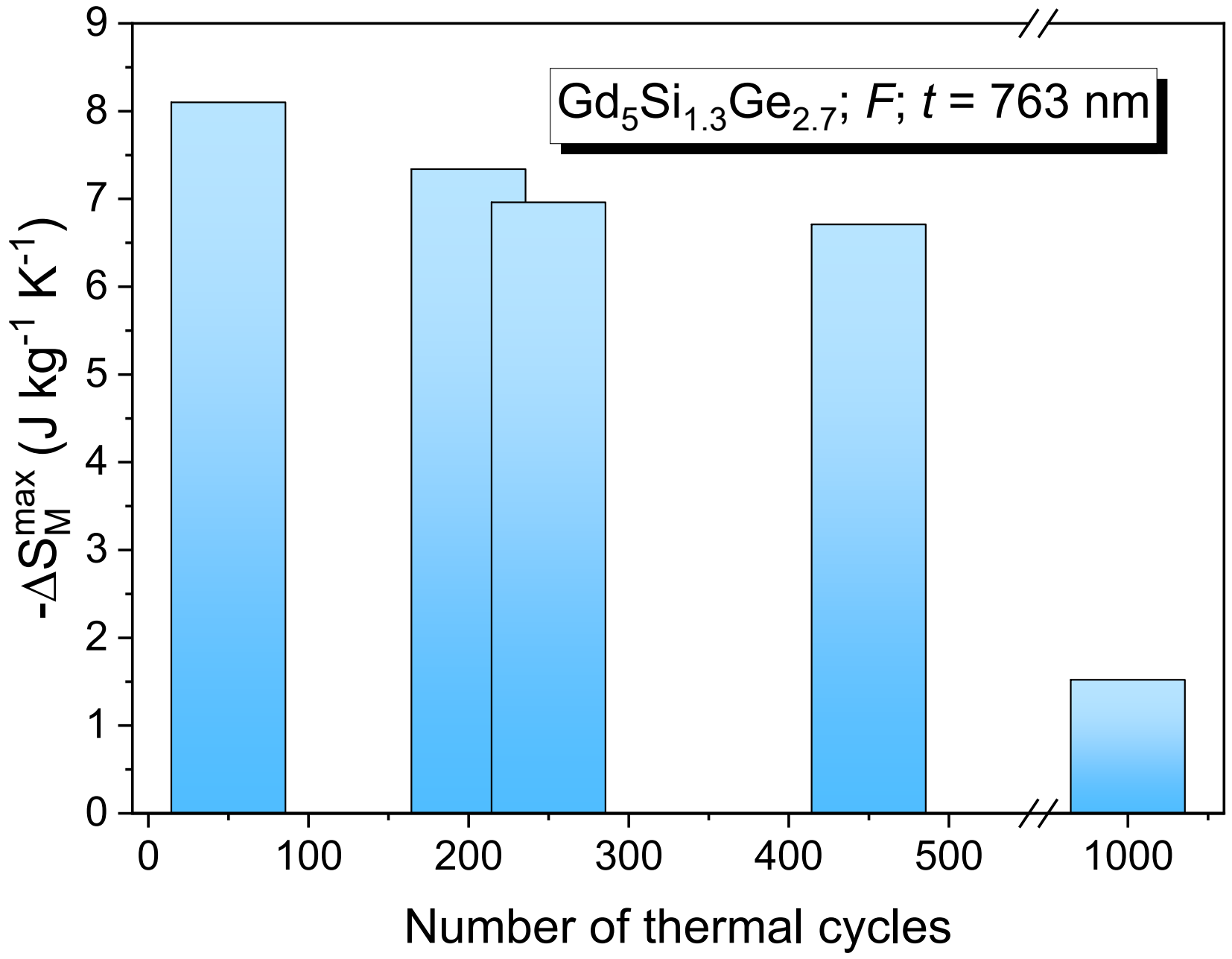

**Figure 5.** Maximum magnetic entropy change ($-\Delta S_M^{max}$) as a function of the number of thermal cycles, showing a clear decrease in $-\Delta S_M^{max}$ with increasing thermal cycling for $Gd_5Si_{1.3}Ge_{2.7}$ thin films under a field change of 5 T.

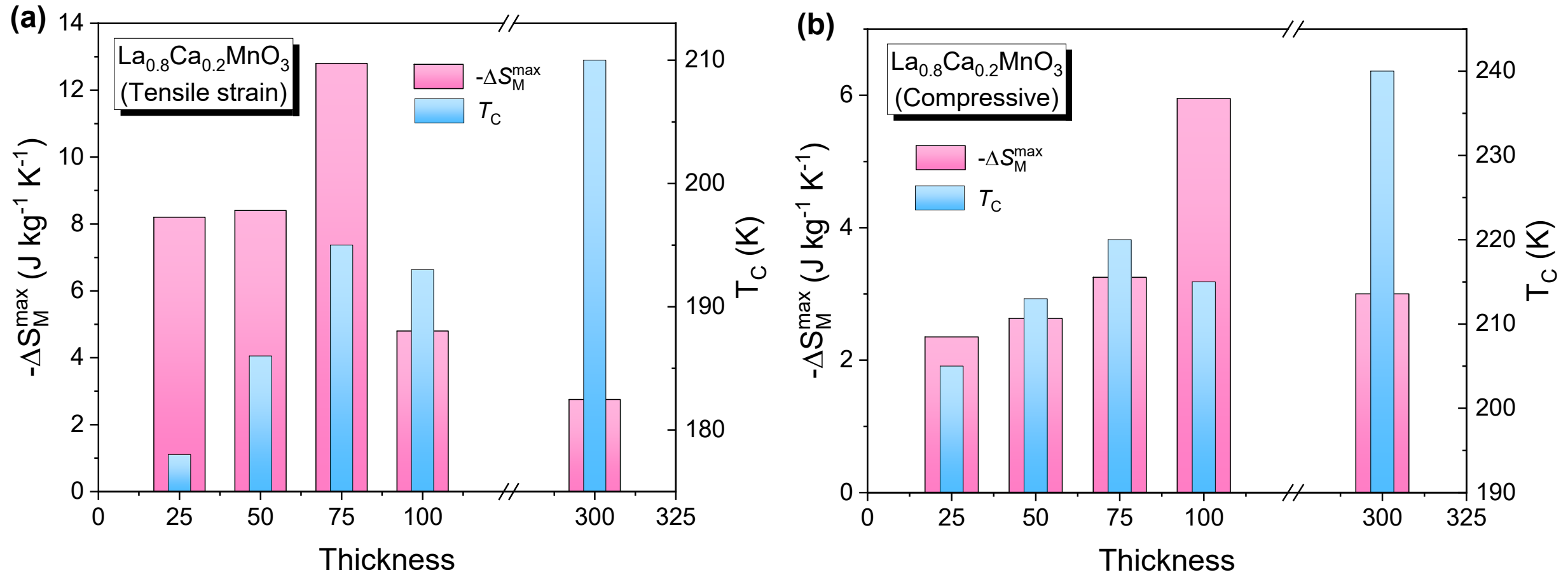

**Figure 6.** Variation of $-\Delta S_M^{max}$ and $T_C$ with the thickness of $La_{0.8}Ca_{0.2}MnO_3$ thin films under (a) tensile and (b) compressive strain, illustrating strain-induced changes in the magnetocaloric effect under a magnetic field change of 6 T.

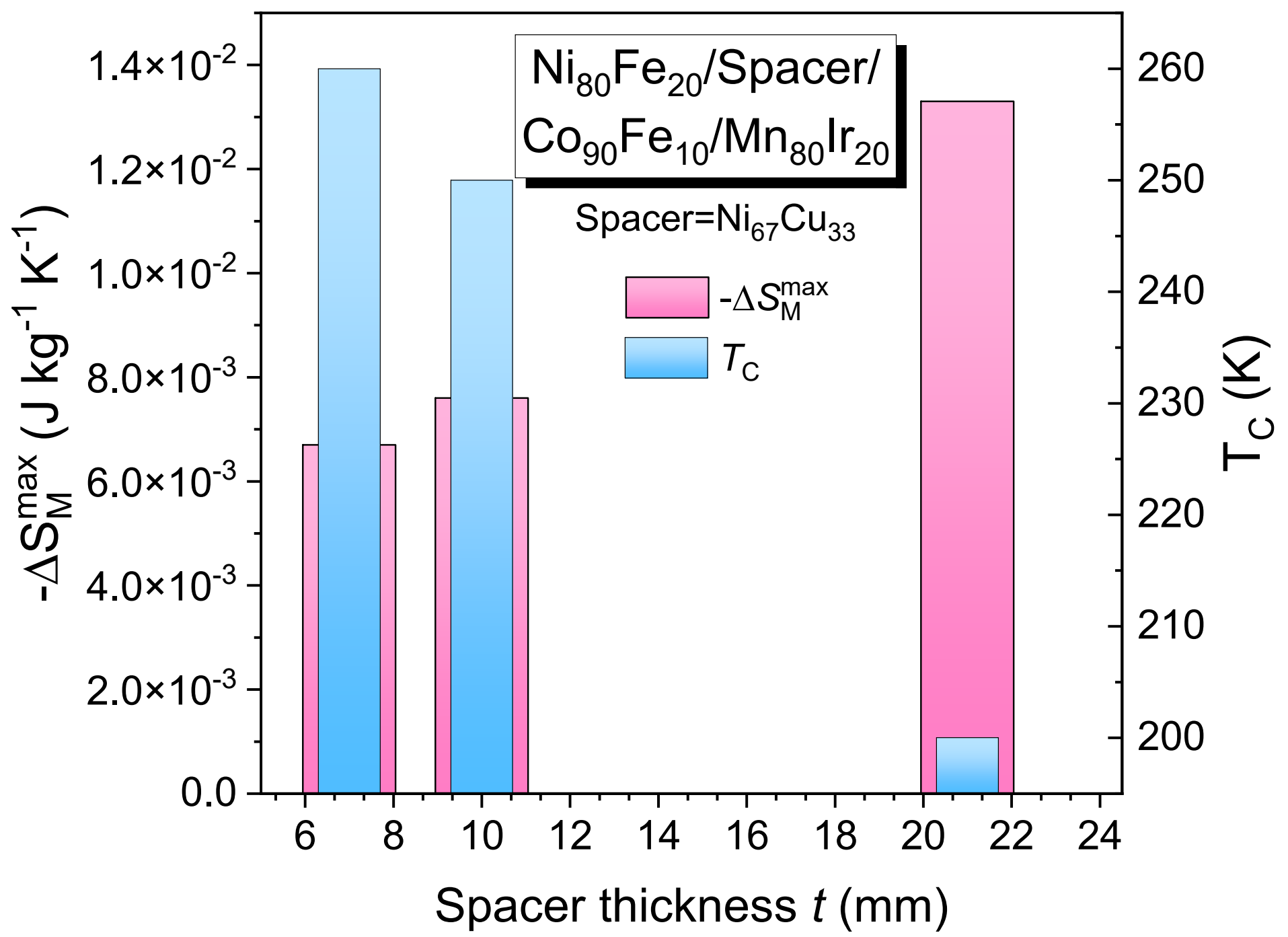

**Firure 7.** Dependence of $-\Delta S_M^{max}$ and $T_C$ on spacer thickness, highlighting the tunability of magnetocaloric behavior through layer design for $Ni_{80}Fe_{20}$/Spacer/$Co_{90}Fe_{10}$/$Mn_{80}Ir_{20}$ (Spacer = $Ni_{67}Cu_{33}$) under a magnetic field change of 2 mT.

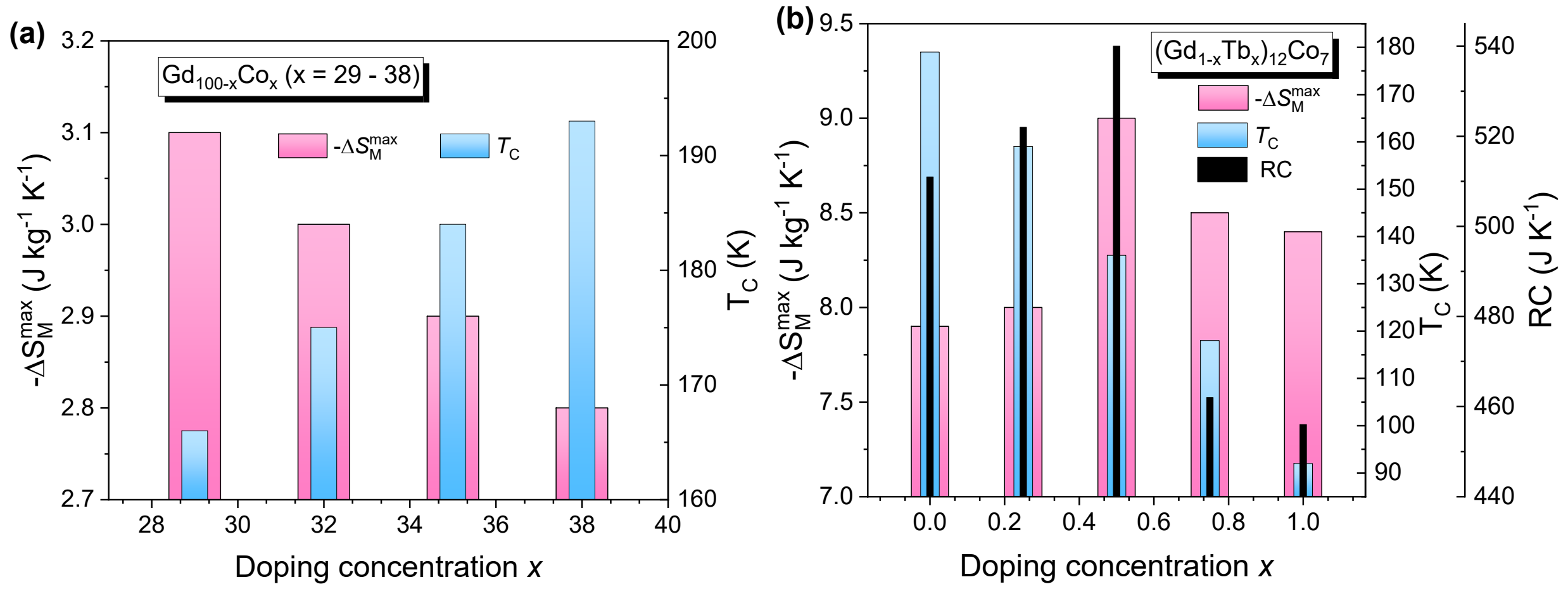

**Figure 8**. (a) Maximum magnetic entropy change ($-\Delta S_M^{max}$) and Curie temperature ($T_C$) as functions of Co doping concentration ($x$) in $Gd_{100-x}Co_x$ ribbons under a field of 1 T; (b) $T_C$, $-\Delta S_M^{max}$, and $RC$ as functions of Tb doping concentration ($x$) in $(Gd_{1-x}Tbx)_{12}Co_7$ alloys under a field of 5 T.

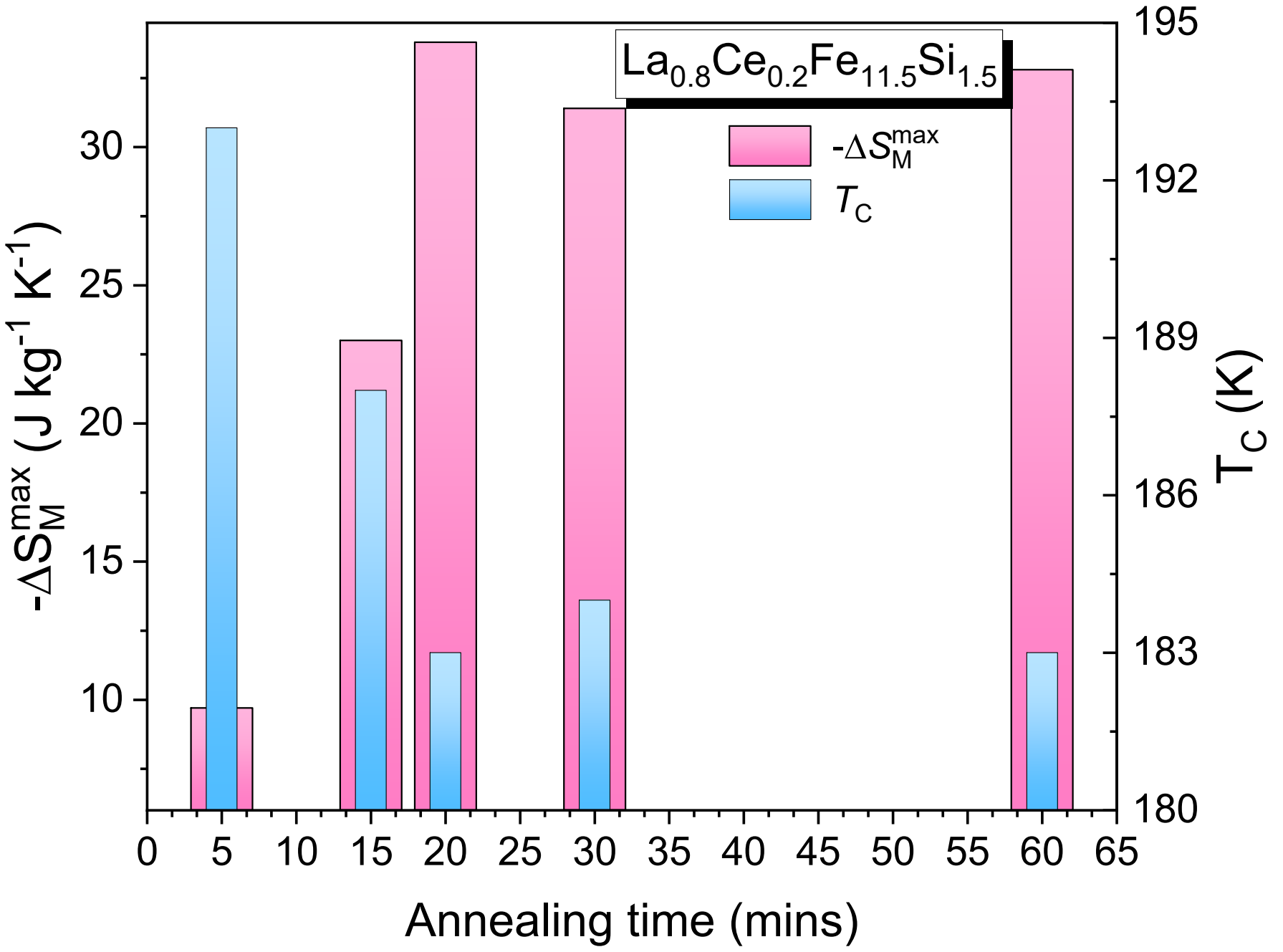

**Figure 9.** Curie temperature ($T_C$) and maximum magnetic entropy change ($-\Delta S_M^{max}$) as functions of annealing time for $La_{0.8}Ce_{0.2}Fe_{11.5}Si_{1.5}$ alloys ribbons annealed at 1273 K under a field of 1.5 T.

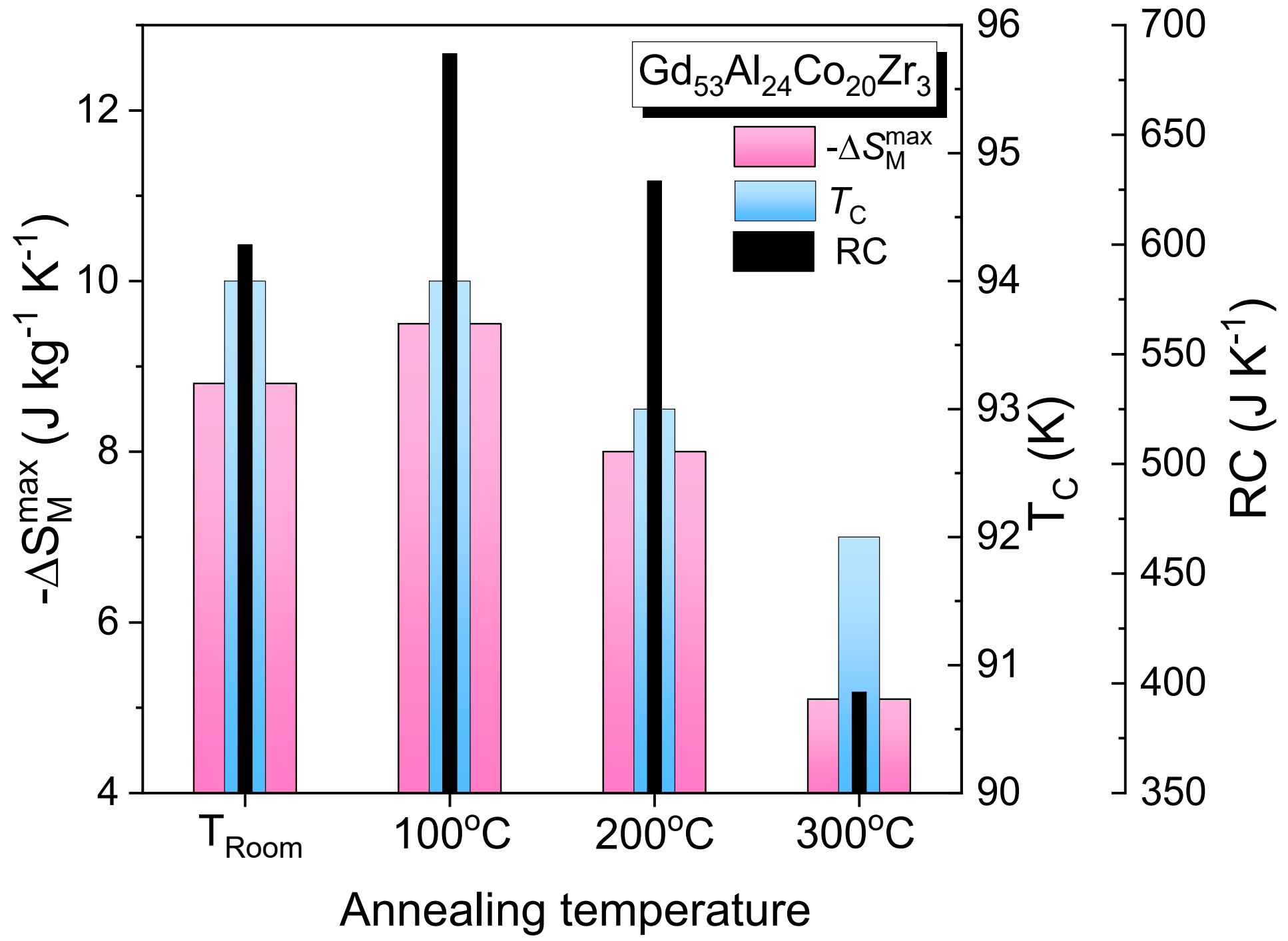

**Figure 10.** Curie temperature ($T_C$), maximum magnetic entropy change ($-\Delta S_M^{max}$), and refrigerant capacity (*RC*) as functions of annealing temperature for $Gd_{53}Al_{24}Co_{20}Zr_3$ alloys wires, including as-spun amorphous ribbon, crystallized ribbons annealed at various temperatures, and bulk sample under a magnetic field change of 5 T.

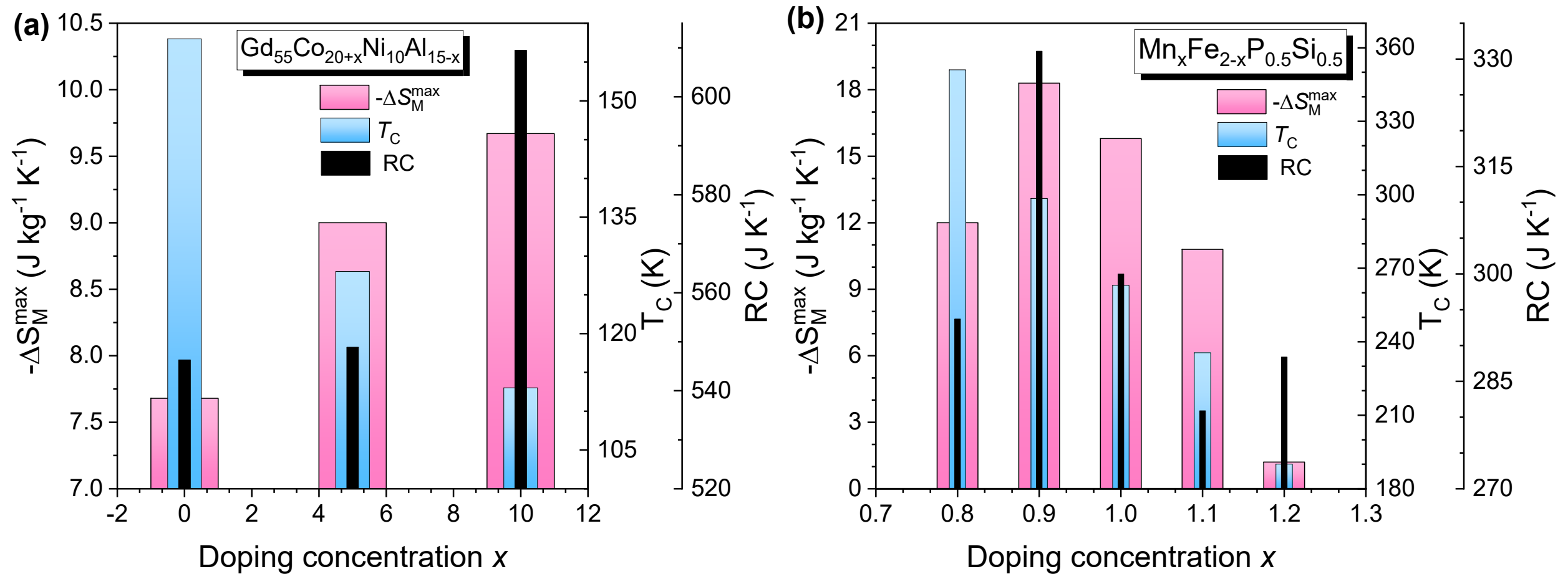

**Figure 11.** Curie temperature ($T_C$), maximum magnetic entropy change ($-\Delta S_M^{max}$), and refrigerant capacity ($RC$) as functions of Co doping concentration ($x$) in (a) $Gd_{55}Co_{20+x}Ni_{10}Al_{15-x}$ wires ($x$ = 10, 5, and 0) and (b) $Mn_xFe_{2-x}P_{0.5}Si_{0.5}$ wires for $x$ = 0.7 to 1.2 under a field change of 5 T.

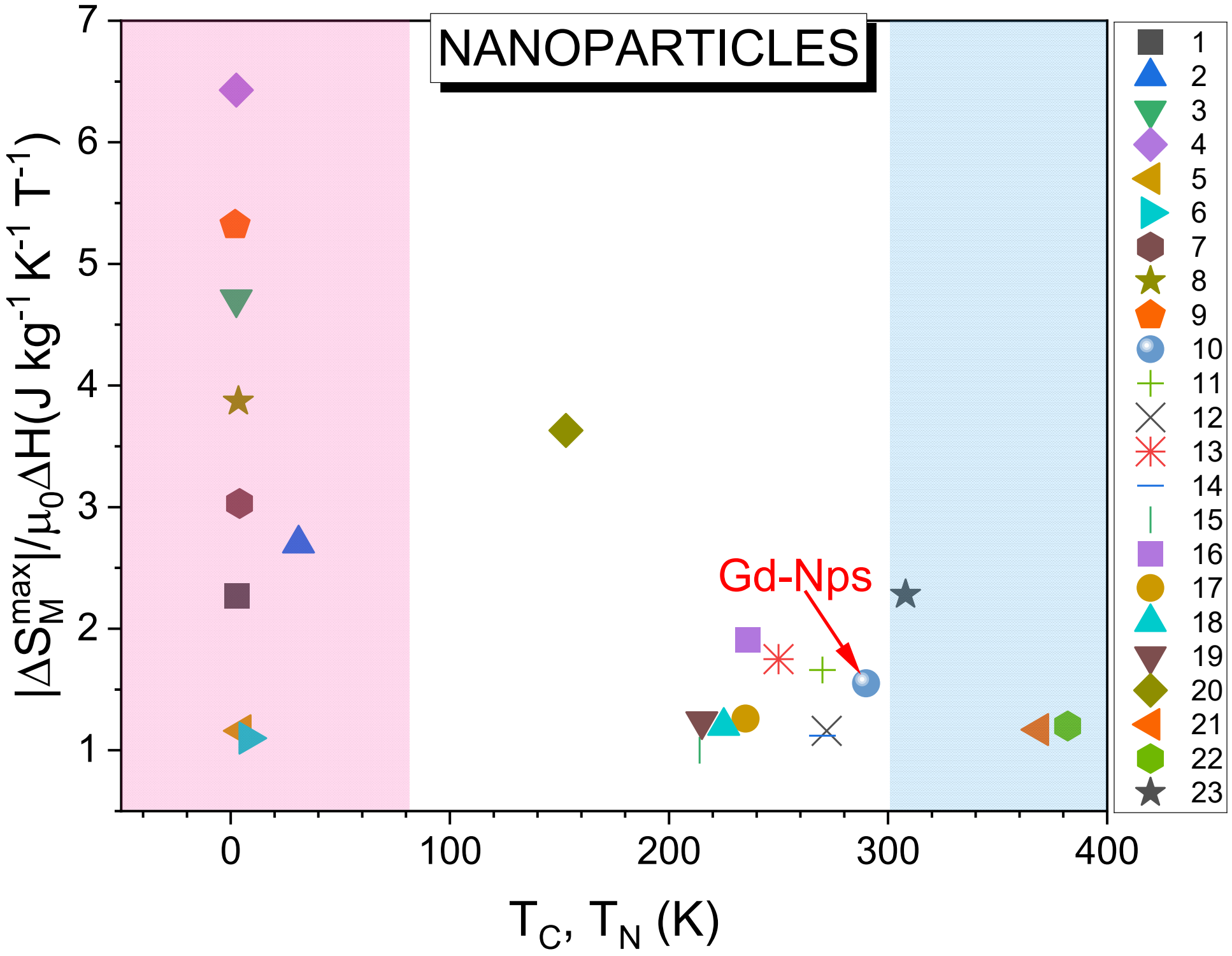

**Figure 12.** Performance coefficients ($|\Delta S_M^{max}|/\mu_0\Delta H_{max}$) of magnetocaloric nanoparticles evaluated at their respective Curie ($T_C$) or Néel ($T_N$) temperatures across three cooling temperature regimes: low ($T < 80$ K), intermediate (80 K $< T <$ 300 K), and high ($T >$ 300 K). *Low-temperature range:* 1-$MnPS_3$ [91]; 2-$GdNi_5$ [66]; 3-$GdVO_4$-30nm [85]; 4-$GdVO_4$-300nm [85]; 5-$Gd_3Fe_5O_{12}$ [58]; 6-$Tb_2O_3$ [61]; 7-$Dy_2O_3$ [61]; 8-$Gd_2O_3$ [61]; 9-$Ho_2O_3$ [61]; *Intermediate-temperature range:* 10-Gd [45]; 11-$La_{0.6}Ca_{0.4}MnO_3$-223nm [47]; 12-$La_{0.6}Ca_{0.4}MnO_3$-122nm [47]; 13-$La_{0.67}Ca_{0.33}MnO_3$ [70]; 14-$La_{0.7}Ca_{0.3}MnO_3$ [71]; 15-$La_{0.8}Ca_{0.2}MnO_3$-28nm [53]; 16-$La_{0.8}Ca_{0.2}MnO_3$-43nm [53]; 17-$Pr_{0.7}Sr_{0.3}MnO_3$ [78]; 18-$Pr_{0.65}(Ca_{0.7}Sr_{0.3})_{0.35}MnO_3$ [81]; 19-$La_{0.35}Pr_{0.275}Ca_{0.375}MnO_3$ [63]; 20-DyCrTiO$_3$ [60]; *High-temperature range:* 21-$La_{0.67}Sr_{0.33}MnO_3$ [50]; 22-$MnFeP_{0.45}Si_{0.55}$ [92]; 23-$La_{0.7}Ca_{0.2}Sr_{0.1}MnO_3$ [84].

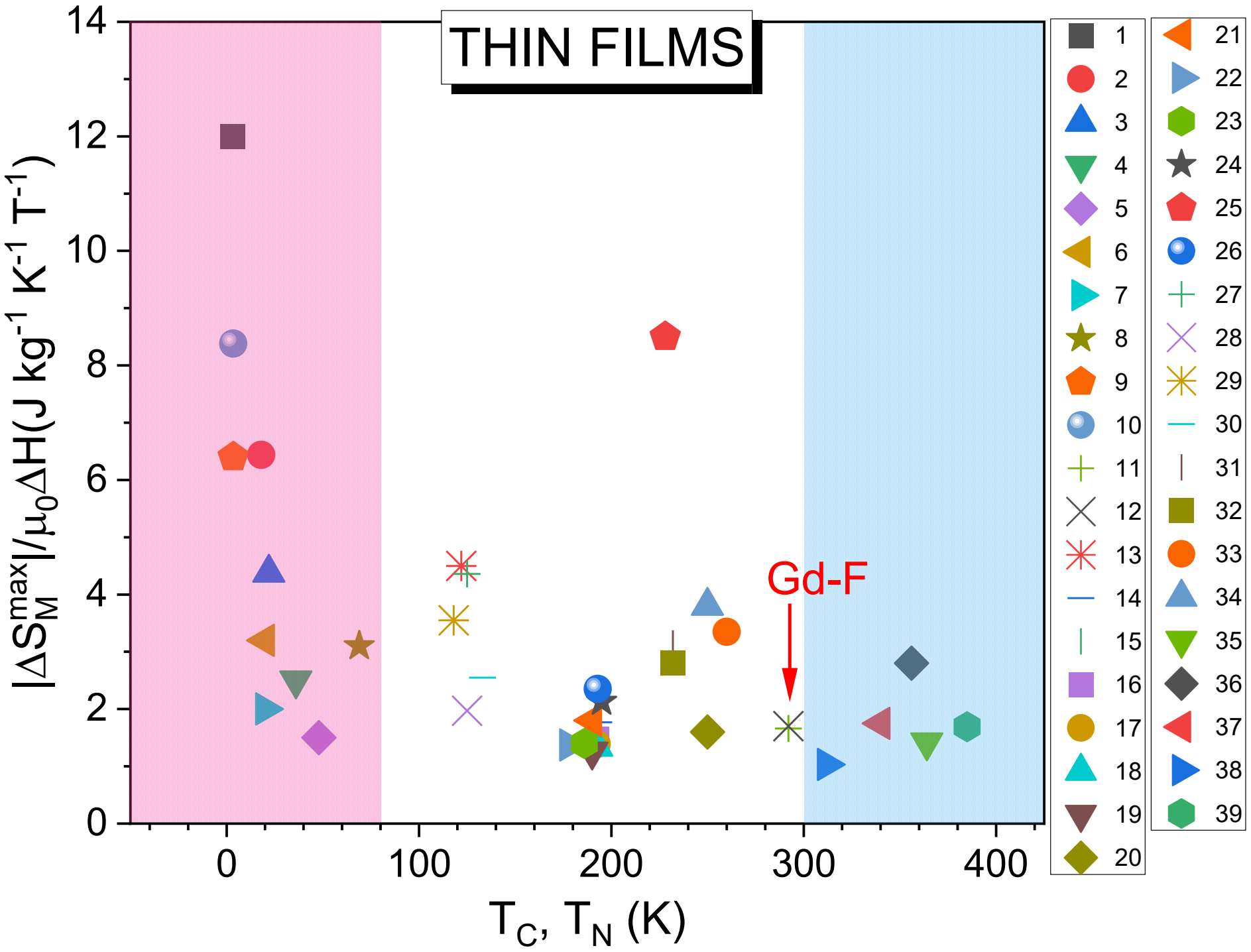

**Figure 13.** Performance coefficients ($|\Delta S_M^{max}|/\mu_0\Delta H_{max}$) of magnetocaloric thin films evaluated at their respective Curie ($T_C$) or Néel ($T_N$) temperatures across three cooling temperature regimes: low ($T$ < 80 K), intermediate (80 K < $T$ < 300 K), and high ($T$ > 300 K). *Low-temperature range:* 1-$EuTiO_3$ [113]; 2-$CrF_3$ [111]; 3-$CrCl_3$ [111]; 4-$CrBr_3$ [111]; 5-$CrI_3$ [111]; 6-$Fe_3[Cr(CN)_6]_2{\cdot}zH_2O$ at 1 T [146]; 7-$Fe_3[Cr(CN)_6]_2{\cdot}zH_2O$ at 5 T [146]; 8-$EuO_1$ [110]; 9-$GdCoO_3$/LAO at 2 T [115]; 10-$GdCoO_3$/LAO at 7 T [115]; *Intermediate-temperature range:* 11-Gd (F, $t$ = 17$\mu$m) [97]; 12-Gd (F, $t$ = 30 nm) annealed at 450 K [25]; 13-$GdSi_2$ [125]; 14-$Gd_5Si_{1.3}Ge_{2.7}$ [101]; 15-$Gd_5Si_{1.3}Ge_{2.7}$ (thermal cycling, 50 cycles) [101]; 16-$Gd_5Si_{1.3}Ge_{2.7}$ (thermal cycling, 200 cycles) [101]; 17-$Gd_5Si_{1.3}Ge_{2.7}$ (thermal cycling, 250 cycles) [101]; 18-$Gd_5Si_{1.3}Ge_{2.7}$ (thermal cycling, 450 cycles) [101]; 19-$Gd_{60}Co_{40}$ [99]; 20-$Ni_{51.6}Mn_{32.9}Sn_{15.5}$ [131]; 21-$La_{0.7}Ca_{0.3}MnO$ (Extrinsic) [103]; 22-$La_{0.8}Ca_{0.2}MnO_3$/STO (tensile strain, $t$ = 25 nm) [106]; 23-$La_{0.8}Ca_{0.2}MnO_3$/STO (tensile strain, $t$ =

50 nm) [106]; 24-$La_{0.8}Ca_{0.2}MnO_3$/STO (tensile strain, $t$ = 75 nm) [106]; 25-$La_{2/3}Ca_{1/3}MnO_3$ [133]; 26-$Pr_{0.7}Sr_{0.3}MnO_3$/PSMO-7 [136]; 27-$Gd_2NiMnO_6$ (In-plane) [120]; 28-$Gd_2NiMnO_6$ (Out of plane) [120]; 29-$EuO_{0.975}$ [110]; 30-$EuO_{0.91}$ [110]; 31-Epitaxial Tb (H//a axis, in-plane) [121]; 32-Epitaxial Tb (H//b axis, in-plane) [121]; 33-$Ni_{80}Fe_{20}$/$Ni_{67}Cu_{33}$/$Co_{90}Fe_{10}$/$Mn_{80}Ir_{20}$ (Spacer=$Ni_{67}Cu_{33}$, t = 7 nm) [119]; 34-$Ni_{80}Fe_{20}$/$Ni_{67}Cu_{33}$/$Co_{90}Fe_{10}$/$Mn_{80}Ir_{20}$ (Spacer=$Ni_{67}Cu_{33}$, t = 10 nm) [119]; *High-temperature range:* 35-$Ni_{53.5}Mn_{23.8}Ga_{22.7}$ [128]; 36-$Ni_{51}Mn_{29}Ga_{20}$ [129]; 37-$Ni_{43}Mn_{32}Ga_{20}Co_5$ [132]; 38-$La_{0.67}Sr_{0.33}MnO_3$ [105]; 39-$CrO_2$/$TiO_2$ [143]

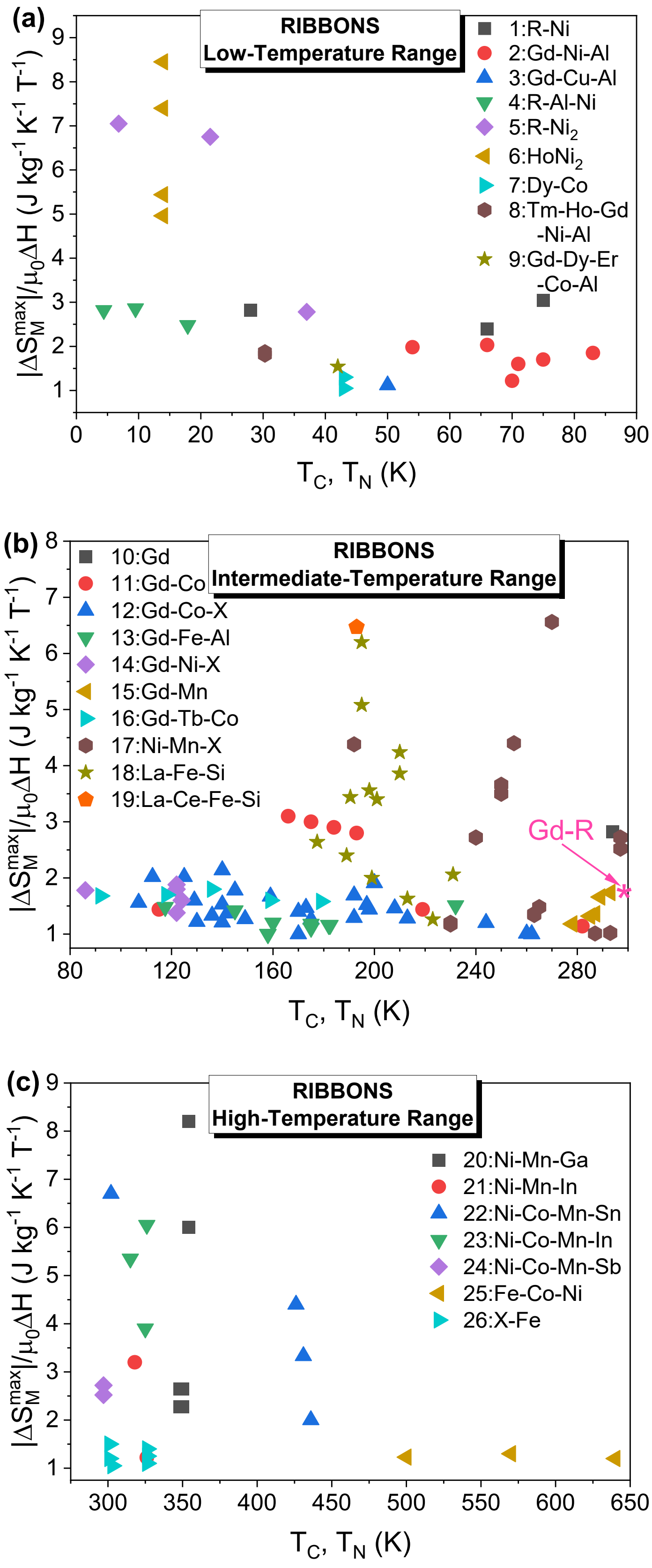
(a)
RIBBONS
Low-Temperature Range
1:R-Ni
2:Gd-Ni-Al
3:Gd-Cu-Al
4:R-Al-Ni
5:R-Ni$_2$
6:HoNi$_2$
7:Dy-Co
8:Tm-Ho-Gd-Ni-Al
9:Gd-Dy-Er-Co-Al
$|\Delta S_M^{max}|/\mu_0\Delta H$ (J kg$^{-1}$ K$^{-1}$ T$^{-1}$)
$T_C$, $T_N$ (K)
(b)
RIBBONS
Intermediate-Temperature Range
10:Gd
11:Gd-Co
12:Gd-Co-X
13:Gd-Fe-Al
14:Gd-Ni-X
15:Gd-Mn
16:Gd-Tb-Co
17:Ni-Mn-X
18:La-Fe-Si
19:La-Ce-Fe-Si
Gd-R
$|\Delta S_M^{max}|/\mu_0\Delta H$ (J kg$^{-1}$ K$^{-1}$ T$^{-1}$)
$T_C$, $T_N$ (K)
(c)
RIBBONS
High-Temperature Range
20:Ni-Mn-Ga
21:Ni-Mn-In
22:Ni-Co-Mn-Sn
23:Ni-Co-Mn-In
24:Ni-Co-Mn-Sb
25:Fe-Co-Ni
26:X-Fe
$|\Delta S_M^{max}|/\mu_0\Delta H$ (J kg$^{-1}$ K$^{-1}$ T$^{-1}$)
$T_C$, $T_N$ (K)

**Figure 14.** Performance coefficients ($|\Delta S_M^{max}|/\mu_0\Delta H_{\max}$) of magnetocaloric ribbons evaluated at their respective Curie ($T_C$) or Néel ($T_N$) temperatures across three temperature cooling regimes: (a) low ($T < 80$ K), (b) intermediate (80 K $< T <$ 300 K), and (c) high ($T >$ 300 K). *Low-temperature range:* 1: R-Ni [189]; 2: Gd-Ni-Al [164,191-1192]; 3: Gd-Cu-Al [196]; 4: R-Al-Ni [206]; 5: R-$Ni_2$ [207-209]; 6: $HoNi_2$ [210]; 7: Dy-Co [211]; 8: Tm-Ho-Gd-Ni-Al [176]; 9: Gd-Dy-Er-Co-Al [155]; *Intermediate-temperature range:* 10: Gd [158]; 11: Gd-Co [158-159,178]; 12: Gd-Co-X [161-162;180-186]; 13: Gd-Fe-Al [163;187-189]; 14: Gd-Ni-X [160,164,189-192]; 15: Gd-Mn [158]; 16: Gd-Tb-Co [195]; 17: Ni-Mn-X [19,213-215,217-228]; 18: La-Fe-Si [172,246-247]; 19: La-Ce-Fe-Si [248]; *High-temperature range:* 20: Ni-Mn-Ga [215,219]; 21: Ni-Mn-In [220,224]; 22: Ni-Co-Mn-Sn [223,227]; 23: Ni-Co-Mn-In [225-2226]; 24: Ni-Co-Mn-Sb [222]; 25: Fe-Co-Ni [241]; 26: X-Fe [156,173,245].

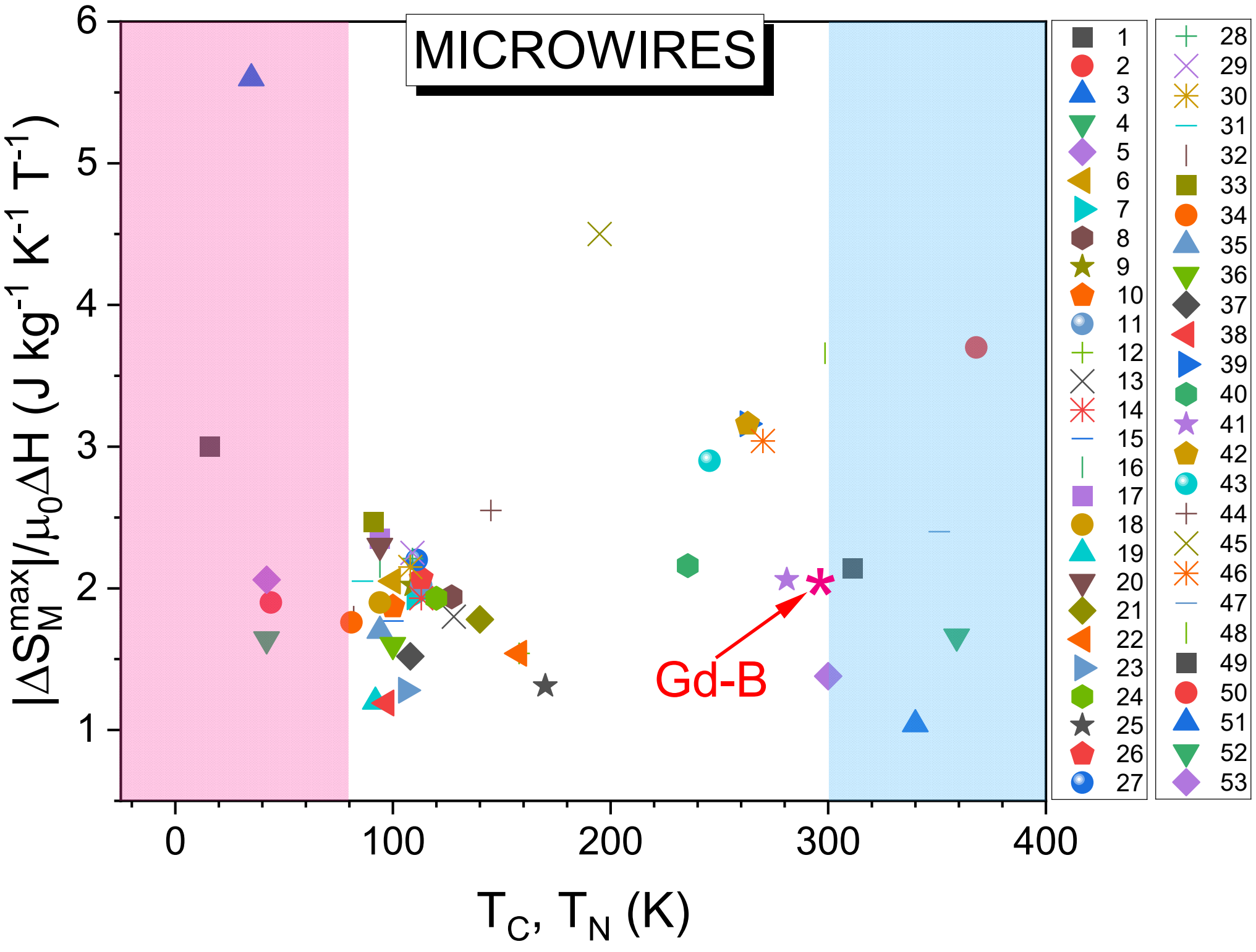

**Figure 15.** Performance coefficients ($-\Delta S_M^{max}/\mu_0\Delta H_{max}$) of magnetocaloric microwires evaluated at their respective Curie ($T_C$) or Néel ($T_N$) temperatures across three temperature cooling regimes: low ($T < 80$ K), intermediate ($80$ K $< T < 300$ K), and high ($T > 300$ K). *Low-temperature range:* 1-HoErCo [260]; 2-HoErFe [261]; 3-DyHoCo [262]; 4-$Dy_{36}Tb_{20}Co_{20}Al_{24}$ [271]; 5-$Ho_{36}Tb_{20}Co_{20}Al_{24}$ [271]; *Intermediate-temperature range:* 6-$Gd_{55}Co_{20}Al_{25}$ [31]; 7-$Gd_{55}Co_{30}Al_{15}$ [264]; 8-$Gd_{55}Co_{25}Al_{20}$ [250]; 9-$Gd_{60}Al_{20}Co_{20}$ [31]; 10-$Gd_{60}Co_{15}Al_{25}$ [265]; 11-$Gd_{60}Al_{20}Co_{20}$ [252]; 12-$Gd_{55}Co_{20+x}Ni_{10}Al_{15-x}$ (x = 10) [254]; 13-$Gd_{55}Co_{20+x}Ni_{10}Al_{15-x}$ ($x$ = 5) [254]; 14-$Gd_{55}Co_{20+x}Ni_{10}Al_{15-x}$ ($x$ = 0) [254]; 15-$Gd_{53}Al_{24}Co_{20}Zr_3$ (SW) [266]; 16-$Gd_{53}Al_{24}Co_{20}Zr_3$ (SW) [251]; 17-$Gd_{53}Al_{24}Co_{20}Zr_3$ (SW, Annealed at 100 °C) [251]; 18-$Gd_{53}Al_{24}Co_{20}Zr_3$ (SW, Annealed at 200 °C) [251]; 19-$Gd_{53}Al_{24}Co_{20}Zr_3$ (SW, Annealed at 300 °C) [251]; 20-$Gd_{53}Al_{24}Co_{20}Zr_3$ (MW) [30]; 21-$Gd_{55}Co_{30}Ni_5Al_{10}$ [264]; 22-$Gd_{55}Co_{30}Ni_{10}Al_5$ [264]; 23-$Gd_{73.5}Si_{13}B_{13.5}/GdB_6$ [35]; 24-

$Gd_3Ni/Gd_{65}Ni_{35}$ [253]; 25-$Gd_{50}$-$(Co_{69.25}Fe_{4.25}Si_{13}B_{13.5})_{50}$ [267]; 26-$Gd_{59.4}Al_{19.8}Co_{19.8}Fe_1$ [268]; 27-$(Gd_{60}Al_{20}Co_{20})_{99}Ni_1$ [252]; 28-$(Gd_{60}Al_{20}Co_{20})_{97}Ni_3$ [252]; 29-$(Gd_{60}Al_{20}Co_{20})_{95}Ni_5$ [252]; 30-$(Gd_{60}Al_{20}Co_{20})_{93}Ni_7$ [252]; 31-$Gd_{50}Co_{20}Al_{30}$ [31]; 32-$Gd_{36}Tb_{20}Co_{20}Al_{24}$ (A+C) [27]; 33-$Gd_{36}Tb_{20}Co_{20}Al_{24}$ (A) [269]; 34-$Gd_{36}Tb_{20}Co_{20}Al_{24}$ (A+C) [269]; 35-$(Gd_{36}Tb_{20}Co_{20}Al_{24})_{99}Fe_1$ [269]; 36-$(Gd_{36}Tb_{20}Co_{20}Al_{24})_{98}Fe_2$ [269]; 37-$(Gd_{36}Tb_{20}Co_{20}Al_{24})_{97}Fe_3$ [269]; 38-$Gd_{19}Tb_{19}Er_{18}Fe_{19}Al_{25}$ [270]; 39-$Mn_xFe_{2-x}P_{0.5}Si_{0.5}$ ($x$ = 1) [256]; 40-$Mn_xFe_{2-x}P_{0.5}Si_{0.5}$ ($x$ = 1.1) [256]; 41-$MnFe_xP_{0.5}Si_{0.5}$ ($x$ = 0.95) [258]; 42-$MnFe_xP_{0.5}Si_{0.5}$ ($x$ = 1) [258]; 43-$MnFe_xP_{0.5}Si_{0.5}$ ($x$ = 1.05) [258]; 44-$Mn_{1.3}Fe_{0.6}P_{0.5}Si_{0.5}$, annealed [272]; 45-$LaFe_{11.6}Si_{1.4}$ [273]; 46-$Ni_{45.6}Fe_{3.6}Mn_{38.4}Sn_{12.4}$ [278]; *High-temperature range:* 47-$Mn_xFe_{2-x}P_{0.5}Si_{0.5}$ (x = 0.8) [256]; 48-$Mn_xFe_{2-x}P_{0.5}Si_{0.5}$ ($x$ = 0.9) [256]; 49-$MnFe_xP_{0.5}Si_{0.5}$ ($x$ = 0.9) [258]; 50-$Ni_{50.5}Mn_{29.5}Ga_{20}$ [275]; 51-$Ni_{50.6}Mn_{28}Ga_{21.4}$ [275]; 52-$Ni_{49.4}Mn_{26.1}Ga_{20.8}Cu_{3.7}$ [277]; 53-$Ni_{44.9}Fe_{4.3}Mn_{38.3}Sn_{12.5}$ [280].

**Table 1.** Maximum entropy change, $\left|\Delta S_M^{max}\right|$, Curie temperature, $T_C$, refrigerant capacity (*RC*), and relative cooling power (*RCP*) for the nanoparticle samples.

| **Samples** | **Size (nm)** | **$T_C$ (K)** | **$\mu_0\Delta H$ (T)** | **$\left|\Delta S_M^{max}\right|$ (J/kg K)** | ***RC* (J/kg)** | ***RCP* (J/kg)** | **Ref.** |
|---|---|---|---|---|---|---|---|
| ***Gadolinium and its alloys*** | | | | | | | |
| Gd (*B*) | *B* | 294 | 1<br>2<br>5 | 2.8<br>5.07<br>9.45 | -<br>-<br>- | 63.4<br>187<br>690 | [44] |
| Gd (*NPs; C*) | 100<br>15 | 290<br>288 | 5<br>5 | 7.73<br>4.47 | -<br>- | 234<br>140 | [45] |
| $Gd_5Si_2Ge_2$ (*NPs; C*) | 85 | 225 | 2 | 0.45 | - | - | [65] |
| $GdNi_5$ (*NPs; C*) | 15 | 31 | 5 | 13.5 | - | - | [66] |
| $Gd_5Si_4$<br>(*B*)<br>*NPs; milled 2 h*<br>*NPs; milled 3 h* | <br>*B*<br>420<br>360 | <br>340<br>320<br>320 | <br>3<br>3<br>3 | <br>~6.5<br>~3<br>< 3 | <br>-<br>-<br>- | <br>~200<br>~340<br>~340 | [46] |
| ***Oxides*** | | | | | | | |
| $LaMnO_3$ (*B*) | *B* | 124 | 5 | 2.69 | 170 | 250 | [62] |
| $LaMnO_3$ (*NPs; C*)<br>Annealed at 1000 $^0$C | 200 | 135 | 5 | 2.67 | 282 | 355 | [62] |
| $LaMnO_3$ (*NPs; C*)<br>Annealed at 800 $^0$C | 40 | 150 | 5 | 2.4 | 284 | 369 | [62] |
| $La_{0.125}Ca_{0.875}MnO_3$ (*B*) | *B* | 123 | 7 | 6.3 | - | 63.1 | [67] |
| $La_{0.125}Ca_{0.875}MnO_3$ (*NPs; C*) | 70 | 113 | 7 | 1.32 | - | 22.8 | [67] |
| $La_{0.4}Ca_{0.6}MnO_3$ (*NPs; C*) | 130 | 260 | 5 | 2.81 | 240.7 | - | [51] |
| $La_{0.4}Ca_{0.6}MnO_3$ (*NPs; C*) | 50 | 100 | 5 | 0.33 | 46.6 | - | [51] |
| $La_{0.4}Ca_{0.6}MnO_3$ (*NPs; C*) | 25 | 80 | 5 | 0.13 | 11 | - | [51] |
| $La_{0.4}Ca_{0.6}MnO_3$ (*NPs; A*) | 10 | 45 | 5 | 0.46 | 8.1 | | [51] |
| $La_{0.5}Ca_{0.5}MnO_3$ (*NPs; C*) | 8.3 | 245 | 2 | 0.75 | - | 93 | [68] |
| $La_{0.6}Ca_{0.4}MnO_3$ (*B*) | *B* | 264 | 5 | 5.5 | - | 139 | [69] |
| $La_{0.6}Ca_{0.4}MnO_3$ (*NPs; C*) | 223 | 270 | 5 | 8.3 | - | 508 | [47] |
| $La_{0.6}Ca_{0.4}MnO_3$ (*NPs; C*) | 122 | 272 | 5 | 5.8 | - | 374 | [47] |
| $La_{0.6}Ca_{0.4}MnO_3$ (*NPs; C*) | 70 | 269 | 5 | 3.5 | - | 251 | [47] |
| $La_{0.6}Ca_{0.4}MnO_3$ (*NPs; C*) | 45 | 258 | 5 | 2.3 | - | 228 | [47] |
| $La_{0.6}Ca_{0.4}MnO_3$ (*NT; C*) | 23 | 280 | 5 | 0.3 | - | 40 | [47] |
| $La_{0.6}Ca_{0.4}MnO_3$<br>*NPs; C; sol-gen* | 45 | 258 | 1 | 0.6 | - | 50 | [47] |
| $La_{0.67}Ca_{0.33}MnO_3$ (*B*) | *B* | 258 | 1 | 5 | 56.2 | 55 | [70]<br>(*) |
| $La_{0.67}Ca_{0.33}MnO_3$ (*NPs; C*) | 60 | 250 | 1 | 1.75 | 33.5 | - | [70] |
| $La_{0.67}Ca_{0.33}MnO_3$ (*NPs; C*) | 20 | 260 | 1 | 0.2 | - | 25.6 | [62] |
| $La_{0.7}Ca_{0.3}MnO_3$ (*B*) | *B* | 264 | 5 | 7.8 | 187 | ~280 | [52] |
| | | | | | | | (*) |
| $La_{0.7}Ca_{0.3}MnO_3$ (*NPs; C*) | 33 | 260 | 5 | 4.9 | 146 | ~170 | [52] |
| $La_{0.7}Ca_{0.3}MnO_3$ (*NPs; C*) | 15 | 241 | 5 | 2.4 | 162.5 | ~180 | [52] |
| $La_{0.7}Ca_{0.3}MnO_3$ (*F; C*) | 150 | 235 | 5 | 2.75 | 200 | ~260 | [52] |
| $La_{0.7}Ca_{0.3}MnO_3$ (*B*) | *B* | 235 | 4.5 | 6.99 | 243.1 | - | [71] |
| $La_{0.7}Ca_{0.3}MnO_3$ (*NPs; C*) | 160 | 270 | 4.5 | 5.02 | 218.4 | - | [71] |
| $La_{0.7}Ca_{0.3}MnO_3$ (*NPs; C*) | 65 | 266 | 1.5 | 1.2 | - | 44 | [72] |
| $La_{0.8}Ca_{0.2}MnO_3$ (*NPs; C*) | 17 | 234 | 4.5 | 0.6 | - | 150 | [53] |
| $La_{0.8}Ca_{0.2}MnO_3$ (*NPs; C*) | 28 | 214 | 4.5 | 4.5 | - | 350 | [53] |
| $La_{0.8}Ca_{0.2}MnO_3$ (*NPs; C*) | 43 | 236 | 4.5 | 8.6 | - | 200 | [53] |
| $La_{0.67}Sr_{0.33}MnO_3$ (*B*) | *B* | 377 | 2 | 2.02 | - | 101 | [73] |
| $La_{0.67}Sr_{0.33}MnO_3$ (*B*) | *B* | 370 | 1 | 1.5 | - | 42 | [74] |
| $La_{0.67}Sr_{0.33}MnO_3$ (*NPs; C*) | 80 | 354 | 2<br>5 | 1.15<br>2.49 | -<br>- | 88<br>225 | [44] |
| $La_{0.67}Sr_{0.33}MnO_3$ (*NPs; C*) | 85 | 369 | 1.5 | 1.74 | - | 52 | [50] |
| $La_{0.67}Sr_{0.33}MnO_3$ (*NPs; C*) | 51 | 367 | 1.5 | 1.3 | - | 48 | [50] |
| $La_{0.67}Sr_{0.33}MnO_3$ (*NPs; C*) | 32 | 362 | 1.5 | 0.32 | - | 20 | [50] |
| $La_{0.67}Sr_{0.33}MnO_3$ (*F*) | 2400 | 348 | 5 | 1.69 | - | 211 | [75] |
| $La_{0.67}Sr_{0.33}MnO_3$ (*B*) | *B* | 370<br>370 | 2<br>5 | 2.68<br>5.15 | -<br>- | 85<br>252 | [76] |
| $La_{0.8}Sr_{0.2}MnO_3$ (*B*) | *B* | 301 | 2 | 2.2 | - | 35 | [77] |
| $La_{0.8}Sr_{0.2}MnO_3$ (*NPs; C*) | 23 | 295 | 2 | 0.5 | - | 32 | [77] |
| $Pr_{0.67}Sr_{0.33}MnO_3$ (*NPs; C*) | 80 | 258 | 2<br>5 | 0.82<br>1.94 | -<br>- | 99<br>265 | [44] |
| $Pr_{0.7}Sr_{0.3}MnO_3$ (*NPs; C*) | 35 | 235 | 5 | 6.3 | - | 385 | [78] |
| $Pr_{0.67}Sr_{0.33}MnO_3$ (*B*) | *B* | 281 | 5 | 7.8 | - | 195 | [79] |
| $Pr_{0.67}Sr_{0.33}MnO_3$ (*B*) | *B* | 260 | 1 | 1.75 | - | 49 | [80] |
| $Nd_{0.67}Sr_{0.33}MnO_3$<br>(*NPs; C*) | 80 | 206 | 2<br>5 | 0.35<br>0.93 | -<br>- | 87<br>246 | [44] |
| $Nd_{0.63}Sr_{0.37}MnO_3$<br>(*SC*) | *B* | 300 | 5 | 8.25 | - | 511 | [21] |
| $Pr_{0.65}(Ca_{0.7}Sr_{0.3})_{0.35}MnO_3$ (*B*) | *B* | 215 | 7 | 7.8 | 273 | 312 | [81] |
| $Pr_{0.65}(Ca_{0.7}Sr_{0.3})_{0.35}MnO_3$<br>(*NPs; C*) | 67 | 225 | 5 | 6.0 | 180 | 142 | [82] |
| $La_{0.35}Pr_{0.275}Ca_{0.375}MnO_3$ (*B*) | *B* | 75 | 5 | 4.5 | 34.64 | - | [63] |
| $La_{0.35}Pr_{0.275}Ca_{0.375}MnO_3$<br>(*NPs; C*) | 50 | 215 | 1<br>5 | 2.94<br>6.2 | 37.2<br>225.6 | - | [63] |
| $La_{0.215}Pr_{0.41}Ca_{0.375}MnO_3$<br>(*B*) | *B* | 210 | 5 | 5.3 | 143.1 | - | [63] |
| $La_{0.7}Ca_{0.3}Mn_{0.9}Ni_{0.1}O_3$<br>(*NPs; C; BM*) | 15 | 145 | 1.5 | 0.95 | - | - | [83] |
| $La_{0.7}Ca_{0.2}Sr_{0.1}MnO_3$<br>(*NPs; C; HEBM*) | 150-<br>300 | 308 | 1.8 | 4.11 | - | 61.12 | [84] |
| $DyCrTiO_5$<br>(*NPs; C; exchange bias*) | 37 | 153<br>($T_N$) | 3 | 10.9<br>@10 K | - | ~76.3 | [60] |
| $Tb_2O_3$ (*NPs; C*) | 51 | 8<br>($T_N$) | 6 | 6.6 | 53.9 | - | [61] |
| $Dy_2O_3$ (*NPs; C*) | 68 | 4<br>($T_N$) | 6 | 18.2 | 46.5 | - | [61] |
| $Gd_2O_3$ (*NPs; C*) | 44 | 3.5<br>($T_N$) | 6 | 23.2 | - | - | [61] |
| $Ho_2O_3$ (*NPs; C*) | 56 | 2<br>($T_N$) | 6 | 31.9 | - | - | [61] |
| $GdVO_4$ (*NPs; C*) | 30 | 2.5 | 7 | 33 | - | - | [85] |
| $GdVO_4$ (*NPs; C*) | 300 | 2.5 | 7 | 45 | - | - | [85] |
| $GdVO_4$<br>(*B*) | 2500 | 2.5 | 7 | 43 | - | - | [85] |
| $GdVO_4$<br>(*B*) | 5000 | 2.5 | 7 | 30 | - | - | [85] |
| $Gd_3Fe_5O_{12}$<br>(*B*) | *B* | 35<br>35 | 1<br>3 | 0.78<br>2.45 | -<br>- | -<br>288 | [58] |
| $Gd_3Fe_5O_{12}$<br>(*NPs; C*) | 50 | 25<br>25 | 1<br>3 | 0.31<br>1.49 | -<br>- | -<br>- | [58] |
| $Gd_3Fe_5O_{12}$<br>(*NPs; C*) | 35 | 5<br>5 | 1<br>3 | 0.67<br>3.47 | -<br>- | -<br>- | [58] |
| ***Austenitic alloys*** | | | | | | | |
| $\gamma$-FeNiMn<br>*NPs; C; BM 10 h* | 17 | 340 | 1 | 0.41 | - | 78 | [86] |
| $(Fe_{70}Ni_{30})_{99}Cr_1$<br>*NPs; C; BM* | 12 | 398<br>398 | 1<br>5 | 0.38<br>1.58 | -<br>- | 82<br>548 | [64] |
| $(Fe_{70}Ni_{30})_{97}Cr_3$<br>*NPs; C; BM* | 10 | 323<br>323 | 1<br>5 | 0.27<br>1.49 | -<br>- | 59<br>436 | [64] |
| $(Fe_{70}Ni_{30})_{95}Cr_5$<br>*NPs; C; BM* | 13 | 258<br>258 | 1<br>5 | 0.37<br>1.45 | -<br>- | 77<br>406 | [64] |
| $(Fe_{70}Ni_{30})_{94}Cr_6$<br>*NPs; C; BM* | 12 | 245<br>245 | 1<br>5 | 0.29<br>1.22 | -<br>- | 62<br>366 | [64] |
| $(Fe_{70}Ni_{30})_{93}Cr_7$<br>*NPs; C; BM* | 11 | 215<br>215 | 1<br>5 | 0.28<br>1.11 | -<br>- | 47<br>306 | [64] |
| ***Others*** | | | | | | | |
| Co<br>*NPs; C*<br><br>$Co_{core}Ag_{shell}$<br>*NPs; C* | 50<br><br><br>40<br>*core*<br>28<br>*shell* | 15<br>15<br>15<br>20<br>20<br>20 | 1<br>2<br>3<br>1<br>2<br>3 | 1.00<br>1.75<br>2.35<br>0.82<br>1.16<br>2.28 | ~10.46<br>~19.01<br>~26.85<br>~5.12<br>~10.65<br>~15.78 | ~10.92<br>~20.40<br>~28.20<br>~5.4<br>~11.4<br>~16.8 | [48] |
| $Ni_{100-x}Cr_x$ (*NPs; C*)<br>$x = 0$<br>$x = 5$<br>$x = 10$<br>$x = 15$ | <br>4.7<br>5.1<br>5.6<br>5.9 | <br>614<br>550<br>349<br>147 | <br>0.1<br>0.1<br>0.1<br>0.1 | <br>0.15<br>0.10<br>0.05<br>0.03 | <br>-<br>-<br>-<br>- | <br>23.30<br>33.45<br>21.97<br>13.17 | <br>[87] |
| $Pr_2Fe_{17}$ (*NPs; C; BM 40 h*) | 11 | 285 | 1.5 | 0.6 | | 60 | [88] |
| $Nd_2Fe_{17}$ (*NPs; C; BM 40 h*) | 11 | 337 | 1.5 | 1 | - | 118 | [88] |
| $Co_2FeAl$ (*NPs; C*) | 16 | 1261 | 1.4 | 15 | - | 89 | [89] |
| $Ni_{50}Mn_{34}In_{16}$ (*NPs; C*) | 150 | 226-<br>241 | 6 | 2 | - | 150 | [90] |
| $MnPS_3$<br>*TM thiophosphate* | 20-50 | 2.85<br>2.85 | 3<br>9 | 6.8<br>12.8 | - | - | [91] |
| $MnFeP_{0.45}Si_{0.55}$<br>*B, HEBM 0h*<br>*NPs, HEBM 26h*<br>*NPs, HEBM 26h, 600 $^0$C* | <br>27000<br>31.6<br>31.6 | <br>392<br>390<br>382.1 | <br>1<br>1<br>1<br>2 | <br>2.8<br>0.8<br>1.2<br>2.4 | <br>-<br>-<br>-<br>- | <br>-<br>-<br>29<br>77 | <br>[92] |
| $Fe_{47.5}Ni_{37.5}Mn_{15}$ (*NPs; C*) | 7.5 | 327 | 5 | 1.3 | - | 297.68 | [93] |
| $\gamma$-$(Fe_{70}Ni_{30})_{89}Zr_7B_4$<br>(*NPs; C; BM*) | 20 | 353 | 1.5 | 0.7 | - | 65 | [94] |

*A: Amorphous; C: Crystalline; NPs: Nanoparticles; F: Films; B: Bulk; P: Powder; SC: Single crystal; NT: Nanotubes. HEBM: High energy ball milling. (*) represents FOMT materials.*

**Table 2.** Maximum entropy change, $\left|\Delta S_M^{\max}\right|$, Curie temperature, $T_C$, refrigerant capacity (*RC*), and relative cooling power (*RCP*) for magnetocaloric thin film samples.

| **Samples** | $T_C$ **(K)** | $\mu_0\Delta H$ **(T)** | $\left\|\Delta S_M^{\max}\right\|$ **J/kg K)** | ***RC*** **(J/kg)** | ***RCP*** **(J/kg)** | **Ref.** |
|---|---|---|---|---|---|---|
| ***Gadolinium and its alloys*** | | | | | | |
| Gd (B) | 294 | 1 | 2.80 | - | 63.4 | [7] |
| | | 2 | 5.07 | - | 187.0 | |
| | | 5 | 10.20 | - | 410.0 | |
| Gd (F, $t$ = 17$\mu$m) | 292 | 1 | 2.70 | ~134.6 | ~140.2 | [97] |
| | | 3 | 5.90 | ~309.1 | ~308.6 | |
| | | 5 | 8.30 | ~452.5 | ~452.2 | |
| | | 7 | 10.50 | ~608.6 | ~627.5 | |
| Gd (F, $t$ = 30 nm) | | | | | | [25] |
| As-deposited | 265 | 1 | 0.60 | - | 20.4 | |
| Annealed at 450 K | 292 | 1 | 1.70 | - | 110.5 | |
| $GdSi_2$<br>(F, $t$ = 20 nm) | 122 | 5 | 22.5 | - | - | [125] |
| $Gd_5Si_2Ge_2$ (B) | 276 | 5 | 18.4 | 360 | - | [45]<br>(*) |
| $Gd_5Si_{1.3}Ge_{2.7}$ (F, $t$ = 780 nm) | 193.5<br>($T_{MS}$) | 5 | 8.83 | 212.0 | - | [100]<br>(*) |
| $Gd_5Si_{1.3}Ge_{2.7}$ (F, $t$ = 763 nm) | 192.5 | | | | - | [101] |
| $Gd_5Si_{1.3}Ge_{2.7}$<br>Thermal cycling, 50 cycles | 192.5 | 5 | 8.10 | 156.8 | - | (*) |
| $Gd_5Si_{1.3}Ge_{2.7}$<br>Thermal cycling, 200 cycles | 192.5 | 5 | 7.34 | - | - | |
| $Gd_5Si_{1.3}Ge_{2.7}$<br>Thermal cycling, 250 cycles | 192.5 | 5 | 6.96 | - | - | |
| $Gd_5Si_{1.3}Ge_{2.7}$<br>Thermal cycling, 450 cycles | 192.5 | 5 | 6.71 | 142.7 | - | |
| $Gd_5Si_{1.3}Ge_{2.7}$<br>Thermal cycling, 1000 cycles | 192.5 | 5 | 1.52 | N/A | | |
| $Gd_{100-x}Co_x$<br>(F, $t$ = 100 nm, $x$ = 0-56) | | | | | | [99] |
| $Gd_{100}$ | 280 | 2 | 1.97 | - | 106 | |
| $Gd_{60}Co_{40}$ | 190 | 2 | 2.51 | - | 139 | |
| $Gd_{56}Co_{44}$ | 205 | 2 | 2.64 | - | 158 | |
| $Gd_{52}Co_{48}$ | 239 | 2 | 1.99 | - | 139 | |
| $Gd_{48}Co_{52}$ | 282 | 2 | 1.71 | - | 152 | |
| $Gd_{44}Co_{56}$ | 337 | 2 | 1.27 | - | 148 | |
| $Gd_x(Fe_{10}Co_{90})_{100-x}$ | | | | | | [126] |
| (F, $t$ = 90 nm, $x$ = 30-70)<br>$Gd_{30}(Fe_{10}Co_{90})_{70}$<br>$Gd_{40}(Fe_{10}Co_{90})_{60}$<br>$Gd_{50}(Fe_{10}Co_{90})_{50}$<br>$Gd_{55}(Fe_{10}Co_{90})_{45}$<br>$Gd_{70}(Fe_{10}Co_{90})_{30}$ | $T_{comp}$ =<br>436<br>540<br>508<br>558<br>598 | <br>1.5<br>1.5<br>1.5<br>1.5<br>1.5 | <br>0.25<br>0.48<br>0.97<br>0.86<br>0.75 | <br>-<br>-<br>-<br>-<br>- | <br>-<br>~28.8<br>~58.2<br>~51.6<br>- | |
| Pt/GdFeCo/Pt (F, $t$ = 80 nm) | 586.8<br>324.1<br>($T_{comp}$) | 1.5 | 1.09 | 38.8<br>@ 560<br>K | - | [127] |
| Ta/GdFeCo/Ta (F, $t$ = 80 nm) | 664.3<br>389.7<br>($T_{comp}$) | 1.5 | 0.78 | 15.84<br>@ 610<br>K | - | [127] |
| ***Heusler alloys*** | | | | | | |
| $Ni_{53.4}Mn_{33.2}Sn_{13.4}$ (F)<br>$t$ = 360 nm<br>$t$ = 700 nm<br>$t$ = 1000 nm | <br>557<br>569<br>570 | <br>1.8<br>1.8<br>1.8 | <br>0.4375<br>1.025<br>1.188 | <br>0.9<br>2.375<br>2.65 | <br>-<br>-<br>- | [102] |
| $Ni_{53.2}Mn_{29.2}Co_{7.0}Sn_{10.6}$ (F)<br>$t$ = 360 nm<br>$t$ = 700 nm<br>$t$ = 1000 nm | <br>860<br>863<br>866 | <br>1.8<br>1.8<br>1.8 | <br>0.165<br>0.176<br>0.187 | <br>7.85<br>9.21<br>9.45 | <br>-<br>-<br>- | [102] |
| $Ni_{53.5}Mn_{23.8}Ga_{22.7}$ (F, $t$ = 400 nm) | 346 | 6 | 8.5 | - | - | [128] |
| $Ni_{51}Mn_{29}Ga_{20}$ (F, $t$ = 250 nm) | 355 | 0.5 | 1.4 | - | - | [129] |
| $Ni_{48}(Co_{5})Mn_{35}In_{12}$ (F, $t$ = 200 nm) | 353 | 9 | 8.8 | - | - | [130] |
| $Ni_{51.6}Mn_{32.9}Sn_{15.5}$ (F, $t$ = 200 nm) | 250 | 1 | 1.6 | - | - | [131] |
| $Ni_{51}Mn_{29}Ga_{20}$ (F, $t$ = 250 nm)<br>*Magnetostructural transition* | 356 | 0.5 | 1.4<br>Cooling<br>1.0<br>Heating | 15.4<br>Cooling<br>15.9<br>Heating | -<br><br>- | [5] |
| $Ni_{53.5}Mn_{23.8}Ga_{22.7}$<br>(F, $t$ = 400 nm) | 346 | 6 | 8.5 | ~65 | ~76.5 | [4] |
| $Ni_{43}Mn_{32}Ga_{20}Co_{5}$<br>(F, $t$ = 350 nm) | 340 | 2 | ~3.5 | ~60 | ~70 | [132] |
| ***Oxides*** | | | | | | |
| $La_{2/3}Ca_{1/3}MnO_{3}$ (F, $t$ = 260 nm)<br>$La_{2/3}Ca_{1/3}Mn_{0.94}Cr_{0.06}O_{3}$ (F, $t$ = 260 nm) on LAO | 228<br>~193 | 1<br>1 | 8.5<br>0.83 | ~346<br>~39 | ~255<br>~23 | [133] |
| $La_{0.7}Ca_{0.3}MnO_{3}$<br>*B*<br>$La_{0.7}Ca_{0.3}MnO_{3}$<br>*F*; $t$ = 150 nm | 264<br><br>235 | 5<br><br>5 | 7.70<br><br>2.75 | ~187<br><br>~200 | -<br><br>- | [134]<br>(*) |
| $La_{0.7}Ca_{0.3}MnO_{3}$<br>*F*; $t$ = 30 nm; Intrinsic<br>*F*; $t$ = 30 nm; Extrinsic | <br>225<br>190 | <br>5<br>5 | <br>0.7<br>9 | <br>-<br>~18 | <br>-<br>~18 | [103] |
| $La_{0.8}Ca_{0.2}MnO_{3}$/STO (tensile strain)<br>$t$ = 25 nm | <br>178 | <br>6 | <br>8.20 | <br>183 | <br>250 | [106] |
| $t$ = 50 nm | 186 | 6 | 8.40 | 221 | 295 | |
| $t$ = 75 nm | 195 | 6 | 12.80 | 255 | 361 | |
| $t$ = 100 nm | 193 | 6 | 4.80 | 85 | 125 | |
| $t$ = 300 nm | 210 | 6 | 2.75 | 80 | 105 | |
| $La_{0.8}Ca_{0.2}MnO_3$/LAO (compressive) | | | | | | [106] |
| $t$ = 25 nm | 205 | 6 | 2.25 | 125 | 180 | |
| $t$ = 50 nm | 213 | 6 | 2.63 | 160 | 225 | |
| $t$ = 75 nm | 220 | 6 | 3.25 | 258 | 339 | |
| $t$ = 100 nm | 215 | 6 | 5.95 | 75 | 105 | |
| $t$ = 300 nm | 240 | 6 | 3.00 | 50 | 95 | |
| $La_{0.88}Sr_{0.12}MnO_3$ (F, $t$ = 100 nm) | 175 | 3 | 1.5 | 172 | - | [104] |
| $La_{0.88}Sr_{0.12}MnO_3$ (F, $t$ = 160 nm) | 154 | 3 | 1.8 | 189 | - | |
| $La_{0.88}Sr_{0.12}MnO_3$ (F, $t$ = 200 nm) | 144 | 3 | 1.7 | 199 | - | |
| $La_{0.67}Sr_{0.33}MnO_3$ (F, $t$ = 20 nm) on LSAT | 321 | 1.5 | 1.47 | - | 32.24 | [105] |
| $La_{0.67}Sr_{0.33}MnO_3$ (F, $t$ = 20 nm) on STO | 312 | 1.5 | 1.54 | - | 50.16 | |
| $La_{0.67}Ba_{0.33}Mn_{0.95}Ti_{0.05}O_3$ (F, $t$ = 97 nm) | 234 | 5 | 2.6 | - | 210 | [135] |
| $Pr_{0.7}Sr_{0.3}MnO_3$/PSMO-7 (F, $t$ = 20 nm) | 193 | 2 | 4.7 | ~131.6 | ~164.5 | [136] |
| $EuTiO_3$ (F, $t$ = 100 nm) | 3 | 2 | 24 | 152 | - | [112] |
| $Gd_2NiMnO_6$ (F, $t$ = 15 nm) | 125 | 5 | 1.40 | - | 75 | [120] |
| In-plane | 11 | 5 | 21.82 | - | - | |
| Out-of-plane | 11 | 5 | 9.84 | - | - | |
| $La_2NiMnO_6$ (F, $t$ = 200 - 250 nm) | | | | | | [137] |
| $La_2NiMnO_6$ (300 mTorr) | 265 | 3 | 1.10 | 55 | 73.3 | |
| | 265 | 5 | 1.60 | 100 | 133.3 | |
| | 265 | 7 | 2.10 | 145 | 193.3 | |
| $La_2NiMnO_6$ (100 mTorr) | 237.5 | 5 | 0.50 | 33.3 | - | |
| $La_2NiMnO_6$ (200 mTorr) | 237.5 | 5 | 0.90 | 50 | - | |
| $La_2NiMnO_6$ (300 mTorr) | 250 | 5 | 1.65 | 100 | 133.3 | |
| $La_2NiMnO_6$ (400 mTorr) | 265 | 5 | 1.60 | 125 | - | |
| $GdCoO_3$/LAO (F, $t$ = 22 nm) | 3.5 ($T_N$) | 2 | 12.79 | ~31.0 | ~19.2 | [115] |
| | | 7 | 58.65 | 319.8 | ~429.0 | |
| $EuO_{1-\delta}$ ($\delta$ = 0, 0.025, 0.09) | | | | | | [110] |
| $EuO_1$ (F, $t$ = 100 nm) | 69 | 2 | 6.2 | 460 | 525 | |
| $EuO_{0.975}$ (F, $t$ = 100 nm) | 118 | 2 | 7.1 | 760 | 880 | |
| $EuO_{0.91}$ (F, $t$ = 100 nm) | 133 | 2 | 5.1 | 670 | 780 | |
| $PrVO_3$/LAO (F, $t$ = 55 nm) | 125 ($T_N$) | 5 | 0.44 | 16.24 | 20.4 | [138] |
| $PrVO_3$/STO | 125 | 5 | 0.26 | 11.36 | 13.4 | |
| (F, $t$ = 100 nm)<br>$PrVO_3$/LSAT<br>(F, $t$ = 41.7 nm) | ($T_N$)<br>125<br>($T_N$) | 5 | 0.31 | 13.27 | 15.7 | |
| ***Others*** | | | | | | |
| $SmCo_3B_2$ (F, as-deposited)<br>$t$ = 90 nm<br>$t$ = 160 nm<br>$t$ = 240 nm | <br>40<br>41<br>43 | <br>5<br>5<br>5 | <br>0.614<br>0.892<br>0.537 | <br>-<br>-<br>- | <br>~2.3<br>~2.4<br>~2.7 | [139] |
| $SmCo_3B_2$ (F)<br>$t$ = 90 nm<br>$t$ = 160 nm<br>$t$ = 240 nm | <br>43<br>44<br>46 | <br>5<br>5<br>5 | <br>0.629<br>0.886<br>1.172 | <br>-<br>-<br>- | <br>~2.7<br>~3.1<br>~5.3 | [139] |
| Ta(20nm)/Er-Co-Al(200-300nm)/<br>Ta(25nm)<br>$ErCo_{1.52}Al_{0.36}$ (as-deposited)<br>$ErCo_{1.69}Al_{0.76}$ (as-deposited)<br>$ErCo_{1.87}Al_{0.16}$ (as-deposited)<br>$ErCo_{1.52}Al_{0.36}$ (annealed at 1073 K)<br>$ErCo_{1.69}Al_{0.76}$ (annealed at 1073 K)<br>$ErCo_{1.87}Al_{0.16}$ (annealed at 1073 K) | <br><br>28<br>17.5<br>17.5<br>12.5<br>12.5<br>12.5 | <br><br>5<br>5<br>5<br>5<br>5<br>5 | <br><br>1.9<br>2.9<br>0.25<br>3.0<br>2.4<br>3.2 | <br><br>17.1<br>26.1<br>2.25<br>27<br>21.6<br>28.8 | <br><br>22.8<br>34.8<br>3.0<br>36<br>28.8<br>38.4 | <br><br>[140] |
| $Tb_{30}Fe_7Co_{63}$ (F, $t$ = 100 nm)<br>P = 50 W<br>P = 60 W<br>P = 70 W<br>P = 80 W<br>P = 90 W<br>P = 100 W | $T_{comp}$<br>=<br>407<br>357<br>330<br>306<br>252<br>224 | <br>1.5<br>1.5<br>1.5<br>1.5<br>1.5<br>1.5 | <br>0.21<br>0.20<br>0.18<br>0.16<br>0.15<br>0.13 | <br>-<br>-<br>-<br>-<br>-<br>- | <br>-<br>-<br>-<br>-<br>-<br>- | [141] |
| Epitaxial Tb (F, $t$ = 100 nm)<br>H//a axis (in-plane)<br>H//b axis (in-plane)<br>H//c axis (out-of-plane)<br>Amorphous Tb (F, $t$ = 100 nm)<br>In-plane<br>Out-of-plane | <br>232<br>232<br>232<br><br>227<br>227 | <br>2<br>2<br>2<br><br>2<br>2 | <br>6.27<br>5.61<br>1.11<br><br>1.98<br>0.67 | | <br>225<br>199<br>18<br><br>86<br>25 | [121]<br><br><br><br>[121] |
| $(Fe_{70}Ni_{30})_{96}Mo_4$<br>(F, $t$ = 30 nm) | 323 | 1<br>2 | 0.77<br>1.38 | 119<br>228 | -<br>- | [142] |
| $CrF_3$ (2D van der Waals)<br>(F, $t$ = 5.19 nm) | 18 | 5 | 32.2 | - | - | [111] |
| $CrCl_3$<br>(F, $t$ = 6.06 nm) | 22 | 5 | 21.9 | - | - | [111] |
| $CrBr_3$<br>(F, $t$ = 6.44 nm) | 36 | 5 | 12.5 | - | - | [111] |
| $CrI_3$<br>(F, $t$ = 7.01 nm) | 48 | 5 | 7.5 | - | - | [111] |
| $CrO_2/TiO_2$ | 385 | 5 | 8.46 | 410 | - | [143] |
| (F, $t$ = 500 nm) | | | | 143<br>(1.5T) | | |
| $Fe_2Ta$ (F, $t$ ~ 100 nm) | 12.5<br>270 | 0.5 | $5.43 \times 10^{-4}$<br>-<br>$1.58 \times 10^{-4}$ | -<br>- | -<br>- | [144]<br>(*) |
| MnCoAs (F, $t$ = 3.58 nm) | 214 – 221 | 1 – 7 | 1.4 – 4.3 | 28.4 – 244.5 | - | [145] |
| $Fe_3[Cr(CN)_6]_2 \cdot zH_2O$<br>(F, $t$ = 1400 nm) | 20 | 1<br>5 | 3.2<br>10 | 21.1<br>273 | -<br>- | [146] |
| $Cr_3[Cr(CN)_6]_2 \cdot zH_2O$<br>(F, $t$ = 1100 nm) | 219 | 1<br>5 | 0.20<br>0.72 | 8<br>44 | -<br>- | [146] |
| ***Heterostructure and multi-layer structures*** | | | | | | |
| Py/Gd/CoFe/IrMn stacks<br>Py=$Ni_{80}Fe_{20}$<br>$Gd_{thick}$ = 20 nm | 120-130 | 5 | 0.0256 | - | - | [147] |
| Py/Gd/CoFe/IrMn stacks<br>Py=$Ni_{80}Fe_{20}$<br>$Gd_{thick}$ = 5 nm | 85 | 3 | 0.0128 | - | - | [147] |
| $La_{1-x}Sr_xMnO_3$ (F, $t$ = 35 nm)<br>($x$ = 0.12)<br>$La_{1-x}Sr_xMnO_3$ (F, $t$ = 35 nm)<br>($x$ = 0.25)<br>$La_{1-x}Sr_xMnO_3$ 12/25<br>$La_{1-x}Sr_xMnO_3$ 25/12 | 170<br><br>295<br><br>170/300<br>170/300 | 3<br><br>295 | 0.2<br><br>0.21<br><br>0.09/0.09<br>0.09/0.09 | 12<br><br>11<br><br>14<br>15 | -<br><br>-<br><br>-<br>- | [148] |
| Gd(30 nm)/W(5 nm) | 280 | 3 | 2.8 | - | - | [98] |
| Quart/ $Ni_{80}Fe_{20}$(10nm)/$Ni_{67}Cu_{33}$(d nm)/$Co_{90}Fe_{10}$(3nm)/<br>$Ir_{20}Mn_{80}$(25nm)/TiO<br>$d$ = 3 – 15 nm | ~330 | 0.003 | 10-15 | ~40-85 | ~50-105 | [149] |
| Si/$Co_{90}Fe_{10}$(20nm)/$Ni_{72}Cu_{28}$(d nm)/<br>$Co_{40}Fe_{40}B_{20}$(15nm)/TiO<br>$d$ = 5 – 20 nm | ~360 | 0.003 | 37.10 | ~180-250 | ~222-297 | [149] |
| $BiFeO_3$(15 nm)/LSMO(40 nm)<br><br>$BiFeO_3$(50 nm)/LSMO(40 nm)<br><br>$BiFeO_3$(120nm)/LSMO(40nm)<br><br>$BiFeO_3$(140nm)/LSMO(40nm) | ~280<br><br>~240<br><br>~260<br><br>~220 | 0.02<br><br>0.02<br><br>0.02<br><br>0.02 | $10 \times 10^{-4}$<br><br>$7 \times 10^{-4}$<br><br>$3 \times 10^{-4}$<br><br>$1.32 \times 10^{-4}$ | 0.21<br><br>0.125<br><br>0.04<br><br>0.01 | -<br><br>-<br><br>-<br><br>- | [118] |
| Cr/Py/$Fe_{30}Cr_{70}$(6nm)/Py/FeMn/Cr (Py=$Ni_{80}Fe_{20}$); $t$ = 50 nm<br>Cr/Py/Cr/Py/FeMn/Cr (Py=$Ni_{80}Fe_{20}$); $t$ = 50 nm | 162<br>160 | 0.025<br>0.025 | ~0.06 - 0.08<br>0.024 | -<br>- | -<br>- | [150] |
| FeRh/$BaTiO_3$ (F, $t$ = 40 nm) | 351 | 2 | 17 | ~272 | ~340 | [114] |
| $Ni_{80}Fe_{20}$/$Ni_{67}Cu_{33}$/$Co_{90}Fe_{10}$/$Mn_{80}Ir_{20}$<br>Spacer $Ni_{67}Cu_{33}$ ($t$ = 7 nm)<br>Spacer $Ni_{67}Cu_{33}$ ($t$ = 10 nm)<br>Spacer $Ni_{67}Cu_{33}$ ($t$ = 21 nm) | <br>260<br>250<br>200 | <br>0.002<br>0.002<br>0.002 | <br>0.0067<br>0.0076<br>0.0133 | <br>~70.1<br>~63.2<br>~21.8 | <br>71.26<br>~66.7<br>~24.1 | [119] |
| Fe/Fe-Cr/Fe simulated (F, $t$ = 6 nm) | ~200-214 | 0.25-1 | >6.4 | - | - | [151] |
| FM/AFM=$MnF_2$/FM (F, $t$ = 30 nm) | ~67 ($T_N$) | 1 | 13 | ~225 | ~300 | [152] |
| Fe/Gd/Fe (F, $t$ = 15 nm) | ~200 | 0.03 | $1.27 \times 10^{-3}$ | - | - | [153] |

*B: Bulk; F: Film; t: thickness; P: Pressure; $T_N$ = Néel temperature; $T_{comp}$ = compensation temperature; $T_{MS}$: Magnetostructural transition temperature. (*) represents FOMT materials.*

**Table 3.** Maximum entropy change, $\left|\Delta S_M^{\max}\right|$, Curie temperature, $T_C$, refrigerant capacity (*RC*), and relative cooling power (*RCP*) for the ribbon samples.

| Samples | $T_C$ (K) | $\mu_0\Delta H$ (T) | $\left\|\Delta S_M^{\max}\right\|$ (J/kg K) | *RC* (J/kg) | *RCP* (J/kg) | Ref. |
|---|---|---|---|---|---|---|
| ***Gadolinium and its alloys*** | | | | | | |
| Gd (*B*) | 294 | 1 | 2.8 | - | 63.4 | [7] |
| | | 2 | 5.07 | - | 187 | |
| | | 5 | 10.2 | ~400 | 410 | |
| Gd (*R*) | 294 | 1.7 | 4.8 | - | - | [177] |
| Gd (*R*) | 293 | 5 | 8.7 | 433.4 | | [158] |
| $Gd_{71}Co_{29}$ (*R*; *A*) | 166 | 1 | 3.1 | 92.3 | - | [159] |
| $Gd_{68}Co_{32}$ (*R*; *A*) | 175 | 1 | 3.0 | 87.4 | - | [159] |
| $Gd_{65}Co_{35}$ (*R*; *A*) | 184 | 1 | 2.9 | 83.6 | - | [159] |
| $Gd_{62}Co_{38}$ (*R*; *A*) | 193 | 1 | 2.8 | 81.4 | - | [159] |
| $Gd_{48}Co_{52}$ (*R*; *A*) | 282 | 1.5 | 1.71 | - | 176 | [178] |
| | 282 | 5 | 4.23 | - | 750 | |
| $Gd_4Co_3$ (*R*; *A*) | 219 | 5 | 7.2 | - | - | [179] |
| $Gd_{60}Co_{25}Al_{15}$ (*R*; *A*) | 125 | 5 | 10.1 | 645 | 860 | [161] |
| $Gd_{55}Co_{25}Al_{20}$ (*R*; *A*) | 112.5 | 5 | 10.1 | 612.6 | 818 | [162] |
| $Gd_{60}Co_{30}Al_{10}$ (*R*; *A*; Sheet parallel) | 140 | 1.9 | 4.06 | 64 | - | [180] |
| $Gd_{60}Co_{30}Al_{10}$ (*R*; *A*; Sheet perpendicular) | 140 | 1.9 | 2.91 | 34 | - | [180] |
| $Gd_{50}Co_{50-x}Fe_x$ (*R*; *A*) | | | | | | [181] |
| $x = 0$ | 267 | 5 | - | - | - | |
| $x = 2$ | 277 | 5 | 4.44 | - | - | |
| $Gd_{50}Co_{50-x}Si_x$ (*R*; *A*) | | | | | | [182] |
| $x = 2$ | 214 | 5 | 5.32 | - | 710 | |
| $x = 5$ | 244 | 5 | 5.98 | - | 740 | |
| $Gd_{48}Co_{50}Zn_2$ (*R*; *A*) | 262 | 5 | 5.02 | - | >700 | [183] |
| $Gd_{50}Co_{48}Zn_2$ (*R*; *A*) | 260 | 5 | 5.04 | - | 700 | [183] |
| $Gd_{55}Co_{35}M_{10}$ (*R*; *A*) | | | | | | [184] |
| *M* = Mn | 197 | 2 | 3.03 | 162 | 224 | |
| *M* = Fe | 268 | 2 | 1.72 | 266 | 337 | |
| *M* = Ni | 192 | 2 | 3.37 | 183 | 253 | |
| $Gd_{55}Co_{20}Fe_5Al_{20-x}Si_x$ (*R*; *A*) | | | | | | [185] |
| $x = 0$ | 130 | 5 | 6.1 | 558 | - | |
| $x = 5$ | 142 | 5 | 6.82 | 665 | | |
| $x = 10$ | 149 | 5 | 6.36 | 700 | | |
| $x = 15$ | 151 | 5 | 4.94 | 519 | | |
| $Gd_{55}Co_{20}Fe_5Al_{20-x}Si_x$<br>($x$=0, 2, 5, 10)<br>($R$; $A$)<br>($x$ = 15, 20, 20; annealed)<br>($R$; $C$) | 129<br>136<br>108<br>137<br>130<br>130<br>-- | 5<br>5<br>5<br>5<br>5<br>5<br>5 | 8.01<br>6.66<br>4.90<br>4.48<br>4.77<br>-<br>4.78 | -<br>-<br>-<br>-<br>-<br>-<br>- | 913<br>719<br>541<br>622<br>596<br>-<br>- | [186] |
| $Gd_{65}Fe_{20}Al_{15}$ ($R$; $A$) | 182 | 5 | 5.8 | 545 | 726 | [163] |
| $Gd_{55}Fe_{15}Al_{30}$ ($R$; $A$) | 158 | 5 | 5.01 | 555 | 741 | [163] |
| $Gd_{55}Fe_{20}Al_{25}$ ($R$; $A$) | 190 | 5 | 4.67 | 651 | 868 | [163] |
| $Gd_{55}Fe_{25}Al_{20}$ ($R$; $A$) | 230 | 5 | 3.77 | 608 | 811 | [163] |
| $Gd_{95}Fe_{2.8}Al_{2.2}$ ($R$; $A$) | 232 | 5 | 4 | 551 | - | [187] |
| $Gd_{95}Fe_{2.8}Al_{2.2}$ ($R$; $C$) | 232 | 5 | 7.53 | 551 | - | [187] |
| $Gd_{55}Fe_{15}Al_{30}$ ($R$; $A$) | 158 | 5 | 5.01 | 741 | - | [163] |
| $Gd_{55}Fe_{20}Al_{25}$ ($R$; $A$) | 182 | 5 | 4.67 | 868 | - | [163] |
| $Gd_{55}Fe_{25}Al_{20}$ ($R$; $A$) | 197 | 5 | 3.77 | 811 | - | [163] |
| $Gd_{55}Fe_{30}Al_{15}$ ($R$; $A$) | 208 | 5 | 3.43 | 857 | - | [163] |
| $Gd_{55}Fe_{35}Al_{10}$ ($R$; $A$) | 228 | 5 | 2.92 | 826 | - | [163] |
| $Gd_{71}Fe_3Al_{26}$ ($R$; $A$) | 117.5 | 5 | 7.4 | 750 | - | [188] |
| $Gd_{65}Fe_{20}Al_{15}$ ($R$; $A$) | 182.5 | 5 | 5.8 | 726 | - | [188] |
| RNi (R=Gd, Tb and Ho)<br>($R$; $C$) | 75<br>66<br>28 | 5<br>5<br>5 | 15.2<br>12<br>14.1 | -<br>-<br>- | 610<br>370<br>550 | [189] |
| $Gd_{63}Ni_{37}$ ($R$; $A$) | 122 | 5 | 9.42 | 600 | 802.6 | [190] |
| $Gd_{71}Ni_{29}$ ($R$; $A$) | 122 | 5 | 9 | 724 | - | [160] |
| $Gd_{68}Ni_{32}$ ($R$; $A$) | 124 | 5 | 8 | 583 | - | [160] |
| $Gd_{65}Ni_{35}$ ($R$; $A$) | 122 | 5 | 6.9 | 524 | - | [160] |
| $Gd_{46}Ni_{32}Al_{22}$ ($R$; $A$) | 66 | 5 | 10.16 | 762 | - | [191] |
| $Gd_{55}Ni_{15}Al_{30}$ ($R$; $A$) | 70 | 5 | 6.12 | 606 | - | [164] |
| $Gd_{55}Ni_{20}Al_{25}$ ($R$; $A$) | 71 | 5 | 7.98 | 782 | - | [164] |
| $Gd_{55}Ni_{25}Al_{20}$ ($R$; $A$) | 75 | 5 | 8.49 | 806 | - | [164] |
| $Gd_{55}Ni_{30}Al_{15}$ ($R$; $A$) | 83 | 5 | 9.25 | 851 | - | [164] |
| $Gd_{34}Ni_{22}Co_{11}Al_{33}$ ($R$; $A$) | 54 | 5 | 9.9 | - | 145 | [192] |
| $Gd_{100-x}Mn_x$ ($R$; $C$)<br>$x$ = 0<br>$x$ = 5<br>$x$ = 10<br>$x$ = 15<br>$x$ = 20 | <br>293<br>289<br>287<br>285<br>278 | <br>5<br>5<br>5<br>5<br>5 | <br>8.7<br>8.3<br>6.8<br>6.6<br>5.9 | <br>433.4<br>451.9<br>353.7<br>354.5<br>321.5 | <br>-<br>-<br>-<br>-<br>- | [158] |
| $Gd_{65}Mn_{35-x}Si_x$ ($R$; $A$)<br>$x$ = 5<br>$x$ = 10 | <br>221<br>218 | <br>5<br>5 | <br>4.6<br>4.7 | <br>625<br>660 | <br>-<br>- | [193] |
| $Gd_{65}Mn_{25}Si_{10}$ ($R$; $A$+$C$) | 288 | 5 | 4.6 | 249 | - | [193] |
| $(Gd_4Co_3)_{1-x}Si_x$ ($R$; $A$)<br>$x$ = 0 | <br>208 | <br>5 | <br>7.3 | <br>547 | <br>- | [194] |
| $x = 0.05$<br>$x = 0.10$ | 198<br>213 | 5<br>5 | 7.2<br>6.4 | 524<br>511 | | |
| $(Gd_{1-x}Tb_x)_{12}Co_7$ (*R*; *A*)<br>$x = 0$<br>$x = 0.25$<br>$x = 0.5$<br>$x = 0.75$<br>$x = 1$ | <br>179<br>159<br>136<br>118<br>92 | <br>5<br>5<br>5<br>5<br>5 | <br>7.9<br>8.0<br>9.0<br>8.5<br>8.4 | <br>511<br>522<br>540<br>462<br>456 | <br>-<br>-<br>-<br>-<br>- | [195] |
| GdCuAl (*R*; *A+C*) | 50 | 5 | 5.6 | 296 | - | [196] |
| $Gd_{55}Co_{25}Ni_{20}$ (*R*; *A*) | 140 | 5 | 6.04 | 450 | - | [197] |
| $Gd_{55}Co_{30}Ni_{15}$ (*R*; *A*) | 175 | 5 | 6.3 | 487 | - | [197] |
| $Gd_{55}Co_{35}Ni_{10}$ (*R*; *A*) | 192 | 5 | 6.47 | 502 | - | [197] |
| $Gd_{60}Mn_{30}Ga_{10}$ (*R*; *A* + *C*) | 177 | 2 | 1.53 | 240 | - | [198] |
| $Gd_{60}Mn_{30}In_{10}$ (*R*; *A+C*) | 190 | 2 | 1.49 | 234 | - | [198] |
| $Gd_{60}Co_{30}In_{10}$ (*R*; *A+C*) | 159 | 4.6 | 7.7 | 406 | - | [199] |
| $Gd_{60}Ni_{30}In_{10}$ (*R*; *A+C*) | 86 | 4.6 | 8.2 | 602 | - | [199] |
| $Gd_{60}Cu_{30}In_{10}$ (*R*; *A+C*) | 115 | 4.6 | 6.6 | 598 | - | [199] |
| $Gd_{60}Fe_0Co_{30}Al_{10}$ (*R*; *A+C*) | 145 | 5 | 8.9 | 539 | - | [200] |
| $Gd_{60}Fe_{10}Co_{20}Al_{10}$ (*R*; *A+C*) | 170 | 5 | 5 | 632 | - | [200] |
| $Gd_{60}Fe_{20}Co_{10}Al_{10}$ (*R*; *A+C*) | 185 | 5 | 4.4 | 736 | - | [200] |
| $Gd_{60}Fe_{30}Co_0Al_{10}$ (*R*; *A+C*) | 200 | 5 | 3.6 | 672 | - | [200] |
| $Gd_{45}RE_{20}Fe_{20}Al_{15}$ (*R*; *A*)<br>(RE=Tb, Dy, Ho, Er) | 138-<br>175 | 5 | 4.46-<br>5.57 | 580-<br>720 | - | [201] |
| $Gd_{55}Co_{19}Al_{24}Si_1Fe_1$ (*R*; *A+C*) | 107 | 5 | 7.8 | 749 | - | [155] |
| $Gd_{55}Co_{35}Mn_{10}$ (*R*; *A+C*)<br>(50 m/s) | 200<br>200 | 2<br>5 | 3.82<br>6.47 | 183.4<br>457.3 | 233.0<br>601.7 | [202] |
| $Gd_{55}Co_{35}Mn_{10}$ (*R*; *A+C*)<br>(600 K / 20 min) | 123/17<br>3 | 2<br>5 | 2.93<br>5.50 | 233.5<br>515.7 | 284.2<br>649 | [202] |
| $Gd_{55}Co_{35}Mn_{10}$ (*R*; *A+C*)<br>(600 K / 30 min) | 123/17<br>0 | 2<br>5 | 2.79<br>5.46 | 242.1<br>536.4 | 284.6<br>671.6 | [202] |
| $Gd_{65}Fe_{10}Co_{10}Al_{15}$ (*R*; *A*) | 160 | 5 | 6.0 | 700 | - | [203] |
| $Gd_{65}Fe_{10}Co_{10}Al_{10}Si_5$ (*R*; *A*) | 175 | 5 | 5.9 | 698 | - | [203] |
| $Gd_{65}Fe_{10}Co_{10}Al_{10}B_5$ (*R*; *A*) | 145 | 5 | 7.1 | 748 | - | [203] |
| $Gd_{55}Co_{35}Ni_{10}$ (*R*; *A*) | 158/21<br>4 | 5 | 5.0 | - | - | [204] |
| $Gd_{50}Co_{45}Fe_5$ (*R*; *A*) | 289 | 5 | 3.8 | - | 673 | [205] |
| $Tm_{60}Al_{20}Ni_{10}$ (*R*; *A*) | 4.4 | 5 | 14.1 | - | 235 | [206] |
| $Er_{60}Al_{20}Ni_{10}$ (*R*; *A*) | 9.5 | 5 | 14.3 | - | 372 | [206] |
| $Ho_{60}Al_{20}Ni_{10}$ (*R*; *A*) | 17.9 | 5 | 12.4 | - | 460 | [206] |
| $ErNi_2$ (*R*; *C*) | 6.8 | 2<br>5 | 14.1<br>20.0 | 146<br>382 | - | [207] |
| $ErNi_2$ (*R*; *C*) | 6.8 | 2<br>5 | 12.4<br>20.2 | 118<br>347 | - | [207] |
| $TbNi_2$ (*R*; *C*) | 37 | 5 | 13.9 | 441 | - | [208] |
| $DyNi_2$ (*R*; *C*) | 21.5 | 2 | 13.5 | 209 | - | [209] |
| $HoNi_2$ (*R*; *C*) | 13.9<br>13.9 | 2<br>5 | 16.9<br>27.2 | 194<br>522 | - | [210] |
| $HoNi_2$ (*R*; *C*) | 13.9<br>13.9 | 2<br>5 | 14.8<br>24.8 | 169<br>465 | - | [210] |
| $Dy_3Co$ (*R*; *C*) | 32<br>43 | 2<br>5 | 2.1<br>6.5 | -<br>- | 83<br>364 | [211] |
| $Tb_{55}Co_{30}Fe_{15}$ (*R*; *A*) | 169 | 5 | 4 | - | - | [212] |
| ***Heusler alloys*** | | | | | | |
| $Mn_{50}Ni_{41}In_9$ (*R*; *C*) | 283 | 3 | 5.7 | 184.2 | 197.8 | [213] |
| $Mn_{50}Ni_{40}In_{10}$, *H*‖<br>Melt-spun ribbons,<br>$Mn_{50}Ni_{40}In_{10}$, *H*⊥<br>Melt-spun ribbons<br>(*R*; *C*) | 230<br>310<br>230<br>310 | 3<br><br>3 | 3.6<br>1.3<br>3.5<br>1.3 | 71<br>89<br>71<br>86 | -<br>-<br>-<br>- | [214]<br>(*)<br><br>(*) |
| $Ni_{50}Mn_{50-x}Sn_x$ (*R*; *C*)<br>*x* = 13 | 255<br>300 | 5 | 22<br>4 | 160<br>75 | - | [19]<br>(*) |
| $Ni_{52}Mn_{26}Ga_{22}$ (*R*; *C*)<br>As-melt<br>Annealed | 350<br><br>354 | 2<br>5<br>2<br>5 | 5.3<br>11.4<br>16.4<br>30 | -<br>-<br>32<br>70 | -<br>-<br>-<br>- | [215] |
| $Mn_3Sn_{2-x}B_x$ (*R*; *C*)<br>(*x* = 0 - 0.5) | 240-<br>250 | 5 | 13.6 –<br>18.3 | - | - | [216] |
| $Mn_3Sn_{2-x}C_x$ (*R*; *C*)<br>(*x* = 0 - 0.5) | 240-<br>250 | 5 | 13.6 –<br>17.5 | - | - | [216] |
| $Ni_{51.1}Mn_{31.2}In_{17.7}$ (*R*; *C*)<br>Annealed at 1073 K / 10 min<br>Annealed at 1073 K / 2 h | 276<br>288<br>288 | 5<br>5<br>5 | 3.1<br>4.1<br>4.4 | 345<br>268<br>294 | -<br>-<br>- | [217] |
| $Ni_{43}Mn_{46}In_{11}$ (*R*; *C*)<br><br>Slow cooled<br><br>Quenched | 245<br>304<br>260<br>320<br>263<br>311 | 5<br><br>5<br><br>5 | 1.31<br>1.45<br>3.48<br>2.05<br>6.79<br>2.72 | -<br>-<br>-<br>-<br>-<br>- | 38.5<br>95.7<br>97.4<br>114.8<br>142.6<br>152.3 | [218]<br>(*)<br><br>(*)<br><br>(*) |
| $Ni_{52}Mn_{26}Ga_{22}$ (*R*; *C*) | 348<br>348 | 2<br>5 | 5.3<br>11.4 | -<br>- | -<br>- | [219] |
| $Mn_{50}Ni_{40.5}In_{9.5}$ (*R*; *C*) | 295 | 5 | 3.7 | 80.5 | - | [220] |
| $Mn_{50}Ni_{40.5}In_{9.5}$ (*R*; *C*, Annealed) | 326 | 5 | 6.1 | 126.6 | - | [220] |
| $Ni_{45}Co_5Mn_{31}Al_{19}$ (*R*; *C*) | 265<br>291 | 1.35 | 2<br>1 | - | - | [221]<br>(*) |
| $Ni_{46}Co_4Mn_{38}Sb_{12}$ (*R*; *C*)<br>(In-plane) | 297 | 5 | 13.5<br>12.6 | - | - | [222] |
| $Ni_{46}Co_4Mn_{38}Sb_{12}$ (*R*; *C*)<br>(Out of plane) | 297 | 5 | 12.6 | - | - | [222] |
| $Ni_{42.9}Co_{6.9}Mn_{38.3}Sn_{11.9}$ (*R*; *C*) | 302 | 1 | 6.7 | 45.3 | - | [223] |
| $Ni_{42.9}Co_{6.9}Mn_{38.3}Sn_{11.9}$<br>(*R*; *C*; Annealed) | 308 | 1 | 25.3 | 55.8 | - | [223] |
| $Ni_{50}Mn_{35}In_{14.25}B_{0.75}$ (*R*; *C*) | 318 | 5 | 16 | 85 | - | [224] |
| $Ni_{48}Co_{2}Mn_{35}In_{15}$ (*R*; *C*) | 326 | 5 | 12.1 | 78 | - | [225] |
| $Ni_{50-x}Co_{x}Mn_{35}In_{15}$ (*R*; *C*) | | | | | | [226] |
| $x = 0$ | 305 | 1 | 0.92 | 6.97 | - | |
| $x = 1$ | 315 | 1 | 5.35 | 31.87 | - | |
| $x = 2$ | 325 | 1 | 3.90 | 42.05 | - | |
| $Ni_{45}Co_{5}Mn_{40}Sn_{10}$ (*R*; *C*) | 436 | 1 | 2 | - | - | [227] |
| | 431 | 3 | 10 | - | - | |
| | 426 | 5 | 22 | - | - | |
| | 424 | 7 | 27 | - | - | |
| $Ni_{42.7}Mn_{40.8}Co_{5.2}Sn_{11.3}$ (*R*; *C*) | 263 | 5 | 6.8 | 24 | - | [228] |
| | 377.5 | | 1.3 | 67 | | (*) |
| $Ni_{42.7}Mn_{40.8}Co_{5.2}Sn_{11.3}$ (*R*; *C*) | 270 | 5 | 32.8 | 44 | - | [228] |
| annealed at 1123 K/10 min | 379 | | 1.6 | 91 | | (*) |
| ***Fe-based alloys*** | | | | | | |
| $Fe_{90}Zr_{10}$ (*R*; *A*) | 230 | 2 | 1.3 | 194 | - | [166] |
| | | 5 | 2.7 | 497 | - | |
| | | 8 | 3.9 | 801 | - | |
| $Fe_{90}Zr_{9}B_{1}$ (*R*; *A*) | 210 | 2 | 1.3 | 198 | - | [166] |
| | | 5 | 2.7 | 492 | - | |
| | | 8 | 3.8 | 795 | - | |
| $Fe_{91}Zr_{7}B_{2}$ (*R*; *A*) | 215 | 2 | 1.2 | 177 | - | [166] |
| | | 5 | 2.5 | 462 | - | |
| | | 8 | 3.6 | 755 | - | |
| $Fe_{90}Zr_{8}B_{2}$ (*R*; *A*) | 240 | 2 | 1.3 | 198 | - | [166] |
| | | 5 | 2.6 | 514 | - | |
| | | 8 | 3.7 | 830 | - | |
| $Fe_{88}Zr_{8}B_{4}$ (*R*; *A*) | 280 | 2 | 1.3 | 201 | - | [166] |
| | | 5 | 2.8 | 551 | - | |
| | | 8 | 4.0 | 905 | - | |
| $Fe_{87}Zr_{6}B_{6}Cu_{1}$ (*R*; *A*) | 300 | 2 | 1.6 | 208 | - | [166] |
| | | 5 | 3.0 | 590 | - | |
| | | 8 | 4.3 | 953 | - | |
| $Fe_{86}Zr_{7}B_{6}Cu_{1}$ (*R*; *A*) | 320 | 2 | 1.6 | 205 | - | [166] |
| | | 5 | 3.1 | 582 | - | |
| | | 8 | 4.4 | - | - | |
| $Fe_{89}Zr_{7}B_{4}$ (*R*; *A*) | 275 | 5 | 3.19 | - | - | [229] |
| $Fe_{87}Zr_{7}B_{4}Dy_{2}$ (*R*; *A*) | 308 | 5 | 3.14 | - | - | [229] |
| $Fe_{87}Zr_{7}B_{4}Tb_{2}$ (*R*; *A*) | 319 | 5 | 3.25 | - | - | [229] |
| $Fe_{87}Zr_{7}B_{4}Gd_{2}$ (*R*; *A*) | 342 | 5 | 3.24 | - | - | [229] |
| $Fe_{89}Zr_{8}B_{3}$ (*R*; *A*) | 271 | 5 | 2.75 | - | - | [230] |
| $Fe_{88}Zr_{8}B_{4}$ (*R*; *A*) | 291 | 5 | 3.04 | - | 644.9 | [230] |
| $Fe_{87}Zr_{8}B_{5}$ (*R*; *A*) | 306 | 5 | 3.25 | - | - | [230] |
| $Fe_{88}Zr_{9}B_{3}$ (*R*; *A*) | 286 | 5 | 3.17 | - | 686.7 | [231] |
| $Fe_{87}Zr_{9}B_{4}$ (*R*; *A*) | 304 | 5 | 3.29 | - | - | [231] |
| $Fe_{86}Zr_9B_5$ (*R*; *A*) | 327 | 5 | 3.34 | - | - | [231] |
| $Fe_{88}Gd_2Zr_{10}$ (*R*; *A*) | 285 | 5 | 4.03 | - | 282 | [232] |
| $Fe_{65}Mn_{15}B_{20}$ (*R*; *A*) | 328 | 1.5 | 0.89 | 72.5 | 99.7 | [233] |
| $Fe_{60}Mn_{20}B_{20}$ (*R*; *A*) | 200 | 1.5 | 0.6 | 62.5 | 84.5 | [233] |
| $Fe_{56}Mn_{24}B_{20}$ (*R*; *A*) | 170 | 1.5 | 0.55 | 51 | 66.7 | [233] |
| $Fe_{70}Mn_{10}B_{20}$ (*R*; *A*) | 450 | 1.5 | 1.01 | 84.4 | 117 | [233] |
| $Fe_{80}Cr_8B_{12}$ (*R*; *A*) | 328 | 1.5 | 1.00 | - | 130 | [234] |
| $Fe_{88}Zr_7B_4Ni_1$ (*R*; *A*) | 285 | 1.5<br>5 | 1.32<br>3.24 | -<br>- | 132<br>- | [235] |
| $Fe_{88}Zr_7B_4Al_1$ (*R*; *A*) | 280 | 1.5 | 1.37 | - | - | [235] |
| $Fe_{88}Zr_9B_1Co_2$ (*R*; *A*) | 285 | 1.5 | 1.61 | - | 149.7 | [236] |
| $Fe_{87}Zr_{11}B_1Co_1$ (*R*; *A*) | 280 | 1.5 | 1.38 | - | 133.9 | [236] |
| $Fe_{88}Ce_7B_5$ (*R*; *A*) | 287 | 1.5<br>5 | 1.52<br>3.83 | -<br>- | -<br>700.9 | [237] |
| $Fe_{88}La_2Ce_5B_5$ (*R*; *A*) | 293 | 1.5<br>5 | 1.53<br>3.85 | -<br>- | -<br>656.7 | [237] |
| $Fe_{90-x}Ni_xZr_{10}$ (*R*; *A*)<br>$x = 0$<br>$x = 5$<br>$x = 10$<br>$x = 15$ | <br>245<br>306<br>356<br>403 | <br>4<br>4<br>4<br>4 | <br>3.04<br>3.26<br>3.30<br>3.10 | <br>334<br>290<br>-<br>- | <br>-<br>-<br>-<br>- | [167] |
| $Fe_{90-x}Sn_xZr_{10}$ (*R*; *A*)<br>$x = 0$<br>$x = 2$<br>$x = 4$ | <br>247<br>269<br>293 | <br>5<br>5<br>5 | <br>3.6<br>4.1<br>3.4 | <br>320<br>255<br>228 | <br>410<br>337<br>280 | [238] |
| $Fe_{82}B_4Mn_4Zr_8Nb_2$ (*R*; *A*) | 237 | 1<br>3 | 0.97<br>2.19 | - | - | [239] |
| $Fe_{78}B_8Mn_4Zr_8Nb_2$ (*R*; *A*) | 259 | 1<br>3 | 0.88<br>1.97 | - | - | [239] |
| $Fe_{74}B_{12}Mn_4Zr_8Nb_2$ (*R*; *A*) | 282 | 1<br>3 | 0.73<br>1.63 | - | - | [239] |
| $Fe_{70}B_{16}Mn_4Zr_8Nb_2$ (*R*; *A*) | 313 | 1<br>3 | 0.68<br>1.58 | - | - | [239] |
| $Fe_{66}B_{20}Mn_4Zr_8Nb_2$ (*R*; *A*) | 328 | 1<br>3 | 0.62<br>1.38 | - | - | [239] |
| $Fe_{64}Mn_{16}P_{10}B_7C_3$ (*R*; *A*) | 266 | 1.5<br>2 | 0.78<br>0.98 | 74.7<br>101.5 | 101.05<br>139.74 | [240] |
| $Fe_{65}Mn_{15}P_{10}B_7C_3$ (*R*; *A*) | 292 | 1.5<br>2 | 0.91<br>1.12 | 79.8<br>109.2 | 117.53<br>147.09 | [240] |
| $Fe_{66}Mn_{14}P_{10}B_7C_3$ (*R*; *A*) | 319 | 1.5<br>2 | 0.91<br>1.12 | 71.9<br>99.8 | 99.84<br>134.25 | [240] |
| $Fe_{67}Mn_{13}P_{10}B_7C_3$ (*R*; *A*) | 339 | 1.5<br>2 | 1.00<br>1.24 | 67.2<br>93.4 | 90.07<br>127.57 | [240] |
| $Fe_{88}Zr_7B_4Cu_1$ (*R*; *A*) | 287 | 1.5 | 1.32 | 121 | 166 | [241] |
| $Fe_{82.5}Co_{2.75}Ni_{2.75}Zr_7B_4Cu_1$ (*R*; *A*) | 400 | 1.5 | 1.4 | 119 | 165 | [241] |
| $Fe_{78}Co_5Ni_5Zr_7B_4Cu_1$ (*R*; *A*) | 500 | 1.5 | 1.85 | 95 | 125 | [241] |
| $Fe_{71.5}Co_{8.25}Ni_{8.25}Zr_7B_4Cu_1$ (*R*; *A*) | 570 | 1.5 | 1.95 | 97 | 130 | [241] |
| $Fe_{66}Co_{11}Ni_{11}Zr_7B_4Cu_1$ (*R*; *A*) | 640 | 1.5 | 1.80 | 98 | 131 | [241] |
| $Fe_{88}Pr_6Ce_4B_2$ (*R*; *A*) | 284 | 5 | 4.15 | - | 725.8 | [242] |
| $Fe_{87}Zr_7B_4Co_2$ (*R*; *A*) | 333 | 5 | 3.42 | - | - | [243] |
| $Fe_{62}Mn_{18}P_{10}B_7C_3$ (*R*; *A*) | 222 | 1.5<br>2 | 0.57<br>0.71 | 48<br>67.2 | 64.57<br>87.68 | [240] |
| $Fe_{60}Co_{12}Gd_4Mo_3B_{21}$ (*R*; *A*) | 387 | 1 | 0.76 | - | - | [244] |
| ***Intermetallic compounds*** | | | | | | |
| $Nd_2Fe_{17}$ (*R*; *C*) | 326 | 1<br>2<br>3<br>4<br>5 | 1.4<br>2.5<br>3.3<br>4.1<br>4.8 | -<br>-<br>-<br>-<br>- | 73<br>169<br>271<br>382<br>496 | [156] |
| $Y_2Fe_{17}$ (*R*; *C*) | 301 | 1<br>2<br>5 | 1.5<br>2.4<br>4.4 | -<br>-<br>- | 75<br>178<br>533 | [173] |
| $Y_2Fe_{17}$ (*R*; *C*) | 305 | 10 | 1.89 | - | - | [174] |
| $Pr_2Fe_{17}$ (*R*; *C*) | 290 | 1<br>2<br>3<br>4<br>5 | 1.0<br>1.8<br>2.5<br>3.1<br>3.7 | -<br>-<br>-<br>-<br>- | 95<br>208<br>328<br>450<br>580 | [156] |
| $NdPrFe_{17}$ (*R*; *C*) | 303/332 | 2 | 2.1 | 175 | - | [245] |
| $Pr_{2-x}Nd_xFe_{17}$ (*R*; *C*)<br>$x = 0.5$<br>$x = 0.7$ | <br>302<br>307 | <br>5<br>5 | <br>3.01<br>4.31 | <br>-<br>- | <br>345<br>487 | [175] |
| $LaFe_{12}Si$ (*R*; *C*)<br>Annealed at 1323 K/2 h | 195 | 5 | 25.4<br>@ 201 K | - | - | [246]<br>(*) |
| $LaFe_{11.8}Si_{1.2}$ (*R*; *C*)<br>Annealed at 1323 K/2 h | 195 | 5 | 31<br>@ 201 K | - | - | [246]<br>(*) |
| $LaFe_{11.2}Si_{1.8}$ (*R*; *C*)<br>Annealed at 1323 K/2 h | 231 | 5 | 10.3<br>@ 240 K | - | - | [246]<br>(*) |
| $LaFe_{11.5}Si_{1.5}$ (*R*; *C*)<br>Annealed at 1273 K/0.033 h | 189 | 5 | 12 | - | - | [172]<br>(*) |
| $LaFe_{11.5}Si_{1.5}$ (*R*; *C*)<br>Annealed at 1273 K/2 h | 201 | 5 | 17 | - | - | [172]<br>(*) |
| $LaFe_{11.6}Si_{1.4}$ (*R*; *C*)<br>Annealed at 1373 K/24 h | 199 | 5 | 10.03 | - | - | [172]<br>(*) |
| $LaFe_{11.6}Si_{1.4}$ (*R*; *C*)<br>Annealed at 1323 K/0.5 h | 223 | 5 | 6.30 | - | - | [172] |
| $LaFe_{11.6}Si_{1.4}$ (*R*; *C*)<br>Annealed at 1323 K/4 h | 213 | 5 | 8.13 | - | - | [172] |
| $LaFe_{11.57}Si_{1.43}$ (*R*; *C*) | 210 | 5 | 21.2 | - | - | [247] |
| Annealed at 1323 K/2 h | | | | | | (*) |
| $LaFe_{11.57}Si_{1.43}$ (*R*; *C*)<br>Annealed at 1273 K/1 h (20 m/s) | 198 | 5 | 17.8 | - | - | [247]<br>(*) |
| $LaFe_{11.57}Si_{1.43}$ (*R*; *C*)<br>Annealed at 1273 K/1 h (40 m/s) | 210 | 5 | 193 | - | - | [247]<br>(*) |
| $LaFe_{11.6*1.1}Si_{1.4}$ (*R*; *C*)<br>Annealed at 1523 K/5 h | 190.5 | 5 | 17.2 | - | 146.2 | [247]<br>(*) |
| $LaFe_{11.6*1.2}Si_{1.4}$ (*R*; *C*)<br>Annealed at 1523 K/5 h | 177.4 | 5 | 13.2 | - | 105.6 | [247]<br>(*) |
| $La_{0.8}Ce_{0.2}Fe_{11.5}Si_{1.5}$ (*R*; *C*)<br>Annealed at 1273 K/10 mins<br>Annealed at 1273 K/15 mins<br>Annealed at 1273 K/20 mins<br>Annealed at 1273 K/30 mins<br>Annealed at 1273 K/60 mins | <br>193<br>188<br>183<br>184<br>183 | <br>1.5<br>1.5<br>1.5<br>1.5<br>1.5 | <br>9.7<br>23<br>33.8<br>31.4<br>32.8 | <br>-<br>-<br>-<br>-<br>- | <br>-<br>-<br>-<br>-<br>- | [248]<br>(*) |
| $La_{0.6}Pr_{0.5}Fe_{11.4}Si_{1.6}$ (*R*; *C*) | 192 | 5 | 21.9 | 458.5 | 481.8 | [249]<br>(*) |
| ***High entropy alloys (HEAs)*** | | | | | | |
| $Tm_{10}Ho_{20}Gd_{20}Ni_{20}Al_{20}$ (*R*; *A*) | 30.3 | 3<br>7 | 5.6<br>12.7 | 223.4<br>637.4 | 282.9<br>793.5 | [176] |
| $Gd_{20}Dy_{20}Er_{20}Co_{20}Al_{20}$ (*R*; *A*) | 42 | 5 | 7.7 | 523 | - | [155] |

*R: Ribbon; A: Amorphous; C: Crystalline; B: Bulk; (*) represents FOMT materials.*

**Table 4.** Maximum entropy change, $\left|\Delta S_M^{\max}\right|$, Curie temperature, $T_C$, refrigerant capacity (*RC*), and relative cooling power (*RCP*) for the microwire samples. Values from microwires of other compositions, bulk glasses and Gd are included for comparison.

| **Microwires** | $T_C$ **(K)** | $\mu_0\Delta H$ **(T)** | $\left\|\Delta S_M^{\max}\right\|$ **(J/kg K)** | ***RC*** **(J/kg)** | ***RCP*** **(J/kg)** | **Ref.** |
|---|---|---|---|---|---|---|
| ***Gadolinium and its alloys*** | | | | | | |
| Gd (*B*) | 294 | 5 | 10.2 | 410 | - | [7] |
| $Gd_{55}Co_{20}Al_{25}$ (*B*) | 103 | 5 | 8.8 | 541 | - | [263] |
| $Gd_{55}Al_{20}Co_{25}$ (*MW; A*) | 110 | 5 | 9.69 | 580 | 804 | [250] |
| $Gd_{55}Co_{20}Al_{25}$ (*MW; A+C*) | 100 | 5 | 10.1 | 653 | 870 | [31] |
| $Gd_{50}Co_{20}Al_{30}$ (*MW; A+C*) | 86 | 5 | 10.1 | 672 | 896 | [31] |
| $Gd_{60}Co_{20}Al_{20}$ (*MW; A+C*) | 109 | 5 | 10.1 | 681 | 908 | [31] |
| $Gd_{55}Co_{30}Al_{15}$ (*MW; A*) | 127 | 5 | 9.71 | 573 | 702 | [264] |
| $Gd_{60}Co_{15}Al_{25}$ (*MW; A*) | 100 | 5 | 9.73 | 732 | 976 | [265] |
| $Gd_{60}Co_{25}Al_{15}$ (*R; A*) | 125 | 5 | 10.1 | 645 | 860 | [161] |
| $Gd_{60}Al_{20}Co_{20}$ (*MW; A+C*) | 113 | 5 | 10.12 | 698 | 936 | [252] |
| $Gd_{60}Fe_{20}Al_{20}$ (*MW; A*) | 202 | 5 | 4.8 | 687 | 900 | [32] |
| $Gd_{53}Al_{24}Co_{20}Zr_3$ *B; A* | 95<br>95 | 5<br>3 | 9.6<br>6.2 | 690<br>340 | -<br>- | [30] |
| $Gd_{53}Al_{24}Co_{20}Zr_3$ (*MW; A*) | 94<br>94 | 5<br>3 | 10.3<br>6.9 | 733<br>420 | -<br>- | [30] |
| $Gd_{53}Al_{24}Co_{20}Zr_3$ (*SW; A*) | 100 | 3 | 5.32 | 467 | 555 | [266] |
| $Gd_{53}Al_{24}Co_{20}Zr_3$ (*SW; A*) | 94<br>94 | 5<br>2 | 8.8<br>4.3 | 600<br>220 | 774<br>296 | [251] |
| $Gd_{53}Al_{24}Co_{20}Zr_3$<br>*SW; C; Annealed at 100 °C* | 94<br>94 | 5<br>2 | 9.5<br>4.7 | 687<br>285 | 893<br>348 | [251] |
| $Gd_{53}Al_{24}Co_{20}Zr_3$<br>*SW; C; Annealed at 200 °C* | 93<br>93 | 5<br>2 | 8.0<br>3.8 | 629<br>243 | 744<br>307 | [251] |
| $Gd_{53}Al_{24}Co_{20}Zr_3$<br>*SW; C; Annealed at 300 °C* | 92<br>92 | 5<br>2 | 5.1<br>2.4 | 396<br>144 | 525<br>184 | [251] |
| $Gd_{55}Co_{25}Ni_{20}$ (*B*) | 78 | 5 | 8.0 | 640 | - | [263] |
| $Gd_{55}Co_{30}Ni_5Al_{10}$<br>(*MW; A*) | 140 | 5 | 8.91 | 532 | 668 | [264] |
| $Gd_{55}Co_{30}Ni_{10}Al_5$<br>(*MW; A*) | 158 | 5 | 7.68 | 523 | 653 | [264] |
| $Gd_{55}Co_{20+x}Ni_{10}Al_{15-x}$ (*MW; A*)<br>$x = 10$<br>$x = 5$<br>$x = 0$ | <br>158<br>128<br>113 | <br>5<br>5<br>5 | <br>7.68<br>9.00<br>9.67 | <br>546.3<br>548.9<br>609.5 | <br>681.0<br>675.0<br>749.5 | [254] |
| $Gd_{73.5}Si_{13}B_{13.5}/GdB_6$<br>(*MW; A+C*) | 106 | 5 | 6.4 | 790 | 885 | [35] |
| $Gd_3Ni/Gd_{65}Ni_{35}$<br>(*MW; A+C*) | 120 | 5 | 9.64 | 742 | - | [253] |
| $Gd_{50}$-$(Co_{69.25}Fe_{4.25}Si_{13}B_{13.5})_{50}$<br>(*MW; A*) | 170 | 5 | 6.56 | 625 | 826 | [267] |
| $Gd_{59.4}Al_{19.8}Co_{19.8}Fe_1$<br>(*MW; A*) | 113 | 5 | 10.33 | 748 | 1006 | [268] |
| $(Gd_{60}Al_{20}Co_{20})_{99}Ni_1$<br>(*MW; A+C*) | 111 | 5 | 10.98 | 725.49 | 970.89 | [252] |
| $(Gd_{60}Al_{20}Co_{20})_{97}Ni_3$<br>(*MW; A+C*) | 109 | 5 | 11.06 | 746.84 | 1000.50 | [252] |
| $(Gd_{60}Al_{20}Co_{20})_{95}Ni_5$ | 109 | 5 | 11.57 | 834.14 | 1138.16 | [252] |
| (*MW; A+C*) | | | | | | |
| $(Gd_{60}Al_{20}Co_{20})_{93}Ni_{7}$ (*MW; A+C*) | 108 | 5 | 10.77 | 733.48 | 977.65 | [252] |
| $Gd_{36}Tb_{20}Co_{20}Al_{24}$ (*MW; A*) | 91 | 5 | 12.36 | 731 | 948 | [269] |
| $Gd_{36}Tb_{20}Co_{20}Al_{24}$ (*MW; A+C*) | 81 | 5 | 8.8 | 500 | 625 | [269] |
| $(Gd_{36}Tb_{20}Co_{20}Al_{24})_{99}Fe_{1}$ (*MW; A+C*) | 94 | 5 | 8.5 | 510 | 635 | [269] |
| $(Gd_{36}Tb_{20}Co_{20}Al_{24})_{98}Fe_{2}$ (*MW; A+C*) | 100 | 5 | 8.0 | 515 | 660 | [269] |
| $(Gd_{36}Tb_{20}Co_{20}Al_{24})_{97}Fe_{3}$ (*MW; A+C*) | 108 | 5 | 7.6 | 520 | 680 | [269] |
| $Gd_{19}Tb_{19}Er_{18}Fe_{19}Al_{25}$ (*MW; A+C*) | 97 | 5 | 5.94 | 569 | 733 | [270] |
| $Gd_{36}Tb_{20}Co_{20}Al_{24}$ (*MW; A+C*) | 82 | 5 | 9 | 518 | 657 | [271] |
| ***Intermetallics compounds*** | | | | | | |
| HoErCo (*MW; A*) | 16 | 5 | 15 | 527 | 600 | [260] |
| HoErFe (*MW; A+C*) | 44 | 5 | 9.5 | 450 | 588 | [261] |
| DyHoCo (*MW; A*) | 35 | 5 | 11.2 | 417 | 530 | [262] |
| $Mn_xFe_{2-x}P_{0.5}Si_{0.5}$ (*M; C*) | | | | | | [256] |
| $x = 0.7$ | >400 | 5 | - | - | - | |
| $x = 0.8$ | 351 | 5 | 12 | 293.7 | - | |
| $x = 0.9$ | 298.5 | 5 | 18.3 | 331.1 | - | |
| $x = 1.0$ | 263 | 5 | 15.8 | 300 | - | |
| $x = 1.1$ | 235.5 | 5 | 10.8 | 280.9 | - | |
| $x = 1.2$ | 190 | 5 | 1.2 | 288.4 | - | |
| $MnFe_xP_{0.5}Si_{0.5}$ (*M; C*) | | | | | | [258] |
| $x = 0.9$<br>$x = 0.95$<br>$x = 1.0$<br>$x = 1.05$ | 311<br>281<br>263<br>245.5 | 5<br>5<br>5<br>5 | 10.7<br>10.3<br>15.8<br>14.5 | 295.8<br>286.2<br>300.0<br>283.9 | 293.7<br>275.3<br>257.1<br>243.1 | (*) |
| $(MnFe)_x(P_{0.5}Si_{0.5})$ (*W*, *C*)<br>$x = 1.85$<br>$x = 1.90$<br>$x = 1.95$<br>$x = 2.00$ | <br>355<br>370<br>340<br>263 | <br>5<br>5<br>5<br>5 | <br>16.3<br>26.0<br>19.4<br>15.8 | <br>~308<br>~367<br>~325<br>~295 | | [257]<br>(*) |
| $Mn_{1.3}Fe_{0.6}P_{0.5}Si_{0.5}$, as-cast<br>(*M; C*)<br>$Mn_{1.3}Fe_{0.6}P_{0.5}Si_{0.5}$, annealed<br>(*M; C*) | 138<br><br>145 | 2<br>5<br>2<br>5 | 1.9<br>4.6<br>5.1<br>10.5 | 160<br>-<br>178<br>440 | -<br>-<br>-<br>- | [272] |
| $Mn_{1.26}Fe_{0.60}P_{0.48}Si_{0.52}$ (*MW; C*) | 141 | 5 | 4.64 | - | - | [272] |
| $Dy_{36}Tb_{20}Co_{20}Al_{24}$ (*MW; A+C*) | 42 | 5 | 8.2 | 301 | 414 | [271] |
| $Ho_{36}Tb_{20}Co_{20}Al_{24}$ (*MW; A+C*) | 42 | 5 | 10.3 | 372 | 474 | [271] |
| $LaFe_{11.6}Si_{1.4}$ (*MW; A*) | 195 | 2 | 9.0 | - | 45 | [273] |
| ***Heusler alloys*** | | | | | | |
| $Ni_2MnGa$ (*GCW; C; Annealed)* | 315 | 3 | 0.7 | - | - | [274] |
| $Ni_{50.5}Mn_{29.5}Ga_{20}$ (*MW; C*) | 368 | 5 | 18.5 | 63 | - | [275]<br>(*) |
| $Ni_{50.6}Mn_{28}Ga_{21.4}$ (*MW; C*) | 340-370 | 5 | 5.2 | 240 | - | [275] |
| $Ni_{48}Mn_{26}Ga_{19.5}Fe_{6.5}$ (*MW; C*) | 361 | 5 | 4.7 | - | - | [276] |
| $Ni_{49.4}Mn_{26.1}Ga_{20.8}Cu_{3.7}$<br>(*MW; C*) | 359 | 5 | 8.3 | 78 | - | [277]<br>(*) |
| $Ni_{45.6}Fe_{3.6}Mn_{38.4}Sn_{12.4}$<br>(*MW; A+C*) | 270<br>300 | 5<br>5 | 15.2<br>4.3 | 146<br>175 | 182<br>215 | [278]<br>(*) |
| $Ni_{48}Mn_{25.6}Ga_{19.4}Fe_{6.5}$ (*MW; A*) | 361 | 5 | 4.7 | - | 18 | [279] |
| $Ni_{48.5}Mn_{26}Ga_{19.5}Fe_{6.5}$ (*MW; C*) | 391 | 5 | 2.91 | - | - | [276] |
| $Ni_{44.9}Fe_{4.3}Mn_{38.3}Sn_{12.5}$ (*MW; C; Annealed*) | 299<br>FOMT | 5<br>5 | 3.7<br>6.9 | ~233<br>78 | -<br>- | [280]<br>(*) |
| $Ni_{45}Mn_{37}In_{13}Co_{5}$ (*GCW; C; Annealed*) | 315 | 5 | 0.5 | - | - | [281] |
| $Ni_{50.95}Mn_{25.45}Ga_{23.6}$ (*GCW; C; Annealed*) | 315 | 3 | 0.7 | - | - | [282] |

*SW: Single wire; MW: Multiple wires; B: Bulk; R: Ribbon; GCW: Glass-coated wires*
*A: Amorphous; C: Crystalline; M: Microwires; (*) represents FOMT materials.*

**Table 5.** Material candidates for energy-efficient magnetic refrigeration applications in the three cooling temperature regimes.

| **Magnetic Cooling Applications** | | |
|---|---|---|
| **Low-Temperature Range (Cryogenic Cooling)** | **Intermediate-Temperature Range** | **High-Temperature Range** |
| $T < 80$ K | $80 \text{ K} < T < 300$ K | $T > 300$ K |
| **Applications:**<br>• Liquefaction of hydrogen and helium<br>• Cryogenics for space technology, superconducting magnets, quantum devices, and sensors | **Applications:**<br>• Near-room-temperature cooling<br>• Biomedical devices (e.g., magnetic hyperthermia)<br>• Electronic component cooling | **Applications:**<br>• Industrial waste heat recovery<br>• Thermomagnetic energy conversion<br>• Magnetic hyperthermia |
| **Material candidates:**<br>• Oxide nanoparticles (e.g., $GdVO_4$, $Gd_2O_3$, $Ho_2O_3$, $Dy_2O_3$, $Gd_3Ga_5O_{12}$)<br>• Intermetallic alloy nanoparticles (e.g., $MnFeP_{0.45}Si_{0.05}$, $Ni_{95}Cr_5$, $MnPS_3$, $GdNi_5$)<br>• Oxide films (e.g., $GdCoO_3$, $EuTiO_3$)<br>• $MnF_2$/FM films<br>• Rare-earth based ribbons (e.g., Gd-Ni-Al, *R*-$Ni_2$, *R*-Al-Ni)<br>• Rare-earth-based microwires (e.g., DyHoCo, HoErCo) | **Material candidates:**<br>• Gd nanoparticles, films and ribbons — $T_C$ ~294 K<br>• Gd-based films ($GdSi_2$, $Gd_5Si_{1.3}Ge_{2.7}$)<br>• Oxide films (e.g., EuO, $Gd_2NiMnO_6$, $La_{0.7}Ca_{0.3}MnO_3$)<br>• Fe-Rh-Pd films<br>• Gd-based ribbons (Gd-Co, Gd-Mn, Gd-Co-*X*)<br>• La-Fe-Si-based ribbons and microwires<br>• Heusler alloy ribbons and microwires (Ni-Mn-*X*) | **Material candidates:**<br>• Manganite nanoparticles and films (e.g., $La_{0.67}Sr_{0.33}MnO_3$, $La_{0.7}Ca_{0.1}Sr_{0.2}MnO_3$)<br>• $CrO_2$/$TiO_2$ films<br>• FeRh and FeRh/$BaTiO_3$ films<br>• Heusler alloy ribbons and microwires (e.g., Ni-Mn-Gd, Ni-Mn-In, Ni-Co-Mn-In, Ni-Co-Mn-Sn, Ni-Mn-Ga-Cu)<br>• *X*-Fe alloy ribbons<br>• Mn-Fe-P-Si microwires |

**Table 6.** The main advantages and key challenges in low-dimensional magnetocaloric materials.

| Form | Advantages | Key Challenges | Material & Engineering Constraints |
|---|---|---|---|
| Nanoparticles | - Suitable for cryogenic & localized cooling<br>- Entropy broadening can improve refrigerant capacity | - Reduced Curie Temperature ($T_C$) and $\Delta S_M$<br>- Require high magnetic fields for AFM types (3–7 T)<br>- Low thermal conductivity & high interfacial resistance<br>- Agglomeration, oxidation, degradation under cycling | - Complex integration into matrices<br>- Thermal/magnetic insulation from binders or coatings<br>- Difficult scalable synthesis with consistent quality<br>- Sensitive to stoichiometry, oxidation<br>- Need protective coatings (e.g., $SiO_2$) that may reduce performance |
| Thin Films | - On-chip & microcooling potential<br>- Integration with other effects (e.g., thermoelectric)<br>- Potential for large $\Delta S_M$ in strained AFM/FM systems | - Reduced $T_C$ and $\Delta S_M$ due to finite-size effects, strain<br>- Low thermal mass and conductivity<br>- $\Delta S_M$ degradation with cycling (e.g., $Gd_5Si_2Ge_2$)<br>- Hysteresis losses in FOMT materials | - Deposition-related defects (grain boundaries, off-stoichiometry)<br>- Limited materials with high film quality<br>- Difficulty maintaining magnetic order during deposition<br>- Volume constraints limit cooling power<br>- Multilayer stacking adds complexity |
| Ribbons | - High surface area<br>- Fast thermal response<br>- Flexible (to some extent) | - Reduced $\Delta S_M$ due to disorder and texture<br>- Mechanical brittleness<br>- Low thermal conductivity<br>- Small volume limits cooling capacity<br>- Hysteresis losses in FOMT materials | - Prone to oxidation<br>- Structural/ compositional uniformity hard to control<br>- Engineering integration (alignment, thermal coupling) is complex |
| Microwires | - High surface-to-volume ratio<br>- Mechanical flexibility<br>- Fast heat exchange | - Fragility under thermal/magnetic cycling<br>- Limited materials can be processed as microwires | - Challenging to fabricate uniform wires<br>- Need dense bundles for sufficient cooling<br>- Poor thermal coupling in bundles/matrices |
| | | - Oxidation/ degradation over time | - Alignment and support design remains unresolved |